\documentclass[%
 aip,
 jmp,%
 amsmath,amssymb,
 preprint,%
]{revtex4-1}

\usepackage[latin1]{inputenc}
\usepackage{graphicx}
\usepackage{amsmath}
\usepackage{amssymb}
\usepackage{color}
\usepackage{bm}
\usepackage{epsfig}
\usepackage{tikz}
\usepackage{setspace}
\begin{document}

\preprint{APS/123-QED}

\title{Transient growth in Rayleigh-B\'{e}nard-Poiseuille/Couette convection}% Force line breaks with \\

\author{J John Soundar Jerome}
	\email{jon@ladhyx.polytechnique.fr}
\author{Jean-Marc Chomaz}
\author{Patrick Huerre}
 
\affiliation{%
 Laboratoire d'Hydrodynamique (LadHyX)\\
 CNRS - \'Ecole Polytechnique, F-91128 Palaiseau cedex, France
}%


%\affiliation{
%\affiliation{
%\affiliation{%

\date{}% It is always \today, today,

\begin{abstract}
An investigation of the effect of a destabilizing cross-stream temperature gradient on the transient growth phenomenon of plane Poiseuille flow and plane Couette flow is presented. Only the streamwise-uniform and nearly-streamwise-uniform disturbances are highly influenced by the Rayleigh number $Ra$ and Prandtl number $Pr$. The maximum optimal transient growth $G_{max}$ of streamwise-uniform disturbances increases slowly with increasing $Ra$ and decreasing $Pr$. For all $Ra$ and $Pr$, at moderately large Reynolds numbers $Re$, the supremum of $G_{max}$ is always attained for streamwise-uniform perturbations (or nearly-streamwise-uniform perturbations, in the case of plane Couette flow) which produce large streamwise streaks and Rayleigh-B\'{e}nard convection rolls ($RB$). The optimal growth curves retain the same large-Reynolds-number scaling as in pure shear flow. A 3D vector model of the governing equations demonstrates that the \textit{short-time} behavior is governed by the classical lift-up mechanism and that the influence of $Ra$ on this mechanism is secondary and negligible. The optimal input for the largest \textit{long-time} response is given by the adjoint of the dominant eigenmode with respect to the energy scalar product: the $RB$ eigenmode without its streamwise velocity component.  These short and long-time responses depict, to leading order, the optimal transient growth $G(t)$. At moderately large $Ra$ (or small $Pr$ at a fixed $Ra$), the dominant adjoint mode is a good approximation to the optimal initial condition for all time. Over a general class of norms that can be considered as growth functions, the results remain qualitatively similar: for example, the dominant adjoint eigenmode still approximates the maximum optimal response.

%\begin{description}
%\end{description}
\end{abstract}

%Use showkeys class option if keyword
%                              %display desired
\maketitle

%\tableofcontents
\section {Introduction}
\label{sec:Intro}

The purpose of this study is to examine transient growth in plane Poiseuille flow and plane Couette flow in the presence of an unstable temperature gradient, under the Boussinesq approximation. The various mechanisms of transient growth in such flows are identified and characterized.

It is well-known that a stationary fluid layer heated from below becomes unstable when buoyancy forces overcome the dissipative forces due to thermal and viscous diffusion. Similar to this problem of natural convection, a secondary motion in the form of streamwise vortex rolls can be set-up in a shear flow with cross-stream temperature gradient. Historically, the linear stability problem was motivated by the observation of cloud streets \cite{Kuettner_1971}. The convection rolls in the atmospheric boundary layer tend to align in the direction of the flow. The moisture in the up-flowing warm air of these rolls condenses to form clouds that are aligned in the streamwise direction, thereby leading to the formation of cloud streets \cite{Kuettner_1971}. This type of fluid motion is commonly encountered in various forms in geophysical flows, heat exchangers, electroplating, chemical vapour deposition, etc. Thus, the linear stability analysis of a horizontal fluid layer heated from below in the presence of laminar shear flow is of fundamental interest.

In general, the onset of transition in fluid flows is via exponential or algebraic growth of disturbances \cite{Schmid_n_Henningson_2001}. Mathematically, an exponentially growing instability can be identified using a normal-mode analysis (what is known as modal stability theory). The formation of Rayleigh-B\'{e}nard convection patterns in a fluid layer heated from below, and the occurrence of Taylor vortices in Taylor-Couette flow, are a few classic examples of transition which occur via an exponential instability that is saturated by non-linear processes to a secondary flow. However, transition to turbulence in wall-bounded flows can occur at Reynolds numbers much smaller than the critical Reynolds number predicted by such modal stability analyses. Experiments in boundary layer flows \cite{Klebanoff_1971, Kendall_1985, Matsubara_and_Alfredsson_2001} show that transition is usually preceded by the presence of streamwise motion in the form of streaks and not via Tollmien-Schlichting waves as predicted by modal stability analysis. The onset of such a transition process is due to the fact that any disturbance, which is otherwise exponentially stable, has the potential to become sufficiently large before eventually decaying. In theory it is attributed to the non-normal nature of the Orr-Sommerfeld and Squire equations \cite{Butler_n_Farrel_1992, Reddy_n_Henningson_1993}. Even though each eigenfunction may decay at its own rate (related to its eigenvalue), a superposition of non-orthogonal eigenfunctions may produce large transient growth before eventually decreasing at the rate of the least decaying eigenfunction. Physically, the source of transient growth of disturbances is related to the inviscid vortex tilting process in the presence of base flow shear whereby a disturbance can feed on the base flow kinetic energy for a short time. The lift-up mechanism$\cite{Butler_n_Farrel_1992, Ellingsen_n_Palm_1975, Landhal_1990}$ and the Orr mechanism$\cite{Orr_1907}$ are two such commonly identified growth phenomena in a shear flow. It is reasonable to assume that the presence of a cross-stream temperature gradient in the base flow would influence this transient growth. If so, what are the dominant physical mechanisms of transient growth in such flows? Is lift-up dominant at all Rayleigh and Prandtl numbers? It is the aim of the present work to examine thoroughly the influence of buoyancy induced by a constant cross-stream temperature gradient on the transient growth phenomenon.

%%For the sake of brevity, we restrict the discussion to RBP only. However, the results of plane Couette flow will be mentioned insofar as it is necessary.

The earliest known stability analysis of plane Poiseuille flow with unstable thermal stratification in a Boussinesq fluid, hereafter referred to as Rayleigh-B\'{e}nard-Poiseuille flow ($RBP$), is due to Gage and Reid$\cite{Gage_n_Reid_1968}$. If $Re_c^{TS} (\approx 5772.2)$ is the Reynolds number (based on the channel half-width) at which Tollmien-Schlichting waves ($TS$) become unstable in plane Poiseuille flow without temperature effects, Gage and Reid showed that for all Reynolds numbers less than a critical value, approximately equal to $Re_c^{TS}$, the dominant eigenmode of $RBP$ is in the form of streamwise-uniform convection rolls due to the Rayleigh-B\'{e}nard instability ($RBI$) above a critical Rayleigh number $Ra_c^{RB} = 1707.78$ (based on the channel width). This value is independent of both Reynolds number and Prandtl number. It was concluded that the effect of a shear flow on the linear stability of a fluid subjected to unstable cross-stream temperature gradient is only to align the rolls along the streamwise direction. Furthermore, the effect of the cross-stream temperature gradient on the Tollmien-Schlichting instability ($TSI$) is negligible for all $Ra < Ra_c^{RB}$: the critical Reynolds number for the onset of $TS$ waves in $RBP$ remains very close to $Re_c^{TS} \approx 5772.2$. The reader is referred to Kelly$\cite{Kelly_1994}$ for a comprehensive review of the major results on the onset and development of thermal convection in $RBP$.

The study of convective-absolute transition in $RBP$ flows was considered by, among others, M\"uller, L\"ucke and Kamps$\cite{Muller_et_al_1992}$ and Carrière and Monkewitz$\cite{Carriere_n_Monkewitz_1999}$. It was based on the long-time behavior of an impulse response wave packet. The base flow is unstable if the wave packet grows exponentially. Furthermore, if a reference frame is singled out by boundary conditions (for example, the reference frame fixed with the stationary wall of Poiseuille flow), the instability is termed as either absolute or convective if the wave packet grows in the same location as the applied impulse or is advected away by the base flow, respectively. The computations showed that it is always the transverse rolls (spanwise-uniform convection rolls) which have the highest absolute growth rate and it is that configuration which should appear at the source location. Carrière et. al.$\cite{Carriere_n_Monkewitz_1999}$ established that the system remains convectively unstable with respect to streamwise-uniform convection rolls and absolutely unstable with respect to transverse convection rolls for all non-zero Reynolds numbers.

More recently, Biau and Bottaro$\cite{Biau_n_Bottaro_2004}$ investigated the effect of \textit{stable} thermal stratification, solely induced by buoyancy, on the spatial transient growth of energy in $RBP$ flow. The analysis showed that the presence of stable stratification reduces the optimal transient growth of perturbations. Perhaps the most akin to the present work is the article by Sameen and Govindarajan$\cite{Sameen_n_Govindarajan_2007}$ who studied the effect of heat addition on the transient growth and secondary instability in channel flow. Here, the effect of heating may be split into three components: the first one is due to the generation of buoyancy forces as in the classical Rayleigh-B\'{e}nard convection problem, the second one is associated with the temperature-dependent base flow viscosity, and the third one results from viscosity variations induced by temperature perturbations. The computations revealed that heat addition gives rise to very large optimal growth. For various control parameter settings, it was demonstrated that viscosity stratification had a very small effect on transient growth. At moderately large Reynolds number ($= 1000$), the optimal disturbances could be either streamwise-uniform vortices (as in pure shear flows) or spanwise-uniform vortices, largely depending on Prandtl number and Grashof number. However, the transient growth mechanisms related to such optimal initial disturbances, and their corresponding response were not examined. Finally, cross-stream viscosity stratification was determined to have a destabilizing influence on the secondary instability of $TS$ waves.

The linear stability characteristics of plane Couette flow with unstable thermal stratification in a Boussinesq fluid, hereafter referred to as Rayleigh-B\'{e}nard-Couette flow ($RBC$), were first computed by Gallagher and Mercer$\cite{Gallagher_n_Mercer_1965}$. As in $RBP$ flow, the dominant eigenmode at all Reynolds numbers is in the form of streamwise-uniform convectional rolls due to RBI.

Clever, Busse and Kelly$\cite{Clever_Busse_Kelly_1977}$ studied the secondary instability of the streamwise-uniform rolls in $RBC$ in an effort to understand the onset of waviness in the rolls and to relate them to the formation of cloud streets in the lower atmosphere. The secondary instabilities of the convection rolls were determined to occur as stationary waves or simply as waves that propagate along the rolls. Clever and Busse$\cite{Clever_Busse_1992}$ later considered the three-dimensional flows arising from these distortions and their stability. They computed the finite-amplitude solutions that evolve from the wavy instability, even at vanishing or negative values of the Rayleigh number.

A comprehensive study of the transient growth in plane Couette flow with cross-stream temperature gradient was performed by Malik, Dey and Alam$\cite{Malik_Dey_n_Alam_2008}$ in the context of a compressible fluid. The optimal energy growth was determined to be strongly impaired by the presence of viscosity stratification in such flows.

In the light of the previous works, the objective of the present investigation is to provide a comprehensive understanding of the effect of buoyancy alone on the transient growth in $RBP$ and $RBC$ flows. Since viscosity stratification was observed to be ineffective for the transient growth in $RBP$ (Sameen and Govindarajan$\cite{Sameen_n_Govindarajan_2007}$), this effect will not be taken into account. A thorough treatment of the non-modal growth in $RBP$ and $RBC$ flows will be given, as a function of the main control parameters, namely, the Reynolds number, Rayleigh number and Prandtl number.

The paper is organized in the following way. Section \ref{sec:Base} describes the base flow configuration and formulates the modal stability analysis. Section \ref{sec:MS} reviews and presents the linear stability characteristics of various exponentially-growing eigenmodes. In section \ref{sec:NMS}, the non-modal stability analysis is introduced and the corresponding results are presented. The dominant transient growth processes are discussed in section \ref{sec:TG_longi}. The issues pertaining to the choice of the norm and to the effect of Prandtl number are also presented in the same section. A brief summary of results and conclusions is given in section \ref{sec:Rem_n_Sum}.

\section{Base Flow and Governing Equations}
\label{sec:Base}
\begin{figure}
\centering
\includegraphics[width=0.6\textwidth,keepaspectratio, trim=-20 0 0 0]{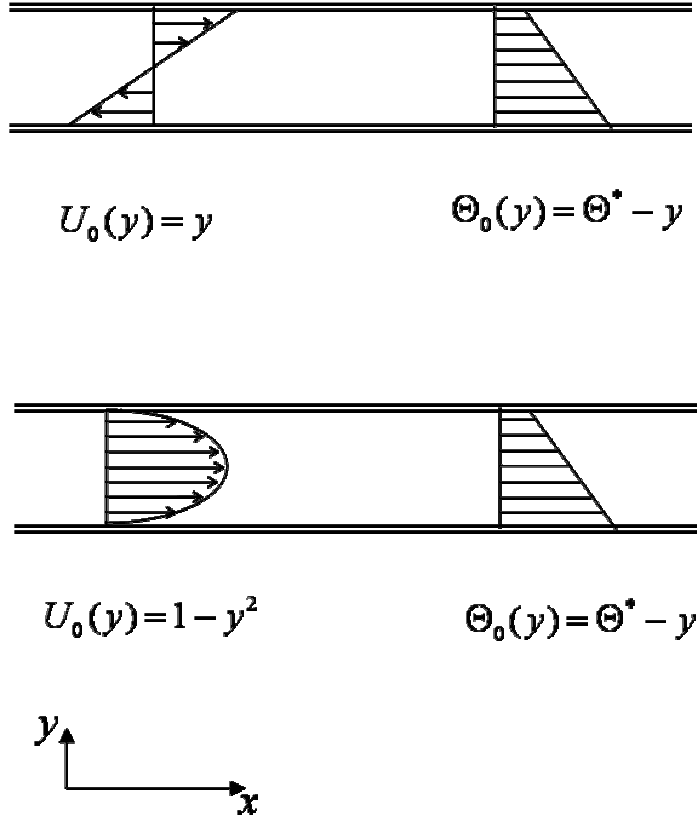}
\caption{Schematic view of Rayleigh-B\'{e}nard-Couette (top) and Rayleigh-B\'{e}nard-Poiseuille (bottom) flows.}
\label{fig:schematic_basic_flow}
\end{figure}

Consider the following normalized velocity and temperature profiles (figure \ref{fig:schematic_basic_flow}) of $RBP$ and $RBC$ flows
\begin{eqnarray}
U_0(y) &= 
	\begin{cases}
	  1-y^2,  & \mbox{plane Poiseuille flow}\\
	 y, & \mbox{plane Couette flow} 
	\end{cases},
\end{eqnarray}
\begin{eqnarray}
\Theta_0(y) &= \Theta^* - y,
\label{eq:basicflow}
\end{eqnarray}
and the non-dimensional parameters
\begin{align}
Re = \frac{U_{max}\frac{h}{2}}{\nu^*},
\mbox{\hspace{10pt}} Ra = \frac{\alpha^* g\Delta T h^3}{\nu^* \kappa^*},
\mbox{\hspace{10pt}} Pr = \frac{\nu^*}{\kappa^*},
\label{eq:NDnumbers}
\end{align}
where $U_{max}$ is the maximum velocity of the base flow, $\Delta T$ is the difference in temperature between the lower and upper wall, $\nu^*$ is the kinematic viscosity, $\kappa^*$ is the thermal diffusion coefficient, $\alpha^*$ is the thermal expansion coefficient, $g$ is the acceleration due to gravity, and $h$ is the channel width. 
Here, the spatial coordinates have been non-dimensionalized with the half-width of the channel. The base flow velocity profiles have been normalized with respect to $U_{max}$, the temperature of the base flow has been normalized with $\frac{\Delta T}{2}$ and $\Theta^*$ is the average non-dimensional temperature of the channel. The base flow profiles in eqn. $\eqref{eq:basicflow}$, namely, plane Poiseuille flow (or plane Couette flow) and the constant temperature gradient are solutions of the Oberbeck-Boussinesq equations.

The governing equations of the disturbance field can be obtained by linearizing the Oberbeck-Boussinesq equations about the base flow. If $\vec{u}(\vec{x},t) = [u, v, w]^T$ and $\theta (\vec{x},t)$ are the disturbance velocity  and temperature field, respectively, they read
\begin{eqnarray}
\nabla\cdot\vec{u}& =0,
\label{eq:LOBE1}
\end{eqnarray}
\begin{eqnarray}
\left(\frac{\partial}{\partial t}+ Re Pr U_0 \frac{\partial}{\partial x} \right)\vec{u} + Re Pr v \frac{dU_0}{dy}\vec{e}_x = -\nabla p + Pr \nabla^2 \vec{u} + Ra_{\mbox{\tiny{\textsl{h/2}}}} Pr \theta \vec{e}_y,
\label{eq:LOBE2}
\end{eqnarray}
\begin{eqnarray}
\left(\frac{\partial}{\partial t}+ Re Pr U_0 \frac{\partial}{\partial x} \right)\theta + v \frac{d \Theta_0}{dy} &= \nabla^2 \theta,
\label{eq:LOBE3}
\end{eqnarray}
where  $Ra_{\mbox{\tiny{\textsl{h/2}}}} = {Ra}/{16}$ is the Rayleigh number based on the half-width of the channel. Here, the thermal diffusive time scale $\frac{(h/2)^2}{\kappa^*}$ has been chosen to non-dimensionalize time, $\vec{u}(\vec{x},t)$ has been scaled with respect to ($\frac{\kappa^*}{h/2}$) while the base flow velocity $U_0(y)$ has been scaled with respect to $U_{max}$. These equations are solved for homogeneous Dirichlet boundary conditions on $\vec{u}(\vec{x},t)$ and $\theta (\vec{x},t)$ at $y = \pm 1$.

The base flow is homogeneous in $x$ and $z$, and hence the perturbation field may be decomposed into independent wave modes,
%\begin{align}
%\vec{u}(\vec{x},t) & = \vec{\tilde{u}}(y, t) \mbox{ exp}{\left[i \left( \alpha x + \beta z \right)\right]} \ +\ c.c.,
%\label{eq:mode1}\\
%\theta (\vec{x},t) & = \tilde{\theta}(y, t) \mbox{ exp}{\left[i \left( \alpha x + \beta z \right)\right]} \ +\ c.c.,
%\label{eq:mode2}
%\end{align}
\begin{align}
\vec{u}(\vec{x},t) & = \vec{\tilde{u}}(y, t) e^{i \left( \alpha x + \beta z \right)} \ +\ c.c.,
\label{eq:mode1}\\
\theta (\vec{x},t) & = \tilde{\theta}(y, t) e^{i \left( \alpha x + \beta z \right)} \ +\ c.c.,
\label{eq:mode2}
\end{align}
where $\alpha$, $\beta$ are the streamwise and spanwise wavenumber, respectively and $c.c.$ stands for the complex conjugate of the preceding expression. Note that this formulation is general and does not assume exponential behavior in time for the state variables.
%%\begin{figure*}
%%\includegraphics{fig_2}% Here is how to import EPS art
%%\caption{\label{fig:wide}Use the figure* environment to get a wide
%%figure that spans the page in \texttt{twocolumn} formatting.}
%%\end{figure*}

\section{Modal Stability Analysis}
\label{sec:MS}
%%by taking $\begin{bmatrix}  \tilde{v},  \tilde{\eta},   \tilde{\theta} \end{bmatrix}^T =\begin{bmatrix}  \hat{v}(y),  \hat{\eta}(y),   \hat{\theta}(y)     \end{bmatrix}^T exp(-i\omega t)$
The modal temporal problem considers disturbances that grow or decay exponentially. If the amplitudes $\vec{\tilde{u}}(y, t)$ and $\tilde{\theta}(y, t)$ in \eqref{eq:mode1} and \eqref{eq:mode2} are taken as
\begin{align}
\vec{\tilde{u}}(y, t) & = \vec{\hat{u}}(y) e^{-i\omega t},
\label{eq:normmode1}\\
\vec{\tilde{\theta}}(y, t) & = \hat{\theta}(y) e^{-i\omega t},
\label{eq:normmode2}
\end{align}
the equations \eqref{eq:LOBE1}, \eqref{eq:LOBE2} and \eqref{eq:LOBE3}, along with the boundary conditions, define a generalized eigenvalue problem for the complex frequency $\omega$ at a given $\left( \alpha, \beta \right)$, $Re$, $Ra$ and $Pr$. Following the classical parallel shear flow analysis, equations \eqref{eq:LOBE1}, \eqref{eq:LOBE2} and \eqref{eq:LOBE3} may be rewritten in terms of reduced variables, namely, the wall-normal velocity $\hat{v}(y)$, wall-normal vorticity $\hat{\eta}(y)$ and temperature perturbations $\hat{\theta}(y)$, leading to the system
%%\begin{align}
%%\left[ \left( \frac{\partial}{\partial t} + i \alpha Re Pr U_0 \right)\left( D^2 - k^2 \right) -  i \alpha Re Pr \frac{d^2U_0}{dy^2} \right] \tilde{v} & = Pr\left( D^2 - k^2 \right)^2 \tilde{v} - k^2Ra Pr \tilde{\theta}, 
%%\label{eq:LOBE_fourier1}\\
%%\left( \frac{\partial}{\partial t} + i \alpha Re Pr U_0 \right) \tilde{\eta} + i \beta Re Pr \frac{dU_0}{dy} \tilde{v} & = Pr\left( D^2 - k^2 \right) \tilde{\eta},
%%\label{eq:LOBE_fourier2}\\
%%\left( \frac{\partial}{\partial t} + i \alpha Re Pr U_0 \right) \tilde{\theta} + \frac{d \Theta_0}{dy} \tilde{v} & = \left( D^2 - k^2 \right) \tilde{\theta},
%%\label{eq:LOBE_fourier3}
%%\end{align}
\begin{eqnarray}
\left[ \left( \frac{-i\omega}{Pr} + i \alpha Re U_0 \right)\left( D^2 - k^2 \right) -  i \alpha Re \frac{d^2U_0}{dy^2} \right] \hat{v} = \left( D^2 - k^2 \right)^2 \hat{v} - k^2 Ra_{\mbox{\tiny{\textsl{h/2}}}} \hat{\theta}, 
\label{eq:LOBE_fourier1}
\end{eqnarray}
\begin{eqnarray}
\left( \frac{-i\omega}{Pr} + i \alpha Re U_0 \right) \hat{\eta} + i \beta Re \frac{dU_0}{dy} \hat{v} = \left( D^2 - k^2 \right) \hat{\eta},
\label{eq:LOBE_fourier2}
\end{eqnarray}
\begin{eqnarray}
\left( -i\omega + i \alpha Re Pr U_0 \right) \hat{\theta} + \frac{d \Theta_0}{dy} \hat{v} = \left( D^2 - k^2 \right) \hat{\theta},
\label{eq:LOBE_fourier3}
\end{eqnarray}
where $D = \frac{d}{dy}$, $k^2 = \alpha^2 + \beta^2$ and the boundary conditions are $\hat{v}( \pm 1) = 0$, $D\hat{v}(\pm 1) = 0$, $\hat{\eta} (\pm 1) = 0$ and $\hat{\theta}(\pm 1) = 0$. Equation \eqref{eq:LOBE_fourier1} is the Orr-Sommerfeld equation forced by buoyancy. The classical Squire equation \eqref{eq:LOBE_fourier2} is simply retained for shear flows in Boussinesq fluids because buoyancy acts normal to the wall and it cannot directly induce wall-normal vorticity. Equation \eqref{eq:LOBE_fourier3} is the linearised temperature equation which is an advection-diffusion equation similar to the one in the linear stability of pure conduction in Boussinesq fluids.

A modified version of Squire's theorem applies for shear flows in the presence of buoyancy $\cite{Gallagher_n_Mercer_1965, Gage_n_Reid_1968}$. The Squire equation has no explicit forcing due to buoyancy. The standard result that Squire modes are always damped holds also at all $\alpha$, $\beta$, $Re$, $Ra$ and $Pr$. Thus, the unstable eigenmodes only come from the coupled equations \eqref{eq:LOBE_fourier1} and \eqref{eq:LOBE_fourier3}. It can be shown that, for every oblique mode ($\alpha \neq 0$, $\beta \neq 0$) at some $Re$, $Ra$, $Pr$, there exists a spanwise-uniform mode $(\beta_{2D} = 0)$ at the same $Ra$ and $Pr$ with the same complex frequency for a smaller $Re$$\cite{Gallagher_n_Mercer_1965, Gage_n_Reid_1968}$ given by
\begin{align}
	i\alpha_{2D} \hat{u}_{2D} &= i\alpha \hat{u} + i\beta \hat{w},
	\label{eq:SQ_transform1}\\
	\hat{p}_{2D} &= \hat{p},
	\label{eq:SQ_transform2}\\
	\alpha_{2D} &= \sqrt{\alpha^2 + \beta^2},
	\label{eq:SQ_transform3}\\
	\alpha_{2D}Re_{2D} &= \alpha Re,
	\label{eq:SQ_transform4}\\
	\omega_{2D} &= \omega,
	\label{eq:SQ_transform5}
\end{align}	
where the subscripts $2D$ refer to variables of the spanwise-uniform mode. Unlike the classical Squire transformation for TS waves, equations \eqref{eq:SQ_transform1}-\eqref{eq:SQ_transform5} preserve the complex frequency $\omega$ due to the choice of a thermal diffusive time scale. It is still true however that, at a given $Ra$ and $Pr$, oblique modes become marginally stable at a larger Reynolds number than spanwise-uniform modes. The Squire transformation has the same implication that $2D$ disturbances are the least stable among all the disturbances. It is sufficient to consider only two-dimensional eigenmodes ($\beta = 0$ or $\alpha = 0$) to find the stability diagram at a fixed $Ra$ and $Pr$.

A spectral collocation method based on Chebyshev polynomials over Gauss-Lobatto collocation points (as given in appendix A.6 of$\cite{Schmid_n_Henningson_2001}$) was implemented in MATLAB to compute the stability characteristics. The computational accuracy depends primarily on the number of polynomial expansion functions $(N + 1)$. When $N \geq 60$, the eigenvalues computed for the cases $Re = 0$ and $Ra = 0$, were observed to match up to eight digits those given in the classical textbooks \cite{Drazin_n_Reid_1981, Schmid_n_Henningson_2001}. The critical Rayleigh numbers of transverse rolls in $RBP$ for small non-zero Reynolds numbers are found to match up to 5 significant digits those from the numerical computations of Fujimura \& Kelly$\cite{Fujimura_n_Kelly_1988}$. Note that all the results for the marginal stability conditions given in the present paper are based on computations with $N = 100$ (or $N = 120$).
\subsection{Dominant modal instability}
\label{sec:MS1}
The results of modal stability analysis of RBP and RBC flows were recovered by numerical computation and the leading eigenmodes are discussed in this section. For a complete description of the linear stability characteristics of plane Poiseuille flow and plane Couette flow with thermal stratification the reader is referred to Kelly$\cite{Kelly_1994}$ and Fujimura et. al.$\cite{Fujimura_n_Kelly_1988}$.
\begin{figure}[h]
%\begin{flushleft}
\epsfig{file=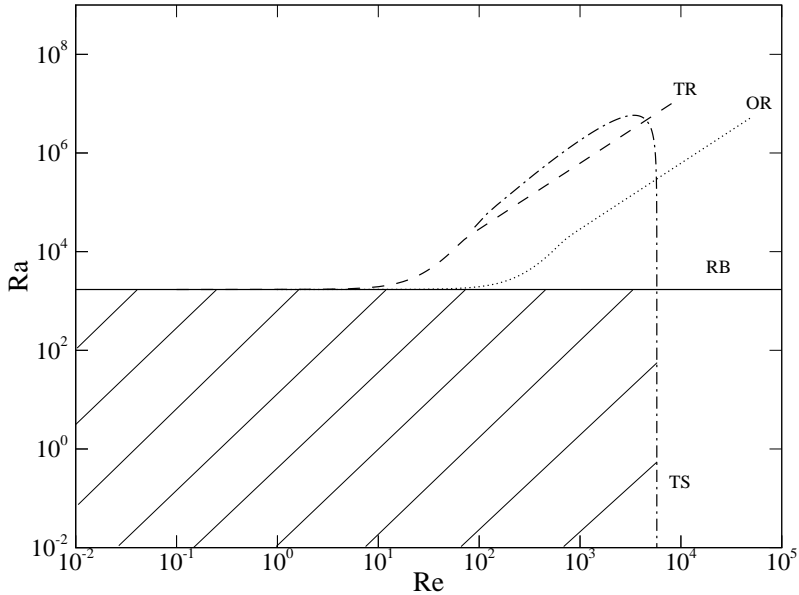,width=0.65\textwidth,keepaspectratio=true}
%\end{flushleft}
\caption{Marginal Stability Diagram of Rayleigh-B\'{e}nard-Poiseuille flow for $Pr = 1$:
$(\textbf{------})$ streamwise-uniform Rayleigh-B\'{e}nard convection rolls ($RB$), $(---)$ Transverse Rolls ($TR$), $(\cdot-\cdot-\cdot)$ Tollmien-Schlichting ($TS$) waves and $(\cdot\cdot\cdot\cdot\cdot)$ Oblique Rolls ($OR$), ${\alpha}/{\beta} = 9.95$. The flow is linearly stable in the hatched rectangular region formed by the lines corresponding to the onset of the $RB$ and $TS$ modes at the lower left of the plot.}
\label{fig:stability_diagram_RBP}
\end{figure}

\begin{figure}[h]
%\begin{flushleft}
\epsfig{file=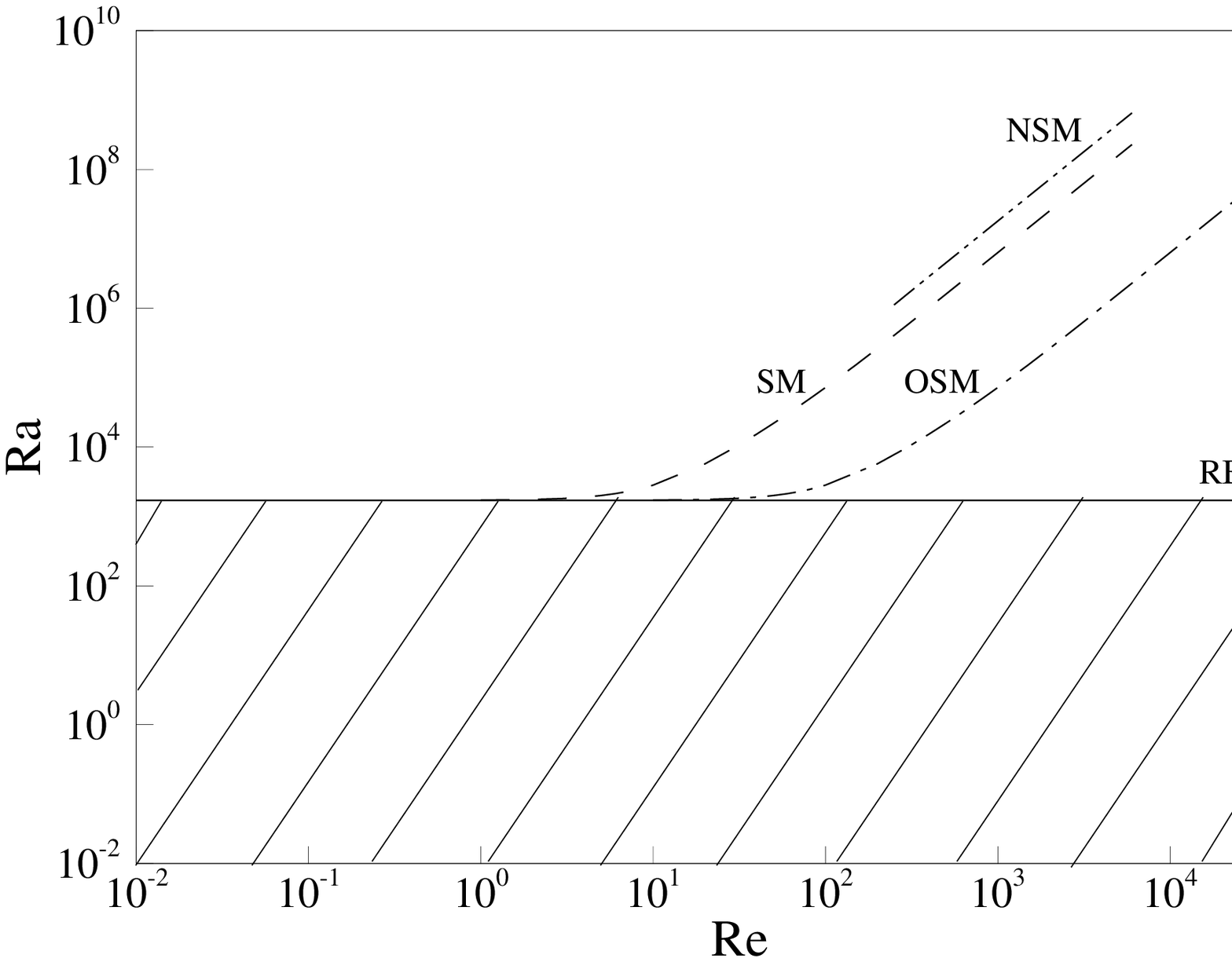,width=0.6\textwidth,keepaspectratio=true}
%\end{flushleft}
\caption{Marginal Stability Diagram of Rayleigh-B\'{e}nard-Couette for $Pr = 1$:
$(\textbf{-----})$ streamwise-uniform Rayleigh-B\'{e}nard convection Rolls ($RB$), $(---)$ Stationary spanwise-uniform mode ($SM$), $(-\cdot\cdot-)$ Non-stationary spanwise-uniform mode ($NSM$), $(\cdot-\cdot-\cdot)$ Oblique stationary spanwise-uniform mode ($OSM$): ${\alpha}/{\beta} = 9.95$. The flow is linearly stable everywhere in the hatched region below the continuous line corresponding to the onset of the $RB$ mode.}
\label{fig:stability_diagram_RBC}
\end{figure}

Figure $\ref{fig:stability_diagram_RBP}$ depicts the stability diagram of  $RBP$ flow when $Pr = 1$. It shows the neutral curves corresponding to the Reynolds number and Rayleigh number at which a given eigenmode, streamwise-uniform ($\alpha = 0$) or spanwise-uniform ($\beta = 0$), is marginally stable ($\omega_i = 0$, where $\omega_i$ is the imaginary part of $\omega$). Note that a streamwise-uniform mode becomes unstable at $Ra_c^{RB}$, independent of Reynolds number and Prandtl number, since in that case all the $U_0$-dependent terms drop-out from equations \eqref{eq:LOBE_fourier1} and \eqref{eq:LOBE_fourier3}. Equations \eqref{eq:LOBE_fourier1} and \eqref{eq:LOBE_fourier3} then reduce to the linear system for pure conduction in Boussinesq fluids. The $\hat{\theta}$ and $\hat{v}$ eigenfunctions of the least stable eigenmode ($RB$) are then identical to those in the no-through flow case ($Re = 0$). Note, however, the presence of $\hat{\eta}$ perturbations governed by \eqref{eq:LOBE_fourier2} corresponding to the tilting of the base flow vorticity by the $\hat{v}$ component as in the lift-up mechanism. Also, two distinct spanwise-uniform modes, namely, Transverse Rolls ($TR$), and Tollmien-Schlichting waves ($TS$) are presented in the stability diagram. The transverse rolls correspond to the $2D$ solutions that evolve continuously from $RB$ rolls at $Re = 0$. They occur at a critical Rayleigh number which increases monotonically with Reynolds number.

All spanwise-uniform modes except for the TS mode become unstable at some $Ra > Ra_c^{RB}$ for all non-zero Reynolds numbers. When $Ra = 0$ the TS mode becomes marginally stable at $Re = Re_c^{TS}$ and for all non-zero Rayleigh numbers below $Ra_c^{RB}$ its critical Reynolds number decreases, however very slowly, monotonically \cite{Gallagher_n_Mercer_1965, Fujimura_n_Kelly_1988}.

These plots suggest that there is only a weak coupling between buoyancy and shear flow stability characteristics. Nevertheless, the TS mode appears to become unstable at Reynolds numbers as low as $Re  =100$, but always at $Ra > Ra_c^{RB}$ which confirms that unstable stratification is favorable to TS instability.

For RBC flow, the marginal stability diagram is shown in figure $\ref{fig:stability_diagram_RBC}$. For the same reasons discussed in the case of $RBP$, the least stable eigenmode ($RB$) at all Reynolds numbers is a streamwise-uniform convection roll \textit{with non-zero streamwise velocity} due to the presence of shear (lift-up mechanism). The critical Rayleigh number at which this eigenmode becomes unstable is always independent of $Re$ and $Pr$. In figure $\ref{fig:stability_diagram_RBC}$, several spanwise-uniform eigenmodes: stationary mode ($SM$), non-stationary mode ($NSM$), and an oblique stationary mode ($OSM$) are also presented (here, the names stationary/non-stationary refer to symmetry preserving or symmetry breaking eigenmodes that have therefore a zero/non-zero phase speed, respectively). As in $RBP$ flow, they become marginally stable at  $Ra > Ra_c^{RB}$ for all non-zero $Re$$\cite{Gallagher_n_Mercer_1965, Fujimura_n_Kelly_1988}$ and the critical Rayleigh number of both $SM$ and $NSM$ increases monotonically with Reynolds number. Thus, the stability diagram essentially remains similar to the case of $RBP$ flow. For all Reynolds numbers, however, the dominant linear instability is $RBI$ in the presence of shear.

Thus, for Rayleigh-B\'{e}nard-Poiseuille flow, the stability boundary consists of two parts \cite{Gage_n_Reid_1968}. One part of the boundary is formed  by the streamwise-uniform Rayleigh-B\'{e}nard mode $\left( RB \right)$ at a constant Rayleigh number equal to $Ra_c^{RB}$ while the other part is due to Tollmien-Schlichting waves occurring at $Re \approx Re_c^{TS}$. In the case of Rayleigh-B\'{e}nard-Couette flow, the second part of the boundary (TS waves) is absent \cite{Gallagher_n_Mercer_1965}. The hatched regions in figures \ref{fig:stability_diagram_RBP} \& \ref{fig:stability_diagram_RBC} represent the domain where $RBP$ and $RBC$ flows, respectively, do not show any exponential instability.

In the case of spanwise-uniform neutral modes, a few scalings laws are evident from the lines of constant slope in figures \ref{fig:stability_diagram_RBP} and \ref{fig:stability_diagram_RBC}. It seems that this remark has not been made in previous studies. In the case of $RBP$ flow, it is observed that, if $Ra_c$ refers to any critical Rayleigh number, it is proportional to $Re_c^{4/3}$ for $TR$. While both $SM$ and $NSM$ in $RBC$ flow obey the scaling law  $Ra_c \propto Re_c^{1/2}$. Similar scaling laws are obtained for the critical wavenumber $\alpha_c$ for all the spanwise-uniform modes except TS waves: $\alpha_c \propto Re_c^{1/3}$ for $TR$ in $RBP$ flow; $\alpha_c \propto Re_c^{-1}$ and $\alpha_c \propto Re_c^{1/2}$ for $SM$ and $NSM$, respectively, in $RBC$ flow.
\section{Non-modal stability analysis}
\label{sec:NMS}
The modal stability analysis gives the conditions for exponential instability. However, a serious shortcoming of this analysis is that there might be a strong transient growth of disturbances before they eventually decay or grow exponentially. Note that this growth can occur in the absence of nonlinear effects \cite{Orr_1907} and is solely due to the non-normality of the Orr-Sommerfeld and Squire equations \cite{Landhal_1990, Butler_n_Farrel_1992, Reddy_n_Henningson_1993}. The main objective of this paper is to investigate this aspect of the perturbation dynamics in $RBP$ and $RBC$ flows.

%%The equations \eqref{eq:LOBE1}, \eqref{eq:LOBE2} and \eqref{eq:LOBE3}, when taken along with the expressions \eqref{eq:normmode1} and \eqref{eq:normmode2}, can be written in matrix-form in terms of the reduced variables $\begin{bmatrix}  \tilde{v},  \tilde{\theta},   \tilde{\eta} \end{bmatrix}^T$
%%\subsection{Adjoint of the Linearised Oberbeck-Boussinesq Equations}
%%\label{sec:adj_eqn_LOBE}
The adjoint of the linear operator \eqref{eq:LOBE_fourier1}, \eqref{eq:LOBE_fourier2} and \eqref{eq:LOBE_fourier3} is now derived. Consider the family of norms that represent a measure of the growth of perturbations,
\begin{align}
	E \left( t; \gamma \right) = \int_\texttt{V}{ \left[ \frac{1}{2} \left( \left| u \right|^2 + \left| v \right|^2 + 
	\left| w \right|^2 \right) + \frac{1}{2} \gamma^2 \left| \theta \right|^2 \right] d\texttt{V}},
	\label{eq:energy_norm_gamma}
\end{align}
where the weight $\gamma$ between the kinetic energy and the temperature perturbations is left arbitrary. Note that $E \left( t; \gamma \right)$ belongs to a class of norms commonly used in the literature; in particular, in \cite{Sameen_n_Govindarajan_2007} and \cite{Biau_n_Bottaro_2004}, $\gamma = 1$ and $\gamma = \sqrt{\left| Ra_{\mbox{\tiny{\textsl{h/2}}}} \right|Pr}$, respectively. It is convenient to rescale the state vector as $\textbf{\textit{q}} = \left[ \hat{v},	\hat{\theta}_*, \hat{\eta} \right]^T$, where $\hat{\theta}_* = \gamma \hat{\theta}$. The direct equations \eqref{eq:LOBE_fourier1}, \eqref{eq:LOBE_fourier2} and \eqref{eq:LOBE_fourier3} can then be written in the matrix form
\begin{equation}
	\left(	\textbf{L}_{OB} +\frac{\partial}{\partial t} \textbf{B}_{OB}\right) \textbf{\textit{q}} = 0,
	\label{eq:LOBE_matrix}
\end{equation}
where
\begin{equation*}
	\begin{array}{lll}
		\textbf{L}_{OB} &= 
			\begin{bmatrix}
				Pr.L_{OS}	&-k^2 \frac{Ra_{\mbox{\tiny{\textsl{h/2}}}} Pr}{\gamma} &0\\
				-\gamma \left( -\frac{d\Theta_0}{dy} \right) &L_{LHE}	&0\\
				i\beta \left( Re Pr \right)\frac{dU_0}{dy}	&0 &Pr.L_{SQ}\\
			\end{bmatrix},
	\end{array}
\end{equation*}
\begin{equation*}
	\begin{array}{ll}
		\textbf{B}_{OB} &=
		\begin{bmatrix}
			k^2-D^2	&0	&0\\
			0	&1	&0\\
			0	&0	&1
		\end{bmatrix}.
	\end{array}
\end{equation*}
The Orr-Sommerfeld and Squire operators, $L_{OS}$ and $L_{SQ}$, respectively, are defined as
\begin{eqnarray}
		L_{OS} = i\alpha Re U_0 \left( k^2-D^2 \right) + i\alpha Re\frac{d^2U_0}{dy^2} +\left( k^2-D^2 											\right)^2,
		\label{eq:L_direct_OS}
\end{eqnarray}
\begin{eqnarray}
		L_{SQ} = i\alpha Re U_0 +\left( k^2-D^2 \right),
		\label{eq:L_direct_SQ}
\end{eqnarray}
and $L_{LHE}$ is the advection-diffusion operator governing the evolution of the rescaled temperature perturbation:
\begin{align}
		L_{LHE} =  i\alpha Re Pr U_0 +\left( k^2-D^2 \right).
		\label{eq:L_direct_LHE}
\end{align}
Homogeneous boundary conditions on the state vector $\textbf{\textit{q}}$ are enforced, as for the reduced variables in \eqref{eq:LOBE_fourier1}, \eqref{eq:LOBE_fourier2} \& \eqref{eq:LOBE_fourier3}. In the case of unstable thermal stratification $\frac{d\Theta_0}{dy} = -1$ and for stable stratification $\frac{d\Theta_0}{dy} = 1$. 
The adjoint of the linear operator, say $\textbf{L}_{OB}^A$, is defined as
%%\begin{equation}
%%\langle \langle \left[\textbf{L}_{OB} + \textbf{B}_{OB} \frac{\partial}{\partial t} \right] \textbf{\textit{q}},\ \textbf{\textit{q}}_A \rangle \rangle = \langle \langle \textbf{\textit{q}},\ \left[ \textbf{L}_{OB}^A + \textbf{B}_{OB} \frac{\partial}{\partial t} \right] \textbf{\textit{q}}_A \rangle \rangle,
%%\label{eq:adj_definition}
%%\end{equation}
\begin{equation}
\langle \langle \textbf{L}_{OB} \textbf{\textit{q}},\ \textbf{\textit{q}}_{A} \rangle \rangle = \langle \langle \textbf{\textit{q}},\ \textbf{L}_{OB}^A \textbf{\textit{q}}_{A} \rangle \rangle,
\label{eq:adj_definition}
\end{equation}
where the angle brackets represent the scalar product
\begin{eqnarray}
	\langle \langle \textbf{\textit{q}}_1,\ \textbf{\textit{q}}_2 \rangle \rangle = 		
	\int_{-1}^{1}{\textbf{\textit{q}}_2^H \textbf{M} \textbf{\textit{q}}_1 dy},
	\label{eq:scalar_product}
\end{eqnarray}
and $\textbf{M} = \mbox{diag} \left( 1, k^2, 1 \right)$. The choice of this weight matrix will become evident by the end of the section where the biorthogonality condition is derived.

Using integration by parts and the boundary conditions on the state vector, the adjoint equations are derived to be
\begin{equation}
\left( \textbf{L}_{OB}^A +\frac{\partial}{\partial t} \textbf{B}_{OB} \right) \textbf{\textit{q}}_{A} = 0,
\label{eq:LOBEadjoint_matrix}
\end{equation}
\begin{equation*}
\textbf{L}_{OB}^A =
	\begin{bmatrix}
		Pr. L_{OS}^A	&-k^2\gamma \left( -\frac{d\Theta_0}{dy} \right) &-i\beta \left( Re Pr \right)\frac{dU_0}{dy}\\
		-\frac{Ra_{\mbox{\tiny{\textsl{h/2}}}} Pr}{\gamma}	&L_{LHE}^A &0\\
		0	&0 &Pr. L_{SQ}^A
	\end{bmatrix},
\end{equation*}
%%\begin{equation*}
%%\textbf{B}_{(OB)}^A =
%%	\begin{bmatrix}
%%		k^2-D^2	&0	&0\\
%%		0	&1	&0\\
%%		0	&0	&1
%%	\end{bmatrix},
%%\end{equation*}
where $L_{OS}^A$, $L_{SQ}^A$ and $L_{LHE}^A $ are the classical adjoint-Orr-Sommerfeld operator \cite{Schmid_n_Henningson_2001}, adjoint-Squire operator \cite{Schmid_n_Henningson_2001}, and the adjoint of the advection-diffusion operator appearing in the linearised temperature equation, respectively:
\begin{eqnarray}
	L_{OS}^A = -i\alpha Re U_0 \left( k^2-D^2 \right) + 2i\alpha Re \frac{dU_0}{dy}D 
	+ \left( k^2-D^2 \right)^2,
	\label{eq:L_adjoint_OS}
\end{eqnarray}
\begin{eqnarray}
	L_{SQ}^A =  -i\alpha Re U_0 + \left( k^2-D^2 \right),
	\label{eq:L_adjoint_SQ}
\end{eqnarray}
\begin{eqnarray}
	L_{LHE}^A = -i\alpha Re Pr.U_0 +\left( k^2-D^2 \right).
	\label{eq:L_adjoint_LHE}
\end{eqnarray}
Note that $\textbf{B}_{OB}$ is a self-adjoint operator and the adjoint state vector $\textbf{\textit{q}}_{A}$ obeys homogeneous boundary conditions similar to the direct state vector $\textbf{\textit{q}}$. 

Let $\textit{q}_{n}$ and $\textit{q}_{Am}$ be any normalized direct and adjoint eigenvectors, respectively, where $n$ and $m$ are indices, such that $\langle \langle \textbf{\textit{q}}_{n},\ 
		\textbf{\textit{q}}_{n} \rangle \rangle = \langle \langle \textbf{\textit{q}}_{Am},\ 
		\textbf{\textit{q}}_{Am} \rangle \rangle = 1$. To find the bi-orthogonality condition, consider the product of the direct operator \eqref{eq:LOBE_matrix} applied to $\textit{q}_{n}$ with $\textit{q}_{Am}$,
\begin{equation*}
	\begin{array}{cc}
		&\langle \langle \left( \textbf{L}_{OB} - i\omega_{n}\textbf{B}_{OB} \right) \textbf{\textit{q}}_{n},\ 
		\textbf{\textit{q}}_{Am} \rangle \rangle = 0,
	\end{array}
\end{equation*}
which up on using the definition of the adjoint \eqref{eq:adj_definition} gives
\begin{equation*}
	\begin{array}{cc}
		&\langle \langle \textbf{\textit{q}}_{n},\ \left( \textbf{L}_{OB}^A + 
		i\omega_{n}^*\textbf{B}_{OB} \right) \textbf{\textit{q}}_{Am}\rangle \rangle = 0.
	\end{array}
\end{equation*}
Since $\textit{q}_{Am}$ is an adjoint eigenvector, it satisfies the eigenfunction formulation of the adjoint operator \eqref{eq:LOBEadjoint_matrix}. Thus, the above equation can be simplified to 
\begin{equation*}
	-i\left( \omega_{n} - \omega_{(m)} \right) \langle \langle \textbf{\textit{q}}_{n},\ \textbf{B}_{OB} \textbf{\textit{q}}_{Am}\rangle \rangle = 0,
	%\label{eq:biorthogonality}
\end{equation*}
which gives the bi-orthogonality condition between any direct eigenvector $\textit{q}_{n}$ and any adjoint eigenvector $\textit{q}_{Am}$ in the form
\begin{equation}
	\langle \langle \textbf{\textit{q}}_{n},\ \textbf{B}_{OB} \textbf{\textit{q}}_{Am}\rangle \rangle = 		2k^2\delta_{nm},
	\label{eq:biorthogonality1}
\end{equation}
or, equivalently,
\begin{equation}
	\langle \textbf{\textit{q}}_{n},\ \textbf{\textit{q}}_{Am}\rangle_\gamma = \delta_{nm},
	\label{eq:biorthogonality}
\end{equation}
where $\delta_{nm}$ is the Kronecker delta and the new scalar product
\begin{equation}
\langle \textbf{\textit{q}}_1,\ \textbf{\textit{q}}_2\rangle_\gamma = \int_{-1}^{1} \left[ \frac{1}{2} \left(\hat{v}_1\hat{v}_2^* + \frac{1}{k^2} \left( D\hat{v}_1D\hat{v}_2^* + \hat{\eta}_1\hat{\eta}_2^* \right) \right) \frac{1}{2} \gamma^2 \hat{\theta}_1 \hat{\theta}_2^* \right]dy
\label{eq:energy_scalar_product}
\end{equation}
has been introduced. The corresponding norm is given by
\begin{equation}
	\left\| \textbf{\textit{q}} \right\|^2_\gamma = \int_{-1}^{1} \left[ \frac{1}{2}\left( \left| \hat{v} \right|^2 + \frac{1}{k^2} \left( \left| D\hat{v} \right|^2 + \left| \hat{\eta} \right|^2\right) \right) + \frac{1}{2} \gamma^2 \left| \hat{\theta} \right|^2 \right] dy,
	\label{eq: norm_general}
\end{equation}
which is precisely the norm $E \left(t; \gamma \right)$ defined in terms of primitive variables in equation \eqref{eq:energy_norm_gamma} but expressed here in reduced variables (see \cite{Schmid_n_Henningson_2001} for a similar derivation).

Taking $\gamma = \sqrt{\left| Ra_{\mbox{\tiny{\textsl{h/2}}}} \right|Pr}$, the direct equations \eqref{eq:LOBE_matrix} and the adjoint equations \eqref{eq:LOBEadjoint_matrix} become identical when $Re = 0$ i.e.,\ the linear operator of the Rayleigh-B\'{e}nard-Poiseuille/Couette problem is then self-adjoint under this specific norm. This property is independent of Prandtl number. In the case of a fluid layer heated from above (stable stratification) the term  $\frac{1}{2}\left| Ra_{\mbox{\tiny{\textsl{h/2}}}} \right|Pr |\hat{\theta}|^2$ in \eqref{eq: norm_general} is the non-dimensional potential energy of the disturbances and the rest of the terms in \eqref{eq: norm_general} denote the kinetic energy. This norm is then identical to the total energy of the perturbations in a Boussinesq fluid in the presence of stable stratification. 

Thus, a relevant measure of perturbation growth is a positive definite norm of the form
\begin{equation}
	E(t) = \int^{1}_{-1} \frac{1}{2} \left[ \left| \hat{v} \right|^2 + \frac{1}{k^2} \left( \left| D\hat{v} \right|^2 
	+ \left| \hat{\eta} \right|^2 \right) + \left|Ra_{\mbox{\tiny{\textsl{h/2}}}}\right|Pr \left| \hat{\theta} \right|^2 \right] dy.
\label{eq:norm_RB}
\end{equation}	
For a more detailed discussion on the norm the reader is referred to section \ref{norm_effect}. The transient growth characteristics are obtained by solving equations \eqref{eq:LOBE_matrix} for an initial disturbance field that would give rise to the maximum possible growth. This disturbance is called the optimal initial condition and it is defined using the growth function
\begin{align}
G(t) = \operatorname*{max}_{\forall \textbf{\textit{q}}(t_0) \neq  \textbf{\textit{0}}} \left[ \frac{E(t)}{E(t_0)} \right],
\label{eq:trans_gwth}
\end{align}
referred to as the optimal transient growth, i.e.\ the maximum possible growth at some time t over all possible non-zero initial conditions. Since different wave vectors are not linearly coupled, $G(t)$ may be considered as a function of $\alpha$ and $\beta$ as well as the control parameters $Re$, $Ra$ and $Pr$. It is convenient to define the quantities
\begin{align}
	G_{max}\left(\alpha, \beta; Re, Ra, Pr \right) &=  \operatorname*{max}_{\forall t \geq 0} G\left(t, \alpha, \beta; Re, Ra, Pr\right),
	\label{eq:Gmax}\\
	S\left(Re, Ra, Pr\right) &=  \operatorname*{\mbox{sup}}_{\alpha, \beta}\ G_{max}\left(\alpha, \beta; Re, Ra, Pr \right),
	\label{eq:Gmaxmax}
\end{align}
where $G_{max}$ is commonly known as the maximum optimal transient growth. Let $t_{max}$ be the time taken to attain the growth $G_{max}$ and $\left( \alpha_{opt},\beta_{opt} \right)$ be the wavenumbers corresponding to the overall optimal growth $S\left(Re, Ra, Pr\right)$.

\begin{figure}[h]
%\begin{center}
\epsfig{file=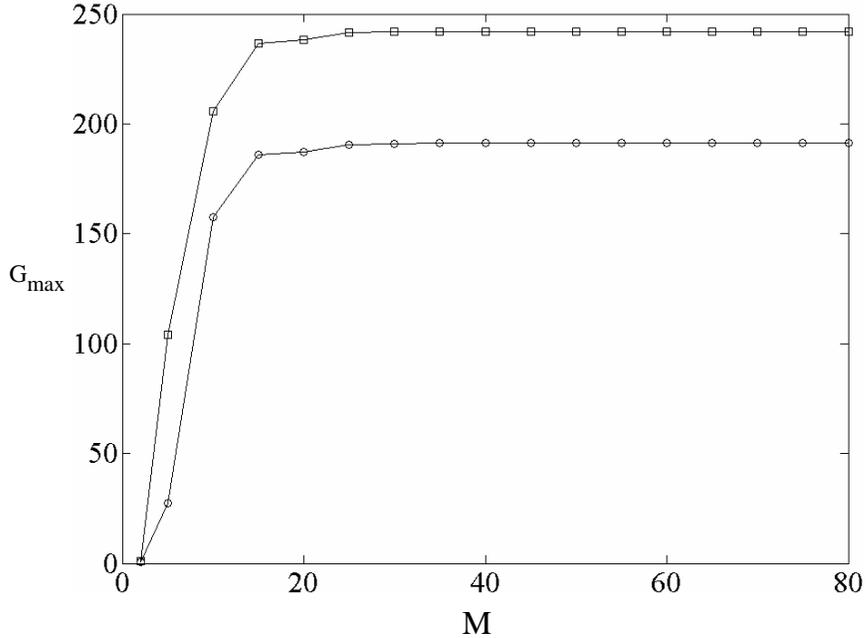,width=0.6\textwidth,keepaspectratio=true}
%\end{center}
\caption{Convergence of the computed value of $G_{max}$ with respect to the number of eigenmodes $M$ considered 
for	computations of transient growth: $Re = 1000$, $Pr = 1$ and $\circ$ $ Ra = 0$, $\square$ $ Ra = 500$ ($\alpha 
=0$, $\beta = 1.8$).}
\label{fig:SVD_convergence}
\end{figure}

To compute $G(t)$ the continuous linear operator, i.e.\ equations \eqref{eq:LOBE_fourier1}, \eqref{eq:LOBE_fourier2} and \eqref{eq:LOBE_fourier3}, was discretized using Chebyshev spectral functions as for the numerical computations in the modal stability analysis. The optimal growth $G(t)$ is then related to the norm of the matrix exponential of the discretized linear operator (gain matrix) and it can be computed using Singular Value Decomposition ($SVD$)$\cite{Reddy_n_Henningson_1993, Schmid_n_Henningson_2001}$. Using an eigenfunction expansion formulation \cite{DiPrima_Habetler_1969, Herron_1980}, the gain matrix can be further approximated by its first few dominant eigenvectors and the norm of the resulting gain matrix is computed using the $SVD$ function in MATLAB (as given in appendix A.6 of the textbook by Schmid and Henningson$\cite{Schmid_n_Henningson_2001}$). Alternatively, one can use the method of power iteration wherein the norm of the matrix exponential, say $\textbf{A}$, is computed by iterative multiplication $q_{k+1} = \frac{\textbf{A} q_k}{\left\|\textbf{A} q_k\right\|}$ (refer to the article by Luchini$\cite{Luchini_2000}$ for more details). The accuracy of both methods depends on the number $N + 1$ of Chebyshev expansion functions and the precision of the former method also depends on the number of eigenmodes $M$ considered in the construction of the gain matrix. Thus, when using the former it must be ensured that all the dominant eigenmodes are taken into account. The computations were successfully validated against those available in Reddy et al.$\cite{Reddy_n_Henningson_1993}$ and Schmid and Henningson$\cite{Schmid_n_Henningson_2001}$. It was observed that $M \approx 60$ is sufficient to compute $G(t)$ up to 5 significant digits over all control parameter values (see figure \ref{fig:SVD_convergence}). Cross-validations were performed between the two methods as a consistency check and in the following sections, only the results from the method of power iteration are presented ($N = 100$).
\subsection{Effect of varying Rayleigh number at constant Reynolds number}
\label{sec:NMS1}

\begin{figure}[h]
%\centering
\epsfig{file=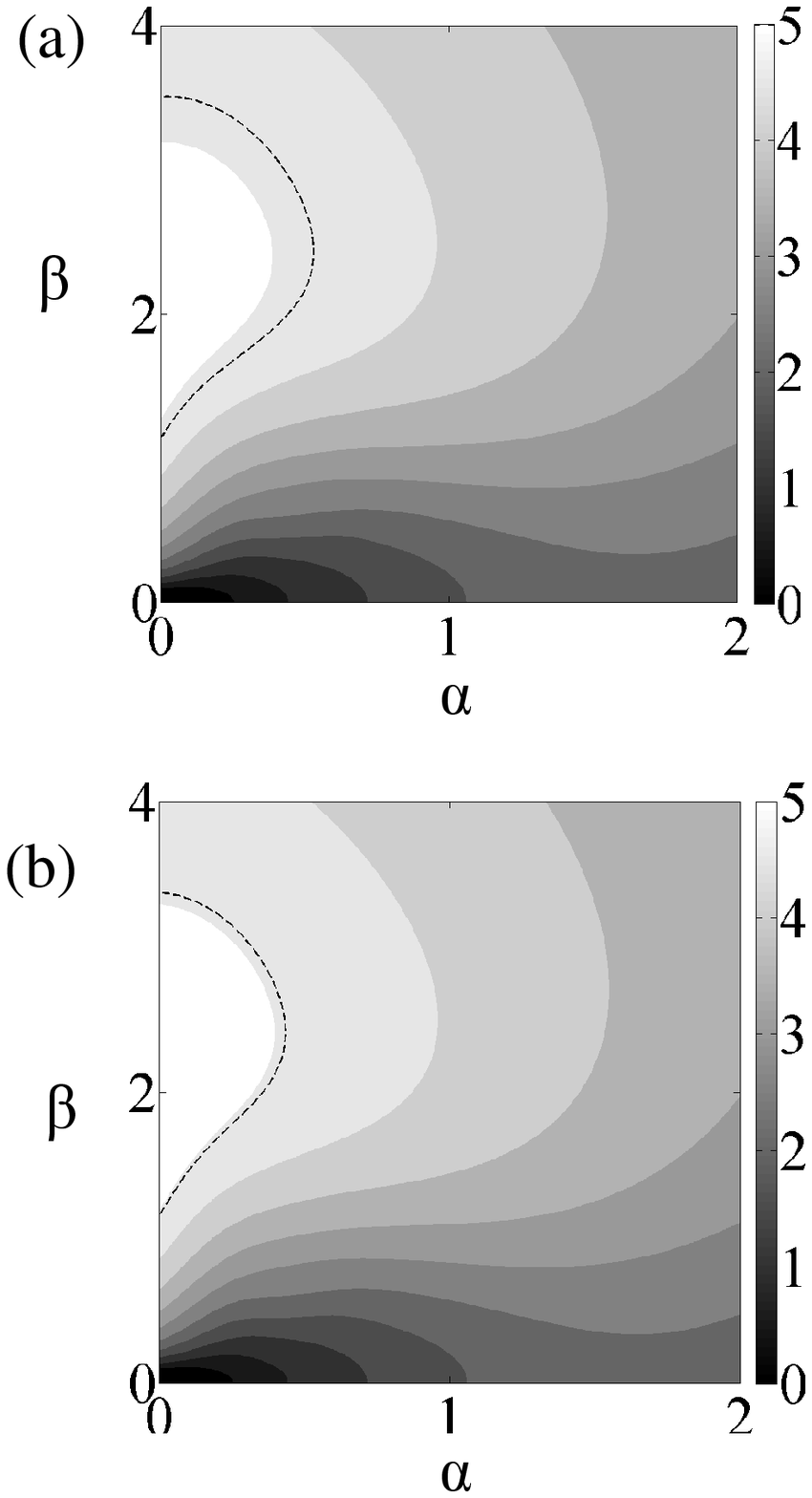,width=0.25\textwidth,keepaspectratio=true}\\
\vspace{5 mm}
\epsfig{file=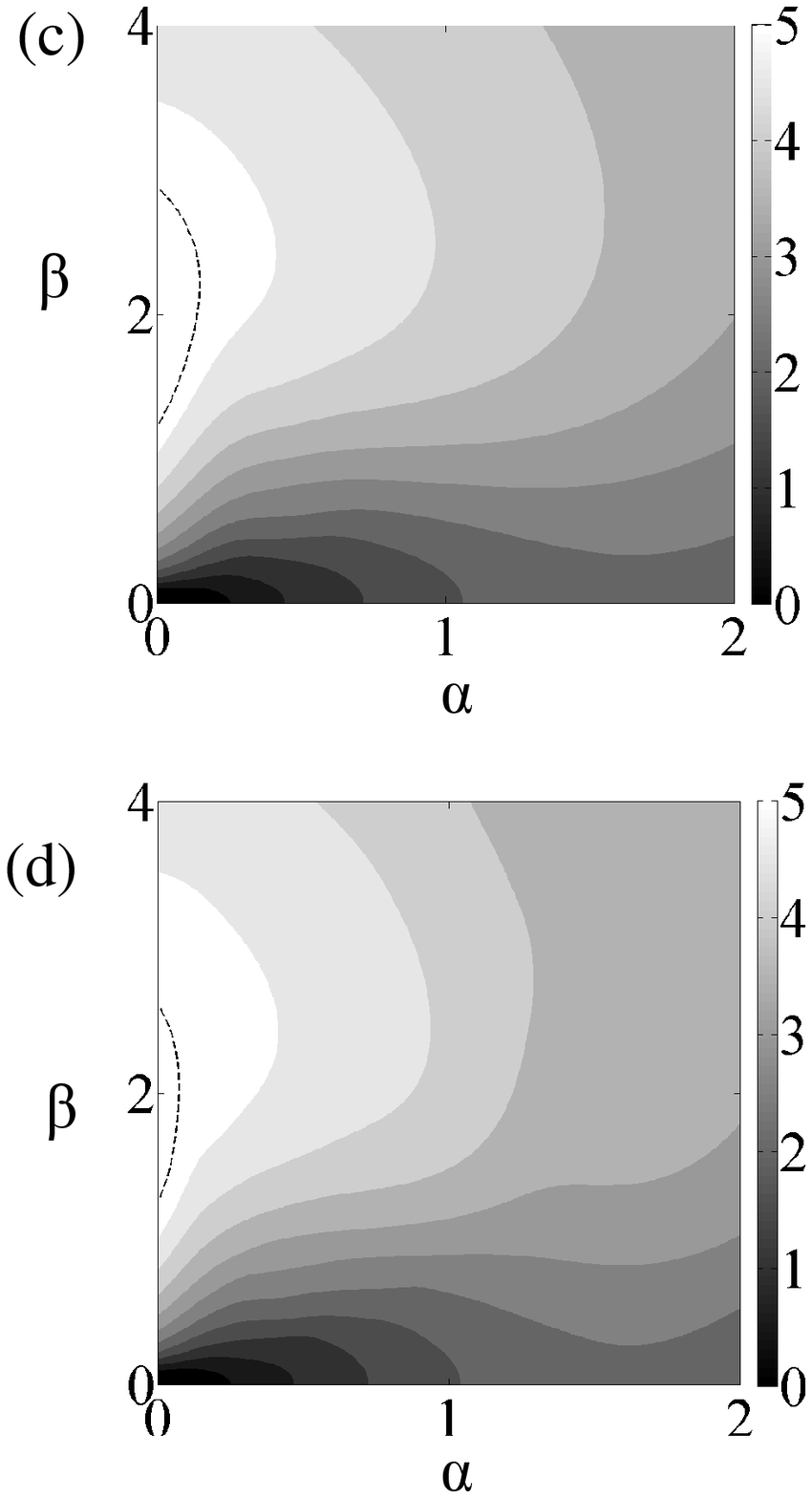,width=0.25\textwidth,keepaspectratio=true}
\caption{Contour plot of $\log G_{max}$ for RBP at $Re = 1000$, $Pr = 1$, and $(a) Ra = 0$, $(b) Ra = 500$, $(c) Ra = 1500$  and $(d) Ra = 1700$. The dashed lines correspond to iso-lines of the growth $G_{max} = \frac{2}{3}S$.}
\label{fig:Gmax_contourplot}
\end{figure}

\begin{figure}[h]
\epsfig{file=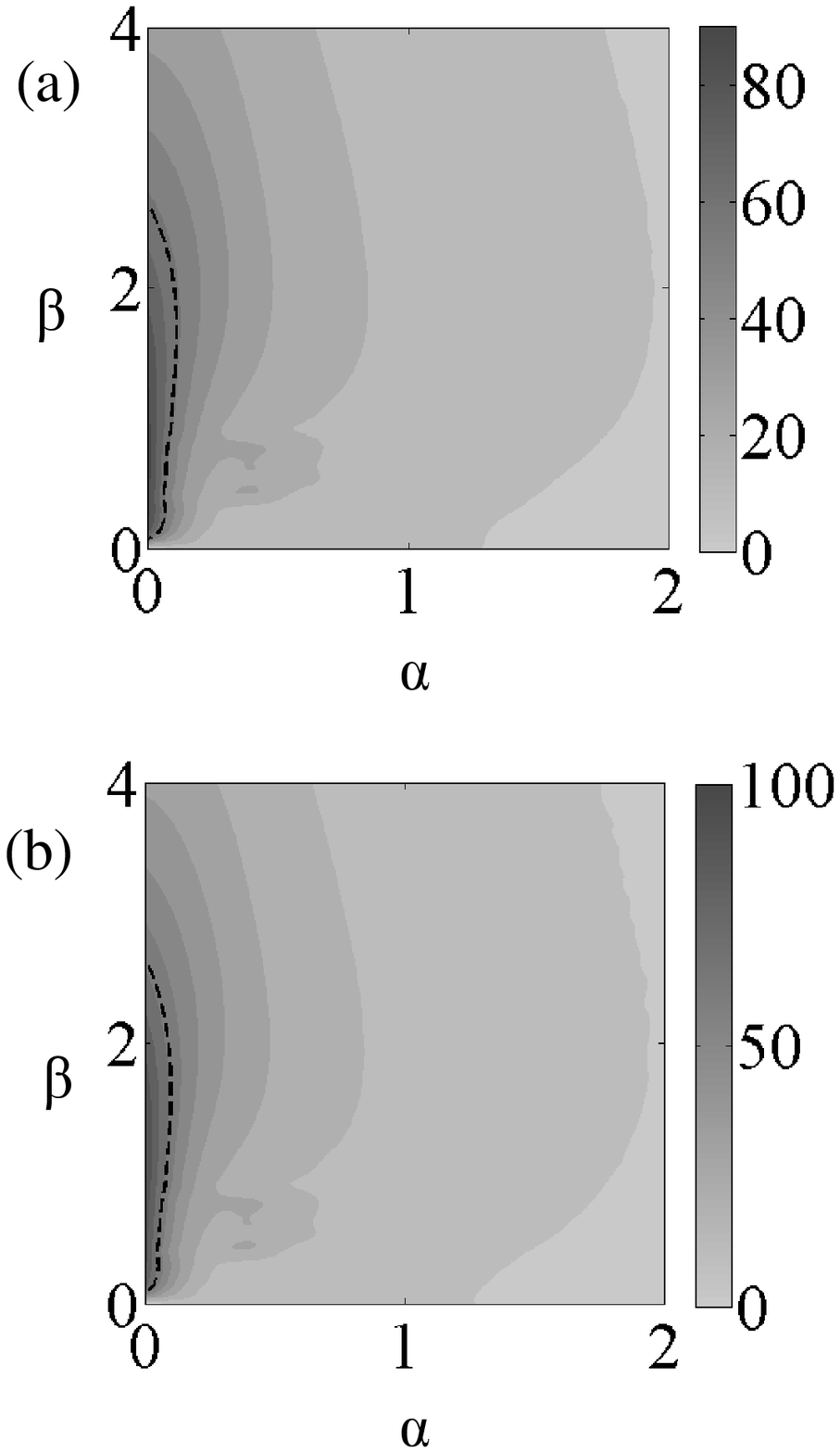,width=0.25\textwidth,keepaspectratio=true}\\
\vspace{5 mm}
\epsfig{file=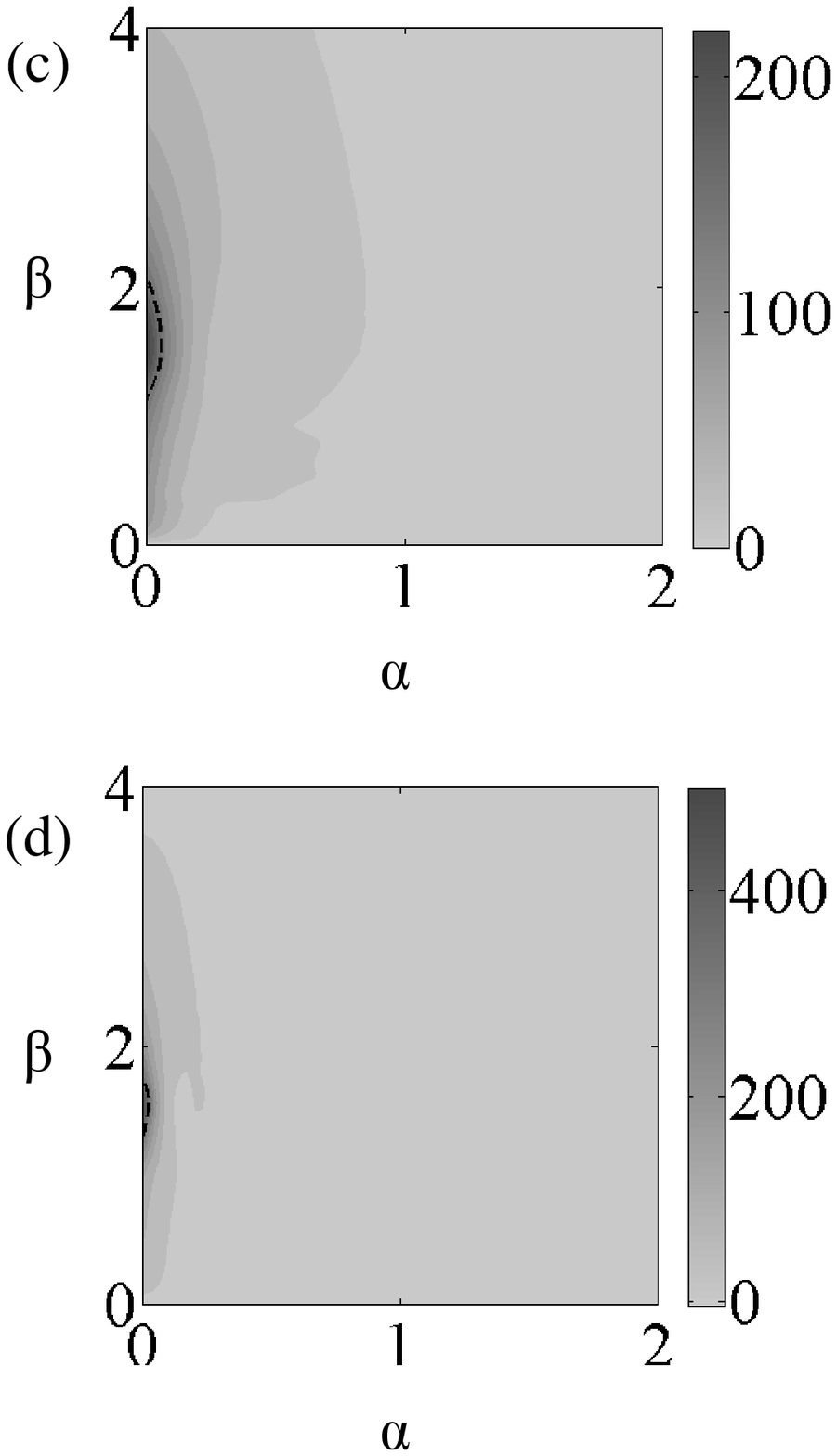,width=0.25\textwidth,keepaspectratio=true}
\caption{Contour plot of $t_{max}$ for RBP at $Re = 1000$, $Pr = 1$, and $(a) Ra = 0$, $(b) Ra = 500$, $(c) Ra = 1500$  and $(d) Ra = 1700$. The dashed lines correspond to iso-lines of $t_{max} = \frac{2}{3}T_{max}$, where $T_{max}$ is the global maximum of all $t_{max}$ in the $\alpha$-$\beta$ plane.}
\label{fig:tmax_contourplot}
\end{figure}

Figures \ref{fig:Gmax_contourplot} and \ref{fig:tmax_contourplot} show the contour level curves of $\log G_{max}$ and $t_{max}$, respectively, with respect to streamwise and spanwise wavenumbers at different Rayleigh numbers $Ra = 0$, $Ra = 500$, $Ra = 1500$, and $Ra = 1700$ for the same Reynolds number $Re = 1000$. In the case when $Ra = 0$ (pure shear flow), the plot (figure \ref{fig:Gmax_contourplot}a) reproduces the results of Reddy et. al. \cite{Reddy_n_Henningson_1993} wherein $S = 196$, $\alpha_{opt} = 0$, $\beta_{opt} = 2.04$ and $t_{max} = 76$. This corresponds to the classical lift-up mechanism \cite{Ellingsen_n_Palm_1975, Landhal_1990} resulting from the interaction between Orr-Sommerfeld and Squire modes due to the presence of the forcing term $-i\beta Re U'$ in the Squire equation \cite{Reddy_n_Henningson_1993, Schmid_n_Henningson_2001}. The contour levels along the $\alpha$-axis ($\beta  =0$) correspond to the lowest values of $G_{max}$, indicating that the spanwise-uniform disturbances which grow via the Orr-mechanism \cite{Orr_1907} are only sub-dominant compared to oblique and streamwise-uniform disturbances for all Reynolds numbers \cite{Butler_n_Farrel_1992, Reddy_n_Henningson_1993, Schmid_n_Henningson_2001}.

The contours of $\log G_{max}$ are remarkably similar for all $Ra$ and the effect of unstable stratification only moderately increases the maximum optimal growth. This variation is evident near the $\beta$-axis, for streamwise-uniform perturbations and some oblique perturbations which are nearly uniform in the direction of the base flow (say, nearly-streamwise-uniform disturbances: $0 \leq \alpha < 0.25$). In figures \ref{fig:Gmax_contourplot}a-d, the gray bars indicate that the overall optimal growth increases from $S = 196$ at $Ra = 0$ to $S = 369.15$ at $Ra = 1700$. The growth $S(1000, Ra, 1)$ is, thus, of the same order of magnitude for all Rayleigh numbers even at the onset of streamwise-uniform convection rolls when $Ra \approx Ra_c^{RB}$ (figure \ref{fig:Gmax_contourplot}d). The optimal wavenumber $\beta_{opt}$ of streamwise-uniform disturbances decreases with the increase in Rayleigh number. The maximum growth via the Orr-mechanism $\left( \mbox{for } \beta = 0\right)$ is not affected by the cross-stream temperature gradient, which is contrary to the observations of Sameen and Govindarajan \cite{Sameen_n_Govindarajan_2007}. Thus, \textit{the global optimal perturbations $S(Re, Ra, Pr = 1)$ are always in the form of streamwise-uniform disturbances for all $Ra$} with an optimal spanwise wavenumber $\beta_{opt}$ varying from $2.04$ to $1.558$ as $Ra$ approaches $Ra_c^{RB}$ (Note that the wavenumber of the most unstable $RB$ mode is $1.558$).

The geometry of the $G_{max}$ contours remains similar for all $Ra$ and $Re$. Figure \ref{fig:tmax_contourplot} indicates that there is a marked difference in the contour geometry of $t_{max}$ with increasing Rayleigh number.  When $Ra = 0$ (figure \ref{fig:tmax_contourplot}a), the maximum of $t_{max}$  lies on the $\beta$-axis and large values of $t_{max}$ occur around this point which is seen by the white contour levels close to that axis. The time taken to attain the growth corresponding to $S(Re, Ra, Pr)$ is $76$  and it is larger than that for any $G_{max}$ along the $\alpha$-axis. This implies that the Orr-mechanism is sustained only for a small time compared to the lift-up mechanism. For $Ra \neq 0$ the plots (figure \ref{fig:tmax_contourplot}b-d) display a small region of white contour levels near the $\beta$-axis and the maximum of $t_{max}$ increases from $76$ for $Ra = 0$ to $366$ for $Ra = 1700$. It can be concluded that the influence of unstable stratification is limited to streamwise-uniform and nearly-streamwise-uniform perturbations (similarly to the contours of $G_{max}$) and the transient growth of these disturbances is sustained over a much longer time than for any other disturbances. The equivalent of the Orr-mechanism in Boussinesq fluids is sustained only over a shorter period of time (as in the case $Ra = 0$).

\begin{figure}[h]
\begin{flushleft}
\epsfig{file=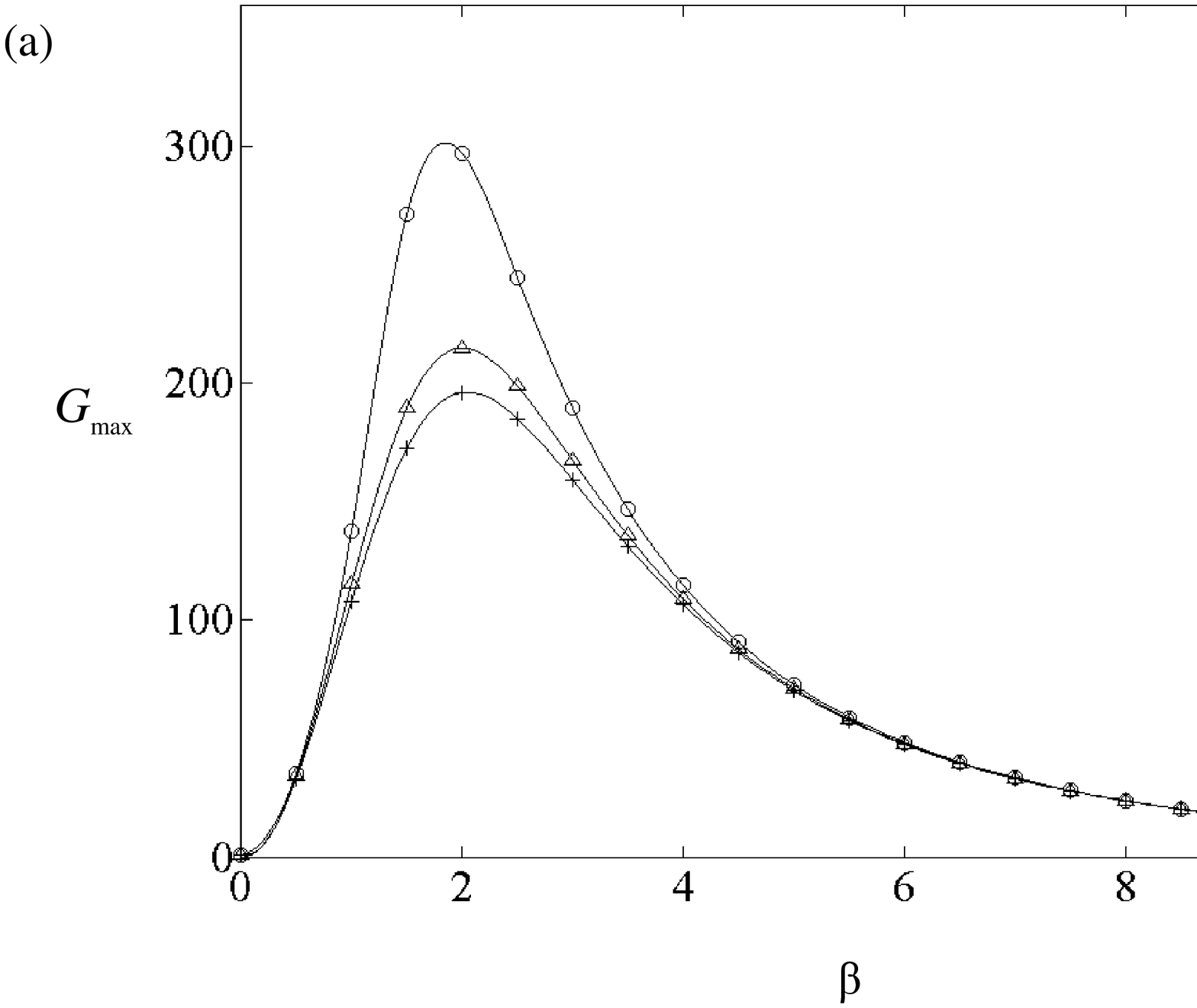,width=0.6\textwidth,keepaspectratio=true} \\
\epsfig{file=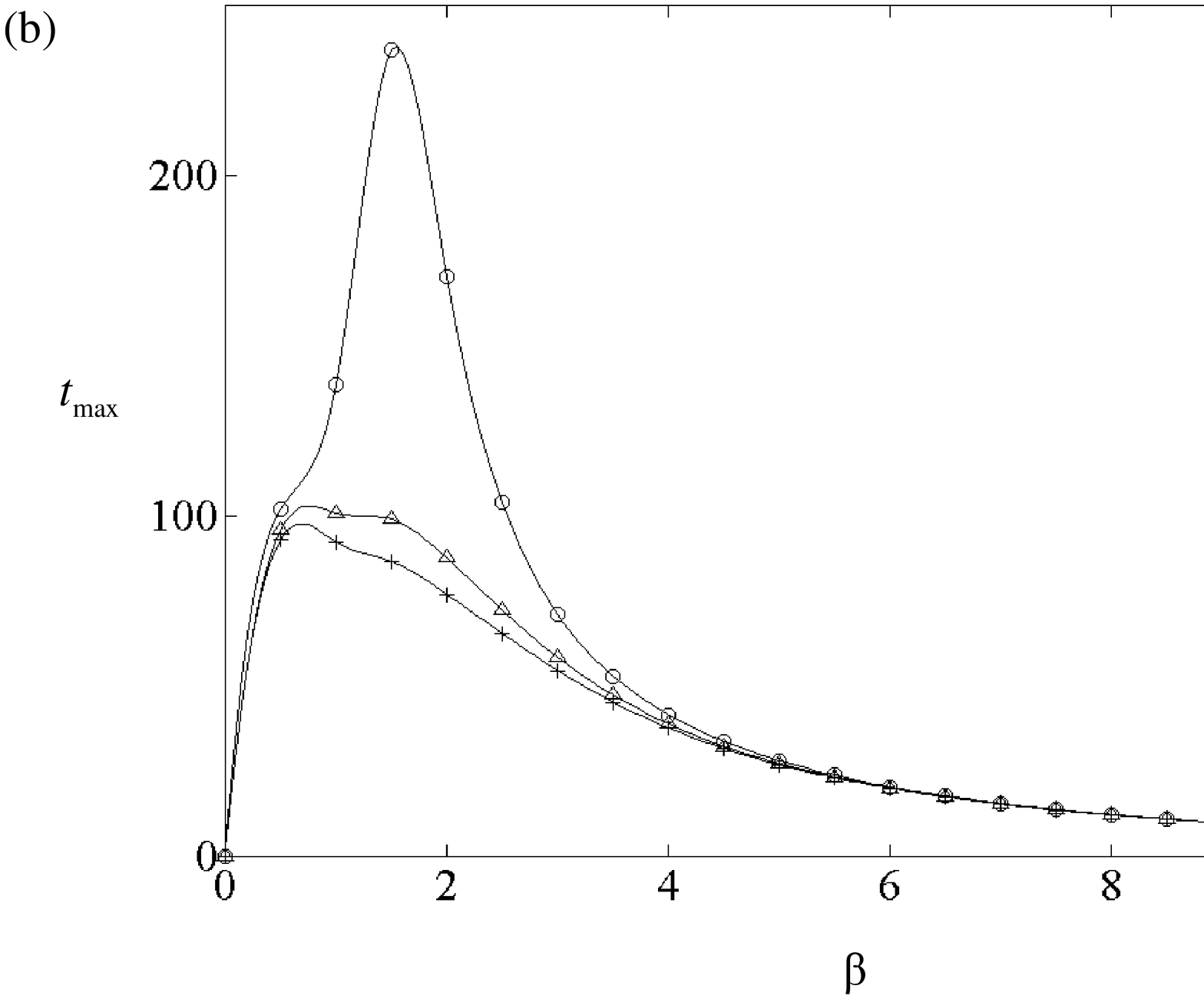,width=0.6\textwidth,keepaspectratio=true}
\end{flushleft}
\caption{Effect of Rayleigh number on $(a)$ $G_{max}$ and $(b)$ $t_{max}$ in $RBP$ for streamwise-independent disturbances of various spanwise wavenumbers $\beta$ at $Re = 1000$, $Pr = 1$, and $+ Ra = 0$, $\triangle Ra = 500$ and $\circ Ra = 1500$.}
\label{fig:Gmax_n_tmax_vs_beta_RBP}
\end{figure}

These observations are more evident in figure \ref{fig:Gmax_n_tmax_vs_beta_RBP} wherein $G_{max}$ and $t_{max}$ are displayed for streamwise-independent disturbances of various spanwise wavenumbers. Here, the results are for $RBP$ flow and different symbols indicate different Rayleigh numbers. Except for a range of spanwise wavenumbers between 1 and 5, the curves are all identical. This shows that the effect of Rayleigh number in $RBP$ is restricted only to a small range of spanwise wavenumbers.

The dashed lines in each of the $G_{max}$ and $t_{max}$ contour plots correspond to iso-lines of the growth $G_{max} = \frac{2}{3}S$ and $t_{max} = \frac{2}{3}T_{max}$, respectively, where $T_{max}$ is the global maximum of all $t_{max}$ in the $\alpha$-$\beta$ plane. The size of the region enclosed by this dashed line constantly decreases with increasing Rayleigh number as it approaches $Ra_c^{RB}$ and this is even more evident in the $t_{max}$ - contour plots. \textit{The presence of a temperature gradient thus sharpens the selection of global optimal perturbations.}

\begin{figure}[h]
\epsfig{file=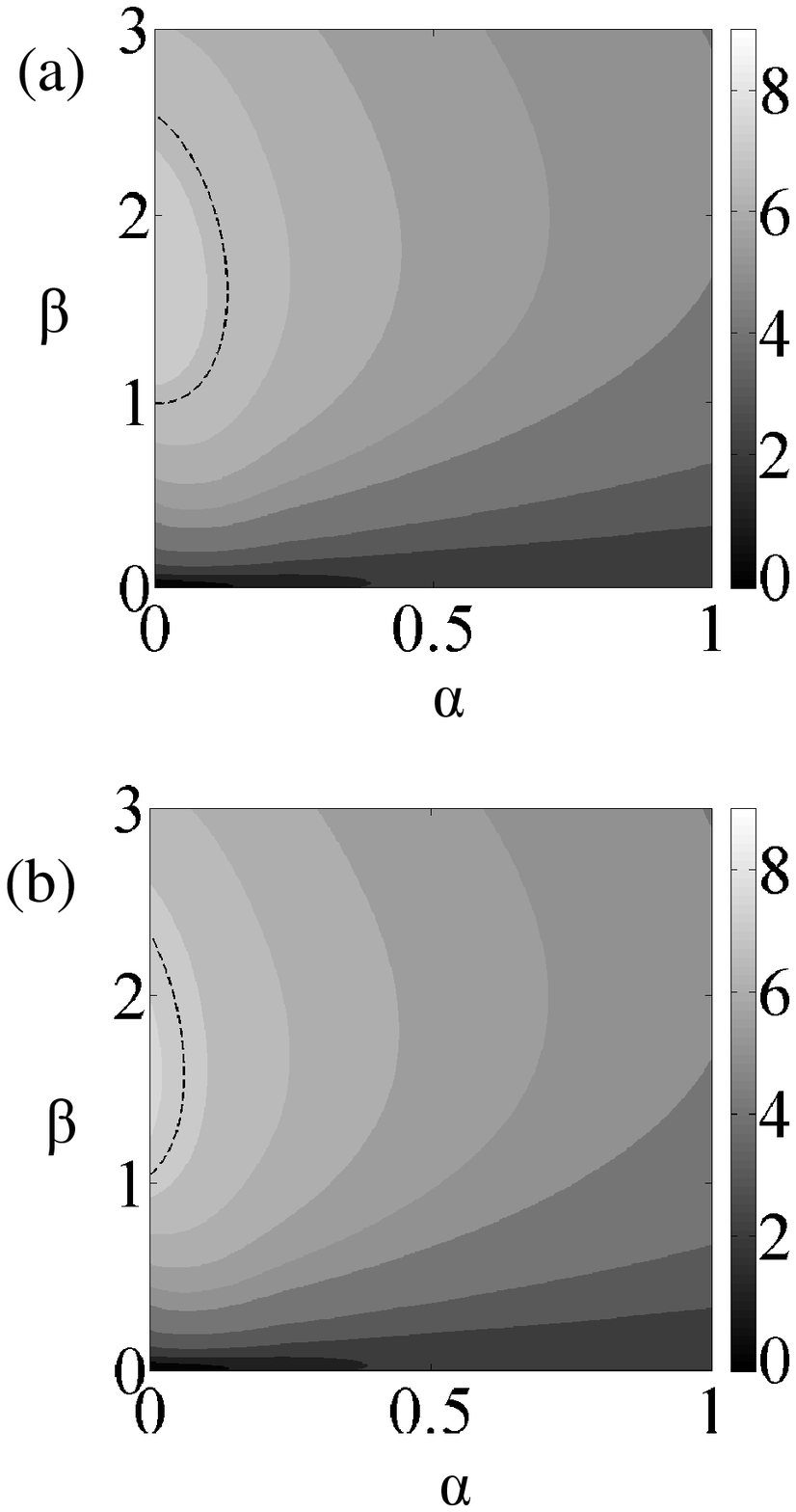,width=0.25\textwidth,keepaspectratio=true}\\
\vspace{5 mm}
\epsfig{file=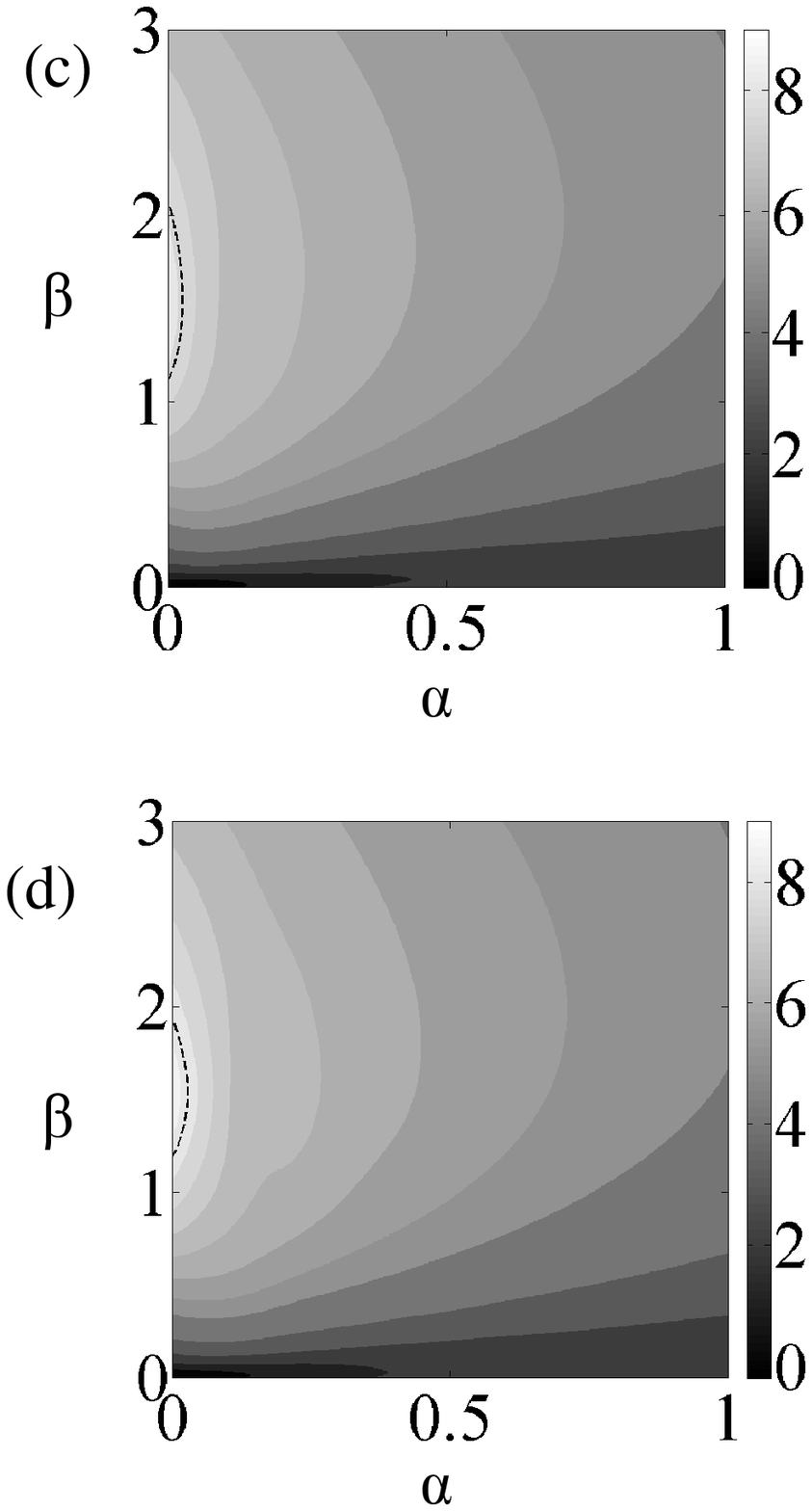,width=0.25\textwidth,keepaspectratio=true}
\caption{Contour plot of $\log G_{max}$ for RBC at $Re = 1000$, $Pr = 1$, and $(a) Ra = 0$, $(b) Ra = 500$, $(c) Ra = 1500$  and $(d) Ra = 1700$. The dashed lines correspond to iso-lines of the growth $G_{max} = \frac{2}{3}S$.}
\label{fig:Gmax_contourplot_RBC}
\end{figure}

\begin{figure}[h]
\epsfig{file=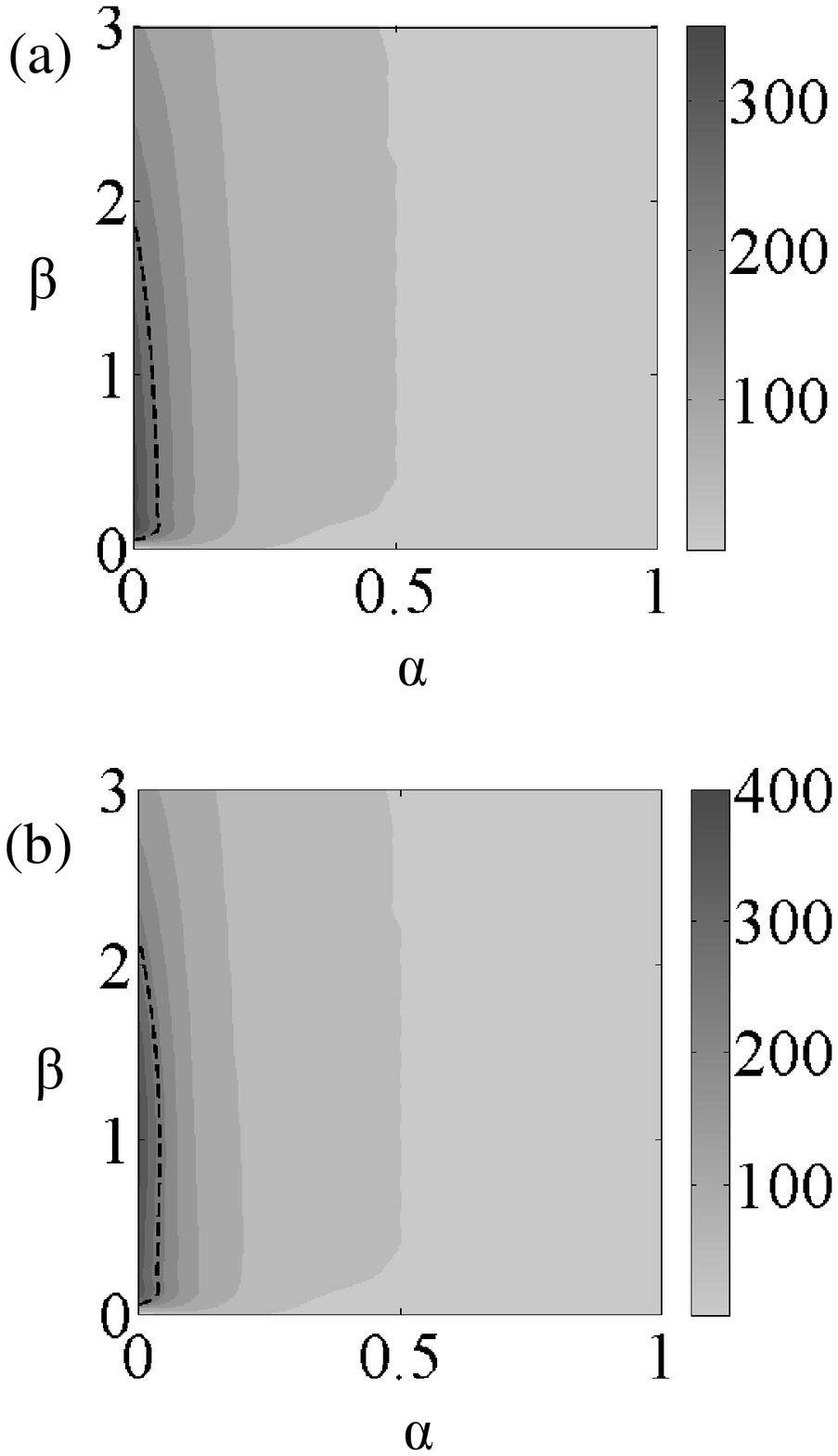,width=0.25\textwidth,keepaspectratio=true}\\
\vspace{5 mm}
\epsfig{file=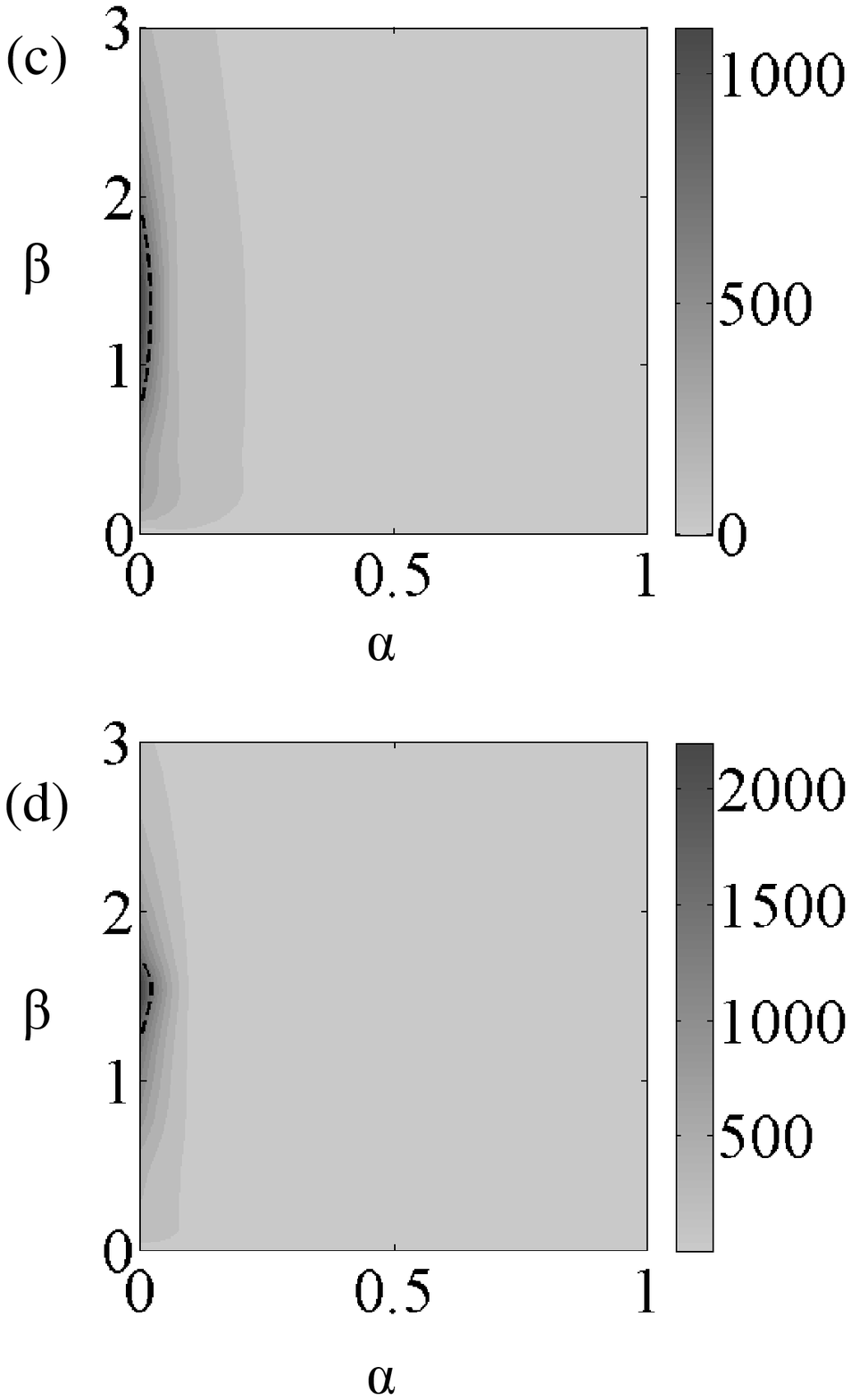,width=0.25\textwidth,keepaspectratio=true}
\caption{Contour plot of $t_{max}$ for RBC at $Re = 1000$, $Pr = 1$, and $(a) Ra = 0$, $(b) Ra = 500$, $(c) Ra = 1500$  and $(d) Ra = 1700$. The dashed lines correspond to iso-lines of $t_{max} = \frac{2}{3}T_{max}$, where $T_{max}$ is the global maximum of all $t_{max}$ in the $\alpha$-$\beta$ plane.}
\label{fig:tmax_contourplot_RBC}
\end{figure}

The equivalent plots for RBC flow are presented in figures \ref{fig:Gmax_contourplot_RBC} \& \ref{fig:tmax_contourplot_RBC}. They are qualitatively similar to figures \ref{fig:Gmax_contourplot} \& \ref{fig:tmax_contourplot}, respectively, except that the growth $S$ corresponds to a nearly-streamwise-uniform disturbance with $\alpha_{opt} << 1$ (Note: at large Reynolds numbers, in plane Couette flow \cite{Schmid_n_Henningson_2001} without a cross-stream temperature gradient, $\alpha_{opt} = \frac{35}{Re}$). As the Rayleigh number increases, however, $\alpha_{opt}$ approaches zero. In comparison with $RBP$, there is a more marked increase in $G_{max}$ and $t_{max}$ with $Ra$. As in $RBP$, this is primarily limited to streamwise-uniform and nearly-streamwise-uniform disturbances.

Thus, it is likely that lift-up remains the most dominant mechanism of transient growth and that the Orr-mechansism is negligibly affected by the presence of unstable stratification. The effect of Prandtl number will be discussed in section \ref{sec:Pr_Effect} but it may already be mentioned that this conclusion holds for all Prandtl numbers too.
%%\begin{figure}[h]
%%\centering
%%\begin{tabular}{c}
%%\includegraphics[width=0.5\textwidth,keepaspectratio]{Couette_Gmax_Ra_0.eps} \\ \includegraphics[width=0.5\textwidth,keepaspectratio]{Couette_Gmax_Ra_500.eps}\\
%%\includegraphics[width=0.5\textwidth,keepaspectratio]{Couette_Gmax_Ra_1500.eps} \\ \includegraphics[width=0.5\textwidth,keepaspectratio]{Couette_Gmax_Ra_1700.eps}\\
%%\end{tabular}
%%\caption{Contour plot of $G_{max}$ for RBP at $Re = 1000$, $Pr = 1$, and $Ra = 0$ (top) and $Ra = 1500$ (bottom); global maximum of $G_{max}$ is $196.04$ and $303.5$, respectively.}
%%\label{fig:Gmax_contourplot}
%%\end{figure}

%%\begin{figure}[h]
%%%\centering
%%\begin{flushleft}
%%\epsfig{file=Poiseuille_Gmax_contours.eps,width=0.6\textwidth,keepaspectratio=true}
%%\end{flushleft}
%%\caption{Contour plot of $G_{max}$ for RBP at $Re = 1000$, $Pr = 1$, and $(a) Ra = 0$, $(b) Ra = 500$, $(c) Ra = 1500$  and $(d) Ra = 1700$.}
%%\label{fig:Gmax_contourplot}
%%\end{figure}
%%
%%\begin{figure}[h]
%%\begin{flushleft}
%%\epsfig{file=Couette_Gmax_contours.eps,width=0.6\textwidth,keepaspectratio=true}
%%\end{flushleft}
%%\caption{Contour plot of $G_{max}$ for RBC at $Re = 1000$, $Pr = 1$, and $(a) Ra = 0$, $(b) Ra = 500$, $(c) Ra = 1500$  and $(d) Ra = 1700$.}
%%\label{fig:Gmax_contourplot_RBC}
%%\end{figure}
%%
%%\begin{figure}[h]
%%\begin{flushleft}
%%\epsfig{file=Poiseuille_tmax_contours.eps,width=0.6\textwidth,keepaspectratio=true}
%%\end{flushleft}
%%\caption{Contour plot of $t_{max}$ for RBP at $Re = 1000$, $Pr = 1$, and $(a) Ra = 0$, $(b) Ra = 500$, $(c) Ra = 1500$  and $(d) Ra = 1700$.}
%%\label{fig:tmax_contourplot}
%%\end{figure}
%%
%%\begin{figure}[h]
%%\begin{flushleft}
%%\epsfig{file=Couette_tmax_contours.eps,width=0.6\textwidth,keepaspectratio=true}
%%\end{flushleft}
%%\caption{Contour plot of $t_{max}$ for RBP at $Re = 1000$, $Pr = 1$, and $(a) Ra = 0$, $(b) Ra = 500$, $(c) Ra = 1500$  and $(d) Ra = 1700$.}
%%\label{fig:tmax_contourplot_RBC}
%%\end{figure}

\subsection{Effect of varying Reynolds number at a constant Rayleigh number}
\label{sec:NMS2}
In wall-bounded shear flows without stratification the optimal transient growth of streamwise-uniform perturbations scales as $Re^2$ at large Reynolds numbers \cite{Gustavsson_1991, Butler_n_Farrel_1992, Reddy_n_Henningson_1993}. This scaling is related to the presence of the large off-diagonal term in the linear operator \eqref{eq:LOBE_matrix}: the coupling term $-i \beta Re$ due to the basic flow shear appearing in the Squire equation for the wall-normal vorticity.
%When $Re >> 1$, the wall-normal vorticity in the normalized Orr-Sommerfeld mode is $\mathcal{O}(1)$ and the wall-normal velocity is $\mathcal{O}(1/Re)$. Whereas the normalized Squire mode, independent of $Re$, has no contribution from wall-normal velocity but an $\mathcal{O}(1)$ wall-normal vorticity. Since any perturbation can be written as a sum of Orr-Sommerfeld and Squire modes, if the initial perturbation has most of its kinetic energy from the wall-normal velocity, its kinetic energy can grow up to $\mathcal{O}(Re^2)$ before eventually decaying exponentially\cite{Reddy_n_Henningson_1993, Schmid_n_Henningson_2001}.
Physically, this transient growth is due to the presence of a non-zero initial wall-normal velocity perturbation in the form of streamwise-uniform vortices that feed the wall-normal vorticity (associated to the streamwise velocity) by the tilting of base flow vorticity through the so-called lift-up mechanism \cite{Ellingsen_n_Palm_1975, Landhal_1990} for all $t\geq0$. The influence of buoyancy on this scaling law is considered in this section and in section \ref{Re_scaling_theory}.

\begin{figure}[h]
\begin{flushleft}
\epsfig{file=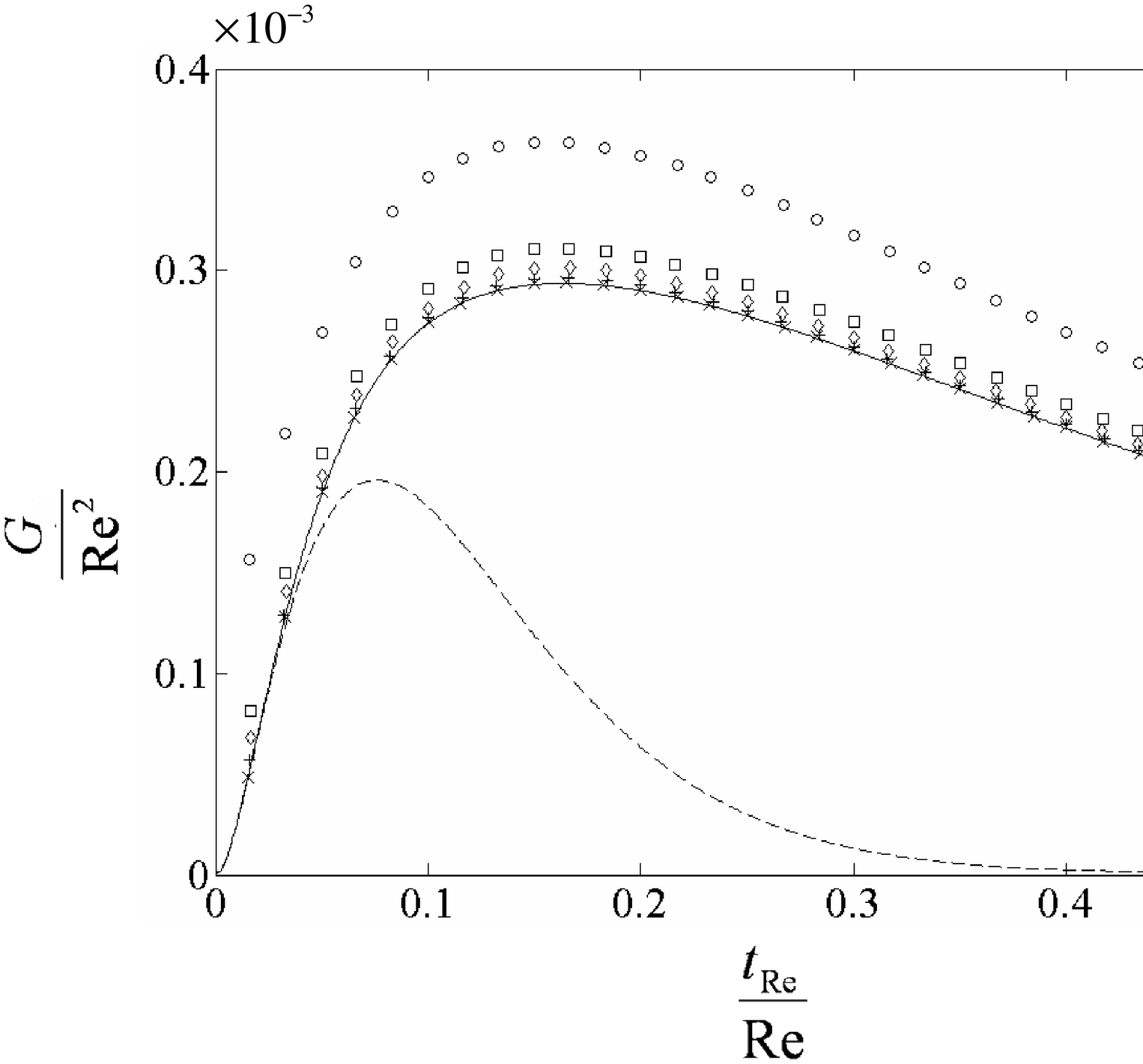,width=0.6\textwidth,keepaspectratio=true}
\end{flushleft}
\caption{Optimal growth curves at various Reynold numbers for streamwise-uniform disturbances in RBP at $Ra = 1500$ ($\alpha = 0$, $\beta = 2.04$): $\textbf{------}$ $Re = 5000$,  $\times$ $ Re = 1000$, $+$ $ Re = 500$, $\diamond$ $ Re = 300$, $\square$ $ Re = 200$, $\circ$ $ Re = 100$ and $---$ $Ra = 0$, $Re = 5000$}
\label{fig:lift_up_scaling_Poiseuille}
\end{figure}

\begin{figure}[h]
\begin{flushleft}
\epsfig{file=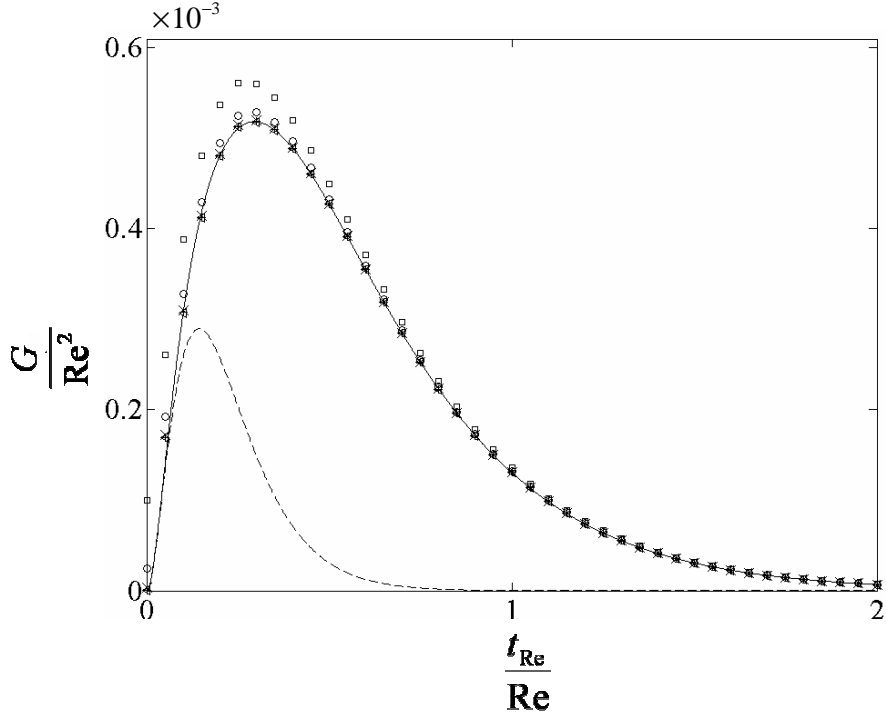,width=0.6\textwidth,keepaspectratio=true}
\end{flushleft}
\caption{Optimal growth curves at various Reynold numbers for streamwise-uniform disturbances in RBC at $Ra = 1000$ ($\alpha = 0$, $\beta = 1.558$): $\textbf{------}$ $Re = 5000$, $\triangleleft$ $ Re = 3000$,  $+$ $ Re = 2000$, $\times$ $ Re = 1000$, $\circ$ $ Re = 400$, $\square$ $ Re = 200$ and $---$ $Ra = 0$, $Re = 5000$}
\label{fig:lift_up_scaling_Couette}
\end{figure}

\begin{figure}[h]
\begin{flushleft}
\epsfig{file=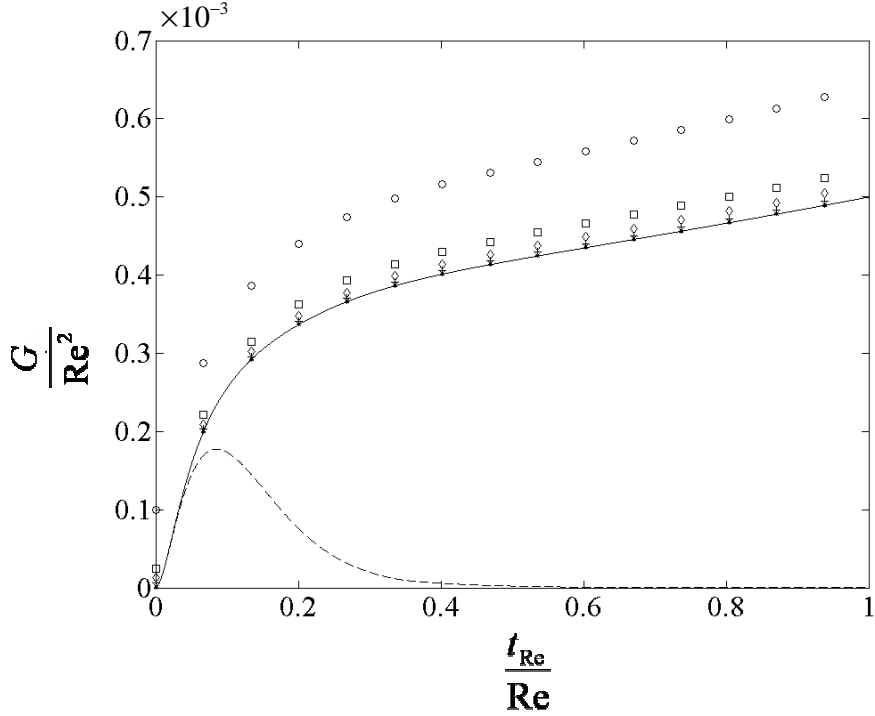,width=0.6\textwidth,keepaspectratio=true}
\end{flushleft}
\caption{Same as figure \ref{fig:lift_up_scaling_Couette} at a super critical Rayleigh number $Ra = 1800$}
\label{fig:lift_up_scaling_Poiseuille_sup_crit}
\end{figure}

In figure \ref{fig:lift_up_scaling_Poiseuille} the optimal transient growth of streamwise-uniform disturbances is displayed at $Ra = 0$ and $Re = 5000$ and also, at $Ra = 1500$ for different Reynolds numbers when $Pr = 1$. The optimal transient growth and the optimization time are scaled as ${G_{opt}}/{Re^2}$ and ${t_{Re}}/{Re}$, respectively, where $t_{Re}$ is the advective time scale scaled with respect to $U_{max}$. It is related to the non-dimensional time $t$ in equations \eqref{eq:LOBE1}, \eqref{eq:LOBE2} \& \eqref{eq:LOBE3} by the relation
\begin{align}
t_{Re} &= \frac{U_{max}}{{h}/{2}}t^{*} = \left( Re Pr \right) t,
\label{eq:time_scaling}
\end{align}
where $t^{*}$ denotes the dimensional time variable. As the Reynolds number increases all the symbols collapse on a single continuous curve, thereby confirming that the large Reynolds number scaling law of pure shear flows, i.e.\
\begin{align}
\frac{G}{Re^2} &= f \left(\frac{t_{Re}}{Re}\right),
\label{eq:large_Re_scaling}
\end{align}
remains valid even when buoyancy is destabilizing the flow. The scaling law fairly predicts the optimal gain even at Reynolds numbers as low as $200$. Biau and Bottaro \cite{Biau_n_Bottaro_2004} presented the same scaling law in their analysis of transient growth in the spatial framework for plane channel flow under the action of stable thermal stratification and, here, a similar result is observed in the temporal framework for an unstable temperature gradient in RBP flows. The scaling law is also valid for $RBC$ flow as seen in figure \ref{fig:lift_up_scaling_Couette}. Similar observations can be made for supercritical Rayleigh numbers $Ra > Ra_c^{RB}$. Figure \ref{fig:lift_up_scaling_Poiseuille_sup_crit} displays $G_{max}/Re^2$ against $t_{Re}/Re$ for various Reynolds numbers (different symbols) at a supercritical Rayleigh number $Ra = 1800$. The continuous line corresponds to $Re = 5000$ and the symbols collapse onto this curve for large Reynolds numbers which indicates that the scaling law \eqref{eq:large_Re_scaling} holds for $Ra > Ra_c^{RB}$ as well. Figure \ref{fig:lift_up_scaling_Poiseuille_sup_crit} corresponds to the case of $RBC$ flow. It still holds for $RBP$ flows (results not presented here). Thus, \textit{optimal transient growth of streamwise-uniform disturbances in  $RBP$ and $RBC$ flow at large Reynolds numbers under both stable and unstable temperature gradient retains the classical scaling law of the lift-up mechanism in pure shear flows at all Rayleigh numbers.}

The short-time transient growth, once rescaled as shown in figures \ref{fig:lift_up_scaling_Poiseuille} \& \ref{fig:lift_up_scaling_Couette}, is remarkably independent of $Re$ and $Ra$. The corresponding maximum optimal transient growth and the time at which it occurs depend only weakly on Rayleigh number. This suggests that the short-time transient growth is predominantly an inviscid process, as further examined in section \ref{sec:inviscid_lift_up}.

\subsection{Domain of Transient Growth}
\label{sec:monotonic_stability}
\begin{figure}[h]
\begin{flushleft}
\epsfig{file=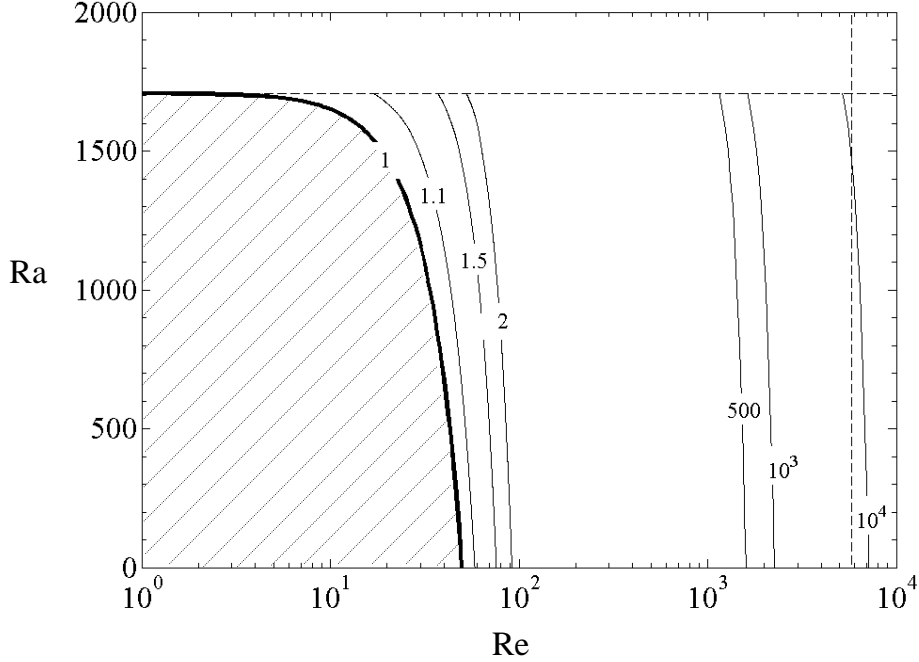,width=0.6\textwidth,keepaspectratio=true}
\end{flushleft}
\caption{Contours of global maximum transient growth $S$ in Rayleigh-B\'{e}nard-Poiseuille flow $\left( Pr = 1 \right)$. For any $Ra$ and $Re$ in the hatched region that is bounded by the thick line and the axes, $S = 1$ wherein the flow is monotonically stable.}
\label{fig:Gmax_contour_ReRa_plane_RBP}
\end{figure}
\begin{figure}[h]
\begin{flushleft}
\epsfig{file=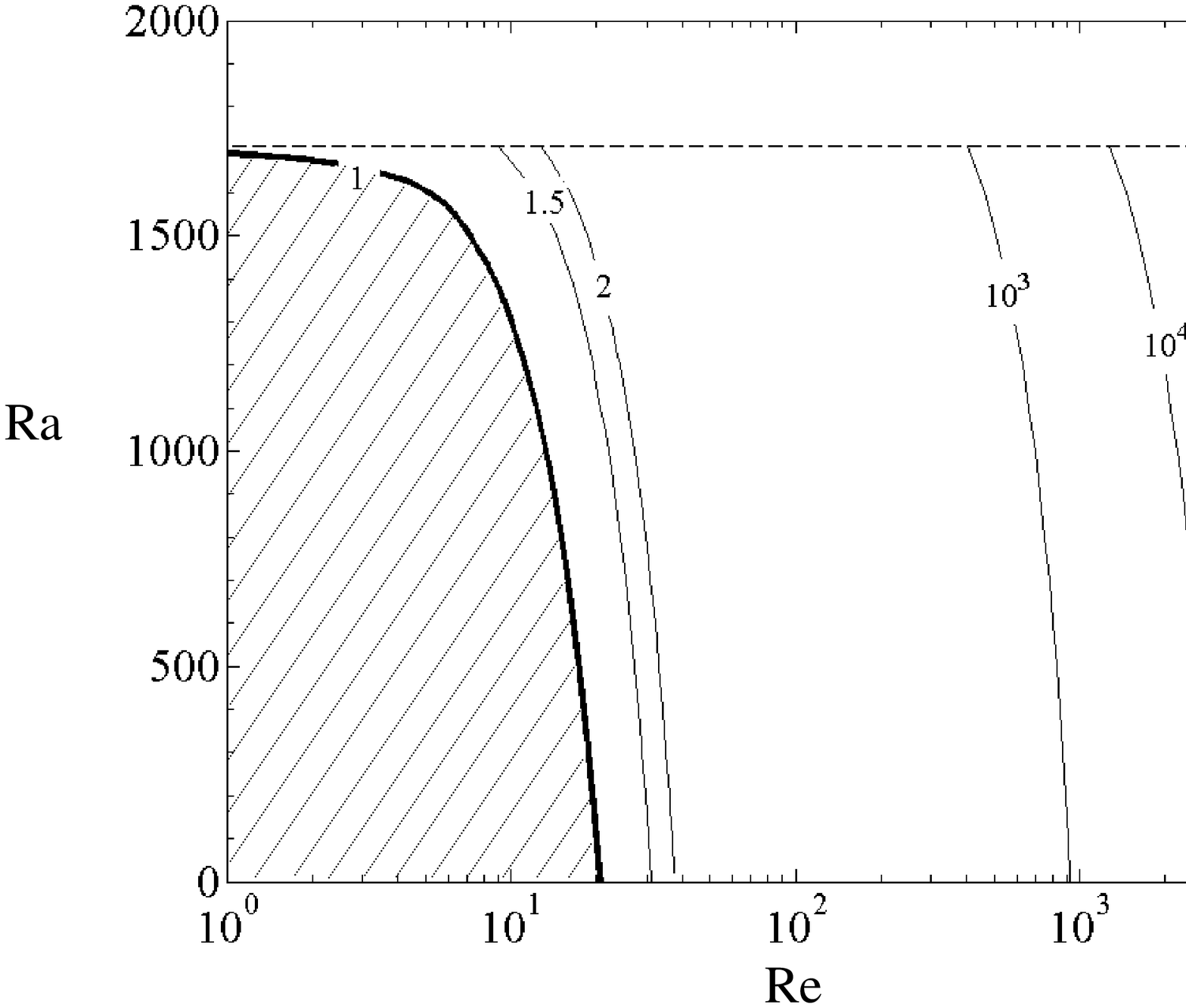,width=0.6\textwidth,keepaspectratio=true}
\end{flushleft}
\caption{Same as in figure \ref{fig:Gmax_contour_ReRa_plane_RBP} but for Rayleigh-B\'{e}nard-Couette flow.}
%\caption{Contours of constant global maximum optimal transient growth $S$ in Rayleigh-B\'{e}nard-Couette flow $\left( Pr = 1 \right)$. For any $Ra$ and $Re$ in the hatched region that is bounded by the thick line and the axes, there exists no disturbance that exhibits transient growth.}
\label{fig:Gmax_contour_ReRa_plane_RBC}
\end{figure}
A state is said to be monotonically stable if the perturbation energy, for any perturbation, decays monotonically to zero \cite{Schmid_n_Henningson_2001}. Along the same line of thought, it is appropriate to look for Rayleigh and Reynolds numbers at which $RBP$/$RBC$ flow does not exhibit transient growth. In terms of the growth function $G(t)$, it is the domain in the $Re$-$Ra$ plane where $G(t)$ is less than unity for all $\alpha$, $\beta$ and $t > 0$.

For $RBP$ and $RBC$ flow, the contours of the global maximum optimal transient growth $S$ in the stable region of the $Re$-$Ra$ plane are displayed in figures \ref{fig:Gmax_contour_ReRa_plane_RBP} and \ref{fig:Gmax_contour_ReRa_plane_RBC}, respectively. The hatched region in both plots represents the domain of no-transient-growth. At $Re = 0$, the flow is monotonically stable for all Rayleigh numbers up to $Ra_c^{RB}$ where the Rayleigh-B\'{e}nard instability occurs, a feature which is consistent with the classical results \cite{Chandra_1961} for the present choice of norm \eqref{eq:norm_RB}. Hence, the thick line meets the $Ra$-axis at $Ra = Ra_c^{RB}$ and the $Re$-axis at the critical Reynolds numbers, $49.6$ and $20.7$ for plane Poiseuille and plane Couette flow, respectively. These numbers match with the critical Reynolds number for monotonic decay of the kinetic energy as computed by Joseph \cite{Joseph_1976}. The iso-contours at large Reynolds numbers are nearly vertical (figures \ref{fig:Gmax_contour_ReRa_plane_RBP} \& \ref{fig:Gmax_contour_ReRa_plane_RBC}) for both $RBP$ and $RBC$ flows indicating that the effect of unstable stratification on the overall optimal transient growth $S$ is negligible.
\section{Transient growth of streamwise-uniform disturbances in RBP and RBC flows}
\label{sec:TG_longi}
\subsection{Lift-up Mechanism in the presence of temperature perturbations}
\label{sec:inviscid_lift_up}
In order to understand the transient dynamics of streamwise-uniform disturbances, the low-order model discussed by Schmid and Henningson\cite{Schmid_n_Henningson_2001} is extended to include temperature effects and buoyancy. Consider the following model of the linear operator \eqref{eq:LOBE_matrix} with 3 degrees of freedom

%It is possible to demonstrate that an inviscid lift-up mechanism accounts for short-time optimal growth in RBP and RBC configuration by making use of a reduced model of the governing equations. Such a model is constructed in this section on the basis of the following observations. There is no explicit temperature term $\tilde{\theta}(y, t)$ in the governing equation \eqref{eq:LOBE_matrix} for the wall-normal vorticity $\tilde{\eta}(y, t)$. The only forcing term in this equation is due to the wall-normal velocity $\tilde{v}(y, t)$ and it is $\mathcal{O}(Re Pr)$ if $\beta$ is of order unity. The coupling between $\tilde{v}(y, t)$ and $\tilde{\theta}(y, t)$ appears as terms $\mathcal{O}(Ra_{\mbox{\tiny{\textsl{h/2}}}}Pr)$ and $\mathcal{O}(1)$ in their respective evolution equations. When $\beta$ is of order unity, the dissipation term in the equation of $\tilde{\theta}(y, t)$ is only $\mathcal{O}(1)$ but it is $\mathcal{O}(Pr)$ in the equations of both $\tilde{v}(y, t)$ and $\tilde{\eta}(y, t)$. If $Re = 0$ the governing equations \eqref{eq:LOBE_matrix} become self-adjoint with respect to the scalar product \eqref{eq:energy_scalar_product}. When $\alpha = 0$ the operators $L_{OS}$ and $L_{LHE}$ are normal and their spectrum depends only on $Ra$ and $Pr$ with an unstable Rayleigh-B\'{e}nard convection roll appearing at $Ra = Ra_c^{RB}$, independently of the Reynolds and Prandtl numbers. Thus, for streamwise-uniform disturbances, can be taken as
\begin{equation}
\frac{d}{dt}
\begin{bmatrix}
  \check{v}      \\
  \check{\theta} \\ 
  \check{\eta}     
\end{bmatrix}
= 
\begin{bmatrix}
  	-bPr      														& \sqrt{Ra_{\mbox{\tiny{\textsl{h/2}}}} Pr}		& 0     \\
  	\sqrt{Ra_{\mbox{\tiny{\textsl{h/2}}}} Pr}		& -b 																		& 0 	\\ 
   	Re Pr																	& 0																			& -aPr
\end{bmatrix}
\begin{bmatrix}
  \check{v}      \\
  \check{\theta} \\ 
  \check{\eta}     
\end{bmatrix},
\label{eq:model}
\end{equation}
%%\begin{figure}[h]
%%\centering
%%\includegraphics[width=0.25\textwidth,keepaspectratio]{xtras/axes_model.eps}
%%\caption{Sketch illustrating the traid of the 3D model for lift-up mechanism}
%%\label{fig:schematic_axes}
%%\end{figure}
where the amplitudes of the field $\check{v}$, $\check{\theta}$ and $\check{\eta}$ are time-dependent only. The coefficients $a$, $b$ are positive and they are related to the eigenvalue of the linear operator. When $Ra_{\mbox{\tiny{\textsl{h/2}}}} = 0$, $\check{v}(t)$ is decoupled from $\check{\theta}(t)$ and the matrix is analogous to the 2D vector model presented in Schmid and Henningson$\cite{Schmid_n_Henningson_2001}$ to illustrate the nature of the lift-up mechanism in pure shear flows. The off-diagonal term $\sqrt{Ra_{\mbox{\tiny{\textsl{h/2}}}} Pr}$ makes the operator self-adjoint at $Re = 0$ so that the state vector $\begin{bmatrix} \check{v}(t),\check{\theta}(t), \check{\eta}(t) \end{bmatrix}^T$ does not exhibit any transient growth. On comparing the dispersion relation of the model with that of the linearised disturbance equations for pure conduction of a static fluid with free-slip boundary conditions, it can be seen that $b^2$ plays the role of the critical Rayleigh number characterizing the linear stability of pure conduction in Boussinesq fluids\footnote[1]{alternatively, one could have used three arbitrary constants, say, $a$, $b$, $c$; with $c$ in the diagonal term of the $\check{\theta}$ equation, thereby relating the critical Rayleigh number to $b$ and $c$}. Thus, the resulting operator \eqref{eq:model} is stable for all $a > 0$ and $Ra_{\mbox{\tiny{\textsl{h/2}}}}<b^2$. 

Such a model is hypothesized on the basis of the following observations. There is no explicit temperature term $\tilde{\theta}(y, t)$ in the governing equation \eqref{eq:LOBE_matrix} for the wall-normal vorticity $\tilde{\eta}(y, t)$. The only forcing term in this equation is due to the wall-normal velocity $\tilde{v}(y, t)$ and it is $\mathcal{O}(Re Pr)$ if $\beta$ is of order unity. Bearing in mind that $\gamma = \sqrt{\mathcal{O}(Ra_{\mbox{\tiny{\textsl{h/2}}}}Pr)}$, the coupling between $\tilde{v}(y, t)$ and $\tilde{\theta}(y, t)$ appears as terms $\mathcal{O}(Ra_{\mbox{\tiny{\textsl{h/2}}}}Pr)^{1/2}$in their respective evolution equations. When $\beta$ is of order unity, the dissipation term in the equation of $\tilde{\theta}(y, t)$ is only $\mathcal{O}(1)$ but it is $\mathcal{O}(Pr)$ in the equations of both $\tilde{v}(y, t)$ and $\tilde{\eta}(y, t)$. When $\alpha = 0$ the operators $L_{OS}$ and $L_{LHE}$ are normal and their spectrum depends only on $Ra$ and $Pr$ with an unstable Rayleigh-B\'{e}nard convection roll appearing at $Ra = Ra_c^{RB}$, independently of the Reynolds and Prandtl numbers. The reduced model \eqref{eq:model}, therefore, appears to be a good representation of the evolution equation of streamwise-uniform disturbances.

The behavior at small time $t$ can be obtained by expanding the solution of system \eqref{eq:model} about $t = 0$. One easily obtains:
%\vspace{2pt}
%\begin{widetext}
%\begin{eqnarray}
\begin{multline}
\begin{bmatrix}  
	\check{v} (t)      \\
  \check{\theta} (t) \\ 
  \check{\eta} (t)    
\end{bmatrix}
=
\begin{bmatrix}
  \check{v}_0\\
  \check{\theta}_0\\ 
  \check{\eta}_0     
\end{bmatrix}
+
\begin{bmatrix}
  -b\check{v}_0 + \sqrt{Ra_{\mbox{\tiny{\textsl{h/2}}}} Pr} \ \check{\theta_0}\\
  \sqrt{Ra_{\mbox{\tiny{\textsl{h/2}}}} Pr} \ \check{v}_0- b\check{\theta}_0\\ 
  RePr\ \check{v}_0 - aPr\ \check{\eta_0}
\end{bmatrix}
t \\
+
\begin{bmatrix}
  (b^2Pr+RaPr)\check{v}_0 - b(1 + Pr)\sqrt{Ra_{\mbox{\tiny{\textsl{h/2}}}} Pr} \ \check{\theta_0}\\
  - b(1 + Pr)\sqrt{Ra_{\mbox{\tiny{\textsl{h/2}}}} Pr} \ \check{v}_0 + (b^2+RaPr)\check{\theta_0}\\
  -(b+a)RePr\ \check{v}_0 + a^2\check{\eta_0} + RePr\sqrt{RaPr}\ \check{\theta_0}       
\end{bmatrix}t^2 
+ O(t^3),
\label{eq:short-time evolution}
\end{multline}
%\end{eqnarray}

%\end{widetext}
%%\begin{equation}
%%	\check{\eta} \left( t \right) =
%%	\begin{cases}
%%	\check{\eta}_0 - aPr\ \hat{\eta_0}t + a^2\check{\eta_0}t^2 \\
%%	+ RePr\ \check{v}_0t +\left( -(b+a)RePr\ \check{v}_0 + RePr\sqrt{\left| Ra \right|Pr}\ \hat{\theta_0} \right)t^2 + O(t^3),
%%	\end{cases}
%%\label{eq:short-time evolution}
%%\end{equation}
where $\left[\check{v}_0, \check{\theta}_0, \check{\eta}_0 \right]^T$ is the initial condition. In \eqref{eq:short-time evolution}, the largest contribution comes from the term proportional to $Re Pr$ in the expression for $\check{\eta}(t)$. It arises from the off-diagonal term in the model \eqref{eq:model} and so $\check{\eta}(t)$ will display algebraic growth in the presence of a non-zero initial condition on $\check{v}(t)$. This is identical to the classical algebraic growth for $t \sim O(\frac{1}{Re})$ of wall-normal vorticity \cite{Schmid_n_Henningson_2001} due to the lift-up mechanism in pure shear flows. Therefore, the growth of the disturbances will be led by $\check{\eta}$, which manifests itself through the appearance of low and high speed streaks. The effect of the initial perturbation temperature field $\check{\theta}_0$ is felt only in the terms $\mathcal{O}\left( t^2 \right)$ because $\check{\theta}$ does not directly force $\check{\eta}$. It affects, however, the decay rate of $\check{v}(t)$ which in turn forces $\check{\eta}$ through the lift-up mechanism. The $t^2$ term in system \eqref{eq:short-time evolution} is $\mathcal{O}(RePr \sqrt{Ra_{\mbox{\tiny{\textsl{h/2}}}} Pr})$ and becomes increasingly important at large Rayleigh numbers. Thus, the small-time expansion \eqref{eq:short-time evolution} suggests that \textit{the influence of buoyancy on the short-time energy growth of streamwise-uniform disturbances in parallel shear flows is only secondary compared to the classical lift-up mechanism.} The initial energy growth, therefore, scales as $Re^2$ at large Reynolds numbers.
\subsection{Reynolds number scaling for $G_{max}(\alpha, \beta; Re, Ra, Pr)$}
\label{Re_scaling_theory}
It is possible to estimate the behavior of $G_{max}(\alpha, \beta; Re, Ra, Pr)$ at $\alpha = 0$ (or small $\alpha Re$) at a fixed Rayleigh and Prandtl number by employing the method followed by Gustavsson \cite{Gustavsson_1991} and Reddy et. al. \cite{Reddy_n_Henningson_1993}. The following analysis is similar to that previously known for pure shear flows\cite{Reddy_n_Henningson_1993, Schmid_n_Henningson_2001, Luchini_2000}. If the wall-normal vorticity in the Squire equation is rescaled as $\bar{\eta} = \hat{\eta}/\beta Re$, equations \eqref{eq:LOBE_fourier1}, \eqref{eq:LOBE_fourier2} \& \eqref{eq:LOBE_fourier3} then depend on only two parameters, namely, $k^2 = \alpha^2 + \beta^2$ and $\alpha Re$ at a fixed Rayleigh number and Prandtl number. The norm \eqref{eq:norm_RB} of the perturbations in the new variables $\begin{bmatrix} \hat{v}, \hat{\theta}, \bar{\eta} \end{bmatrix}^T$ can then be expressed as
%%\begin{eqnarray}
%%E(t) = E_{(\hat{v}, \hat{\theta})}(t) + \frac{\beta^2}{k^2}Re^2 E_{(\hat{\eta})}(t)
%%\label{eq:norm_RB_rescaled2}
%%\end{eqnarray}
\begin{eqnarray}
E(t) = \int^{1}_{-1}{\left( \left| \hat{v} \right|^2 + \frac{1}{k^2} \left| D\hat{v} \right|^2 +  \left| Ra_{\mbox{\tiny{\textsl{h/2}}}} \right| Pr \left| \hat{\theta} \right|^2 \right)}dy + 
\frac{1}{2} \frac{\beta^2}{k^2} Re^2 \int^{1}_{-1}{\left| \bar{\eta} \right|^2} dy.
\label{eq:norm_RB_rescaled}
\end{eqnarray}
The first bracketed term, defined as $E_{(\hat{v}, \hat{\theta})}(t)$, is the contribution to the energy from only the wall-normal velocity $\hat{v}$ and temperature $\hat{\theta}$ and the second integral, defined as $E_{(\hat{\eta})}(t)$, is the contribution from the wall-normal vorticity $\hat{\eta}$ alone. When $\alpha = 0$, the evolution equations for the wall-normal velocity and the temperature perturbations are independent of Reynolds number and they are identical to the linearised Oberbeck-Boussinesq equations in the linear stability analysis of pure conduction in Boussinesq fluids. If $L_{(\hat{v}, \hat{\theta})}$ denotes this coupled linear operator, then $\begin{bmatrix} \tilde{v}(y, t), \tilde{\theta} (y, t) \end{bmatrix}^T = \exp \mbox{\{} -\mbox{i}L_{(\hat{v}, \hat{\theta})}t\mbox{\}} \begin{bmatrix} \tilde{v}(y, 0), \tilde{\theta} (y, 0) \end{bmatrix}^T$. Since the operator is normal for the present choice of norm $E_{(\hat{v}, \hat{\theta})}$ issued from the norm \eqref{eq:norm_RB} and since its spectrum lies in the lower half-plane for all $k$, the Hille-Yosida theorem \cite{Reddy_n_Henningson_1993} implies that $E_{(\hat{v}, \hat{\theta})}(t) \leq E_{(\hat{v}, \hat{\theta})}(0)$. Furthermore, the wall-normal vorticity is governed by the Squire operator $L_{SQ}$ which is forced by $\hat{v}$ but not by $\hat{\theta}$ in \eqref{eq:LOBE_matrix}. If the initial wall-normal velocity were zero, $E_{(\hat{\eta})}(t)$ would decrease monotonically given that the Squire equation is simply a diffusion equation.
%Thus, in order to achieve a large transient growth, the initial perturbation $\begin{bmatrix} \tilde{v}(y, 0), \tilde{\theta} (y, 0), \bar{\eta} (y, 0) \end{bmatrix}^T$ should be chosen so that most of the initial energy is in the velocity and temperature perturbations:
%\begin{equation}
%E_{(\hat{v}, \hat{\theta})}(0) >> Re^2 E_{(\hat{\eta})}(0).
%\label{eq:eng_arg1}
%\end{equation}
%Since, $E_{(\hat{v}, \hat{\theta})}(t)$ does not grow, it follows that the perturbations that experience the maximum growth satisfy
%\begin{equation}
%Re^2 E_{(\hat{\eta})}(\bar{t}) >> E_{(\hat{v}, \hat{\theta})}(\bar{t})
%\label{eq:eng_arg2.1}
%\end{equation}
%where, $\bar{t} \approx t_{max}$, the time taken to achieve the maximum transient growth $G_{max}$. 
The definition of the growth function \eqref{eq:trans_gwth} gives
\begin{align}
	G_{max} &=  \operatorname*{max}_{\forall \textbf{\textit{q}}(t_0) \neq  \textbf{\textit{0}}, t \geq 0}  \left[\frac{E_{(\hat{v}, \hat{\theta})}(\bar{t}) + 
	Re^2 E_{(\hat{\eta})}(\bar{t})}{E_{(\hat{v}, \hat{\theta})}(0) + Re^2 E_{(\hat{\eta})}(0)}\right],
	\label{eq:eng_scaling_proof_blah}
\end{align}
and at large Reynolds numbers, in order to achieve a large transient growth, the initial perturbation should be chosen so that most of the initial energy is in the velocity and temperature perturbations:
\begin{equation}
E_{(\hat{v}, \hat{\theta})}(0) >> Re^2 E_{(\hat{\eta})}(0).
\label{eq:eng_arg1}
\end{equation}
Since, $E_{(\hat{v}, \hat{\theta})}(t)$ does not grow, it follows that, if $\bar{t} \approx t_{max}$, the time taken to achieve the maximum transient growth $G_{max}$, at large $Re$ the perturbations that experience the maximum growth satisfy
\begin{equation}
Re^2 E_{(\hat{\eta})}(\bar{t}) >> E_{(\hat{v}, \hat{\theta})}(\bar{t}),
\label{eq:eng_arg2.1}
\end{equation}
thus, for $Re >> 1$,
\begin{align}
	G_{max} &\approx Re^2 \operatorname*{max}_{\forall \textbf{\textit{q}}(t_0) \neq  \textbf{\textit{0}}, t \geq 0}  \left[\frac{E_{(\hat{\eta})}(\bar{t})}{E_{(\hat{v}, \hat{\theta})}(0)}\right].
	\label{eq:eng_scaling_proved}
\end{align}
The measures $E_{(\hat{v}, \hat{\theta})}$ and $E_{(\hat{\eta})}$ are of order unity or less and they depend on the state variables which in turn depend only on the wavenumber $k = \beta$ (since $\alpha = 0$), $Ra$ and $Pr$. Thus, the above expression simply becomes 
\begin{align}
	G_{max} &\approx Re^2 \zeta(\beta; Ra, Pr),
	\label{eq:eng_scaling_proved_again}
\end{align}
where, $\zeta(\beta; Ra, Pr)$ is some function of the spanwise wavenumber $\beta$, $Ra$ and $Pr$. Note that this scaling relation holds both for $RBP$ and $RBC$ and becomes more accurate at large Reynolds numbers.

\subsection{Long-time Optimal Response}
\label{long_time_opt}
The domain being finite in the eigenfunction direction $y$, the DiPrima-Habetler theorem\cite{DiPrima_Habetler_1969} applies and the spectrum is discrete and complete\cite{Herron_1980}. The solution of the direct equations \eqref{eq:LOBE_matrix} (and also the adjoint equations \eqref{eq:LOBEadjoint_matrix}) may be expanded as
\begin{equation}
\textbf{\textit{q}} \left( t \right) = \sum_j{c_{j}\phi_{j}\exp \left( -i\omega_{j}t \right)},
\label{eq:eig_expn}
\end{equation}
where $\omega_{j}$ and $\phi_{j}$ are the eigenvalues and eigenfunctions of the linear operator \eqref{eq:LOBE_matrix} and $c_{j}$ are complex components of $q(t)$ along $\phi_{j}$. If $\omega_{1}$ is the eigenvalue with the largest imaginary part, it should lead the large time dynamics of $q(t)$: 	
\begin{equation*}
%\begin{array}{cll}
%\lim_{t \to \infty} \textbf{\textit{q}} \left( t \right) = {c_{(1)}\phi_{(1)}\exp \left( -i\omega_{1}t \right)},
 \mathop{}_{t \rightarrow \infty}^{\mbox{lim}} \textbf{\textit{q}} \left( t \right) = {c_{1}\phi_{1}\exp \left( -i\omega_{1}t \right)},
%\end{array}
\end{equation*}
and the constant $c_{1}$ is given by
\begin{equation}
c_{1} = \frac{\langle \textbf{\textit{q}}\left( t = 0 \right),\ \phi_{A1}\rangle_\gamma}{\langle \phi_{1},\ \phi_{A1}\rangle_\gamma},
\label{eq:c_1}
\end{equation}
where $\textbf{\textit{q}}\left( t = 0 \right)$ is the initial condition and $\phi_{A1}$ the adjoint eigenfunction associated with $\phi_{1}$. This demonstrates the classical result that the optimal initial perturbation for the large time dynamics is the adjoint of the leading eigenmode.

As noticed already, in the direct equations \eqref{eq:LOBE_matrix} the coupled linear operator for $\hat{v}$ and $\hat{\theta}$ is independent of $\hat{\eta}$ and the Squire equation for $\hat{\eta}$ is forced by the solution of this coupled operator. Hence, in general, the solution to the direct equations can be written in terms of the eigenfunction expansion \eqref{eq:eig_expn}, splitting modes in two families, namely, Orr-Sommerfeld-Oberbeck-Boussinesq modes  (OSOB modes) and Squire modes (Sq-modes):
%\begin{widetext}
\begin{eqnarray}
\begin{bmatrix}
	\tilde{v}(y, t)\\
	\tilde{\theta}(y, t)\\
	\tilde{\eta}(y, t)\\
\end{bmatrix}
=
\sum_j{
A_{j} \mbox{exp} \left( -i\lambda_{j}t \right)
\begin{bmatrix}
	\hat{v}_j(y)\\
	\hat{\theta}_j(y)\\
	\hat{\eta}^p_j(y)\\
\end{bmatrix}
}
 +
\sum_j{B_{j} \mbox{exp} \left( -i\mu_{j}t \right)
\begin{bmatrix}
	0\\
	0\\
	\hat{\eta}_j(y)\\
\end{bmatrix}},
\label{eq:eig_expn_full}
\end{eqnarray}
%\end{widetext}
where $\left\{\lambda_j\right\}$ are the OSOB eigenvalues of the coupled equations \eqref{eq:LOBE_fourier1} and \eqref{eq:LOBE_fourier2} involving $\hat{v}$ and $\hat{\theta}$ only, $\mbox{\{}\hat{\eta}^p_j\mbox{\}}$ are the forced wall-normal vorticity functions, and  $\left\{\mu_j\right\}$ are the eigenvalues of the Squire equation. The coefficients $\mbox{\{}A_j\mbox{\}}$ and $\mbox{\{}B_j\mbox{\}}$ are complex constants that can be determined from the initial conditions on the state variables.
%Let us denote the vector eigenfunctions in the first sum as the OSOB modes (Orr-Sommerfeld-Oberbeck-Boussinesq modes), wherein the functions $\mbox{\{}\hat{v}_j\mbox{\}}$ and $\mbox{\{}\hat{\theta}_j\mbox{\}}$ are the eigenfunctions of \eqref{eq:LOBE_fourier1} and \eqref{eq:LOBE_fourier2}; the functions $\mbox{\{}\hat{\eta}^p_j\mbox{\}}$ are the forced wall-normal vorticity functions. The eigenfunctions of the Squire equation have been represented by $\mbox{\{}\hat{\eta}_j\mbox{\}}$ and they form the Sq-modes (Squire modes) as given in the second sum. The coefficients $\mbox{\{}A_j\mbox{\}}$ and $\mbox{\{}B_j\mbox{\}}$ are constants that can be determined from the initial conditions on the state variables.
In the case of the adjoint linear operator \eqref{eq:LOBEadjoint_matrix}, it is $\hat{\eta}_{A}$ that forces the adjoint wall-normal velocity and temperature. The adjoint Squire equation is independent of the adjoint wall-normal velocity. Similarly to the expansion \eqref{eq:eig_expn_full}, the solution to the adjoint equations can be written as
%\begin{widetext}
\begin{eqnarray}
\begin{bmatrix}
	\tilde{v}_{A}(y, t)\\
	\tilde{\theta}_{A}(y, t)\\
	\tilde{\eta}_{A}(y, t)\\
\end{bmatrix}
=
\sum_j{
A_{Aj} \mbox{exp} \left( -i\lambda^*_{j}t \right)
\begin{bmatrix}
	\hat{v}_{Aj}(y)\\
	\hat{\theta}_{Aj}(y)\\
	0\\
\end{bmatrix}
}
+
\sum_j{
B_{Aj} \mbox{exp} \left( -i\mu^*_{j}t \right)
\begin{bmatrix}
	\hat{v}_{Aj}^{p}(y)\\
	\hat{\theta}_{Aj}^{p}(y)\\
	\hat{\eta}_{Aj}(y)\\
\end{bmatrix}
},
\label{eq:adj_eig_expn_full}
\end{eqnarray}
%\end{widetext}
where $^*$ on the eigenvalues denotes the complex conjugate and the vector eigenfunctions in the first sum are the adjoint OSOB modes corresponding to the homogeneous part of the coupled linear operator of the adjoint variables $\hat{v}_{A}$ and $\hat{\theta}_{A}$ in equation \eqref{eq:LOBEadjoint_matrix}. The vector eigenfunctions in the second sum are the adjoint Squire modes, wherein the functions $\mbox{\{}\hat{\eta}_{Aj}\mbox{\}}$ are the eigenfunctions of adjoint Squire operator and the functions  $\mbox{\{}\hat{v}_{Aj}^{p}\mbox{\}}$ and $\mbox{\{}\hat{\theta}_{Aj}^{p}\mbox{\}}$ are the corresponding forced wall-normal velocity and temperature functions. The coefficients $\mbox{\{}A_{Aj}\mbox{\}}$ and $\mbox{\{}B_{Aj}\mbox{\}}$ are complex components in the direction of the adjoint eigenmodes. Note that the above eigenfunction formulation is valid only if the eigenvalues $\left\{\lambda_j\right\}$ and $\left\{\mu_j\right\}$ are distinct which is the case except for a set of parameters of zero measure.

If only streamwise-uniform disturbances ($\alpha = 0$) are considered, the direct and adjoint equations of RBP (or RBC) flow given by \eqref{eq:LOBE_matrix} and \eqref{eq:LOBEadjoint_matrix} pertaining to the scalar product \eqref{eq:energy_scalar_product} with $\gamma^2 = |Ra_{\mbox{\tiny{\textsl{h/2}}}}|Pr$ become
%\begin{widetext}
%%\begin{multline}
%%	i\omega
%%		\begin{bmatrix}
%%			-D_\beta^2	&0	&0\\
%%			0	&1	&0\\
%%			0	&0	&1
%%		\end{bmatrix} 
%%		\begin{bmatrix}
%%			\hat{v}\\
%%			\hat{\theta}\\
%%			\hat{\eta}
%%		\end{bmatrix}
%%		\\ =
%%		\begin{bmatrix}
%%			Pr D_\beta^4	&-\beta^2 |Ra_{\mbox{\tiny{\textsl{h/2}}}}|Pr &0\\
%%			-1 &-D_\beta^2 &0\\
%%			i\beta \left( Re Pr \right)\frac{dU_0}{dy}	&0 &-Pr D_\beta^2
%%		\end{bmatrix}
%%		\begin{bmatrix}
%%			\hat{v}\\
%%			\hat{\theta}\\
%%			\hat{\eta}
%%		\end{bmatrix}
%%	\label{eq: L_longi2}
%%\end{multline}
%%and
%%\begin{multline}
%%		-i\omega^*
%%		\begin{bmatrix}
%%			-D_\beta^2	&0	&0\\
%%			0	&1	&0\\
%%			0	&0	&1
%%		\end{bmatrix} 
%%		\begin{bmatrix}
%%			\hat{v}_{A}\\
%%			\hat{\theta}_{A}\\
%%			\hat{\eta}_{A}
%%		\end{bmatrix}
%%		\\=
%%		\begin{bmatrix}
%%			Pr D_\beta^4	&-\beta^2 |Ra_{\mbox{\tiny{\textsl{h/2}}}}|Pr &-i\beta \left( Re Pr \right)\frac{dU_0}{dy}\\
%%			-1 &-D_\beta^2 &0\\
%%			0	&0 &-Pr D_\beta^2
%%		\end{bmatrix}
%%		\begin{bmatrix}
%%			\hat{v}_{A}\\
%%			\hat{\theta}_{A}\\
%%			\hat{\eta}_{A}
%%		\end{bmatrix},
%%	\label{eq: L_longi}
%%\end{multline}
%\end{widetext}
identical except for the coupling term between wall-normal velocity and wall-normal vorticity. This term is dependent on Reynolds and Prandtl numbers and independent of Rayleigh number. In the direct equations, the wall-normal vorticity is forced by the wall-normal velocity and its dominant streamwise-uniform eigenmode, corresponding to the Rayleigh-B\'{e}nard convection roll, has a non-zero streamwise velocity component when $Re \neq 0$ and  $Pr \neq 0$. Whereas, in the adjoint equations, it is the wall-normal vorticity term that forces the wall-normal velocity as seen in the eigenfunction expansion shown in \eqref{eq:adj_eig_expn_full}. Hence, the governing equations \eqref{eq:LOBEadjoint_matrix} corresponding to adjoint streamwise-uniform OSOB modes $\begin{bmatrix}	\hat{v}_{Aj}, \hat{\theta}_{Aj}, 0 \end{bmatrix}^T$ become identical with those of the pure conduction problem where the least stable eigenmode is, indeed, the Rayleigh-B\'{e}nard mode with zero wall-normal vorticity and therefore zero streamwise velocity. Since the eigenvalues of the adjoint modes are complex conjugates of those of the corresponding direct modes, $Ra \neq 0$, the adjoint of the leading eigenmode is the Rayleigh-B\'{e}nard mode without its contribution from the wall-normal vorticity, or more precisely, zero streamwise velocity. 

Thus, \textit{the Rayleigh-B\'{e}nard convection mode without streamwise velocity is the optimal input to obtain the largest long-time response from RBP/RBC flows.} Note that this result is independent of Reynolds number and Prandtl number.

\subsection{Transient Growth at arbitrary time}
\label{sec:domi_T}
\begin{figure}[h]
\begin{flushleft}
\epsfig{file=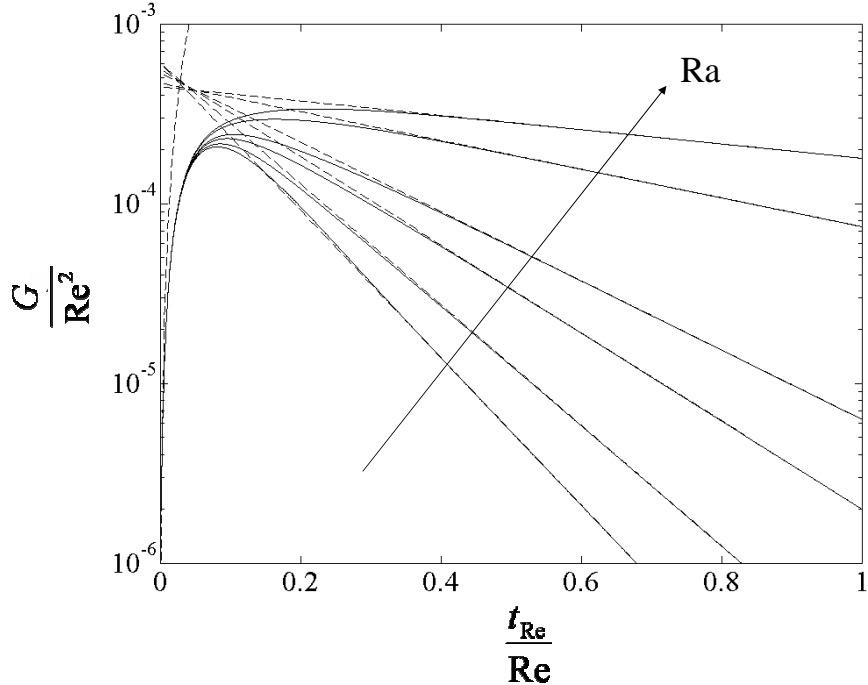,width=0.6\textwidth,keepaspectratio=true}
\end{flushleft}
\caption{$RBP$ flow: Comparison between computed optimal gain curve $\left( \textbf{------} \right)$ and the asymptotic estimates for short and large times $\left( --- \right)$ at various Rayleigh numbers (from inside to outside: $Ra = 300, 500, 800, 1000, 1500, 1700$) at $Re = 1000$, $Pr = 1$, $\alpha = 0$ and $\beta = 2.04$}
\label{fig:shape_of_opt_curve_Re_1000_Ra_various}
\end{figure}

\begin{figure}[h]
\begin{flushleft}
\epsfig{file=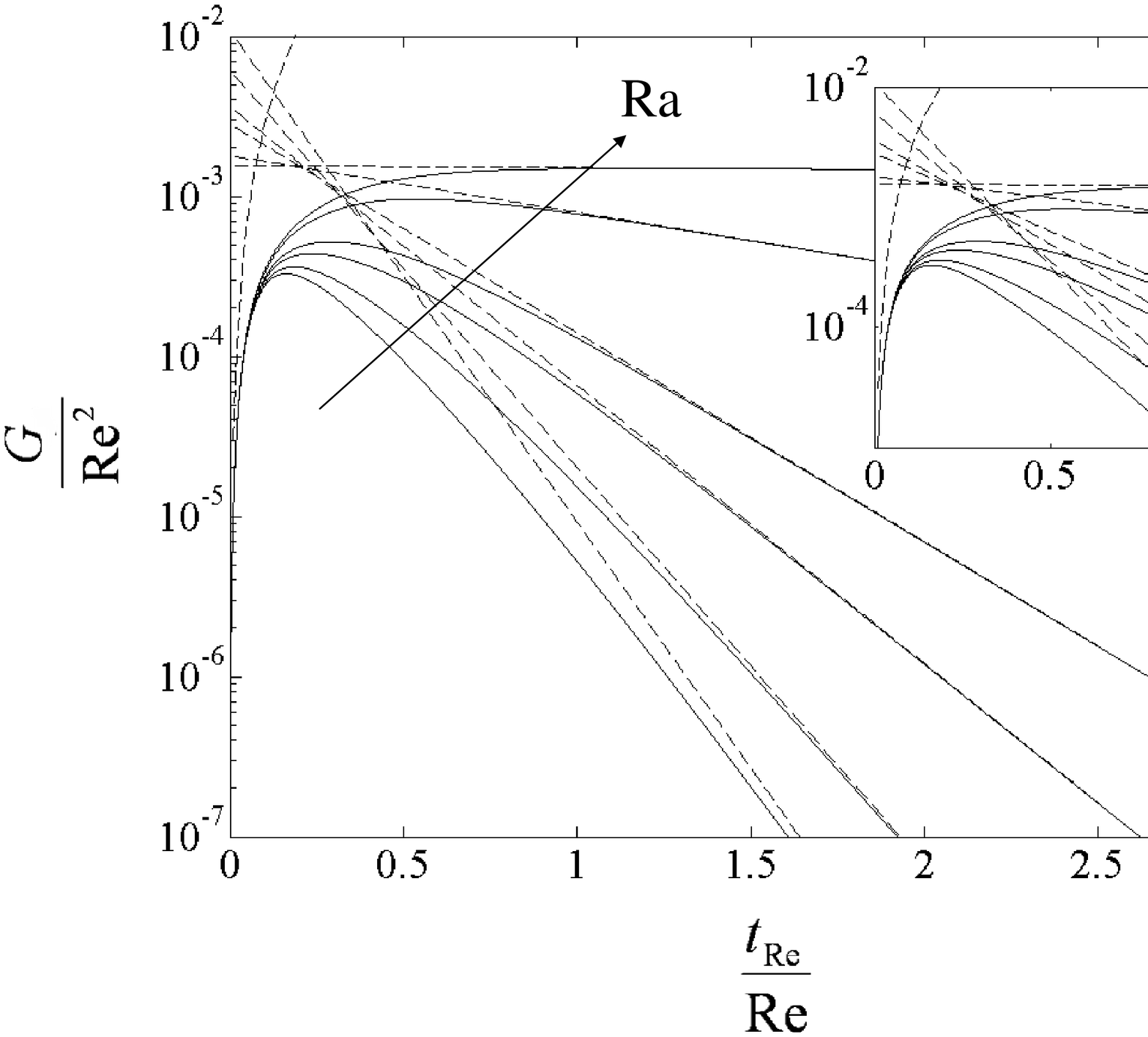,width=0.6\textwidth,keepaspectratio=true}
\end{flushleft}
\caption{$RBC$ flow: Comparison between computed optimal gain curve $\left( \textbf{------} \right)$ and the asymptotic estimates for short and large times $\left( --- \right)$ at various Rayleigh numbers (from inside to outside: $Ra = 300, 500, 800, 1000, 1500, 1700$) at $Re = 1000$, $Pr = 1$, $\alpha = 0$ and $\beta = 1.558$}
\label{fig:shape_of_opt_curve_Re_2000_Ra_various_RBC}
\end{figure}

Figures \ref{fig:shape_of_opt_curve_Re_1000_Ra_various} and \ref{fig:shape_of_opt_curve_Re_2000_Ra_various_RBC} display semi-log plots of the optimal gain (continuous line) versus time for a fixed Reynolds number at various Rayleigh numbers up to $Ra_c^{RB}$. All the continuous curves are identical for small time until close to the maximum optimal gain which is larger for large Rayleigh numbers. At later times, however, they separate and decay at a rate which decreases with increasing $Ra$. The slope of the optimal growth curve at large time corresponds to the exponential decay rate of the least stable eigenmode ($RB$), thereby providing a justification for the slower decay rate at larger $Ra$. The y-intercept (t = 0) of the asymptotic straight line defines the extra gain which according to \eqref{eq:c_1} may be estimated as $\frac{1}{Re^2}\frac{\left\|\phi_{1}\right\|^2 \left\|\phi_{A1}\right\|^2}{\left|\langle \phi_{1},\ \phi_{A1}\rangle_\gamma \right|^2}$ for large times (in figures \ref{fig:shape_of_opt_curve_Re_1000_Ra_various} and \ref{fig:shape_of_opt_curve_Re_2000_Ra_various_RBC}, $\gamma = \sqrt{\left| Ra_{\mbox{\tiny{\textsl{h/2}}}} \right|Pr}$). The dashed curves represent the long-time asymptote as estimated with this extra gain and the slope is obtained from the imaginary part of the dominant eigenvalue ($RB$-mode). It is observed that the prediction at large times is excellent. At short times, the dashed line represents the prediction from the pure lift-up mechanism wherein the wall-normal velocity forces the wall-normal vorticity.  In effect it represents the inviscid optimal growth and it is computed numerically at very large Reynolds numbers up to $10^8$ in the case without thermal stratification (see appendix for justification). It fits remarkably well with the computed short-time optimal gain (continuous curves) for all $Ra$. \textit{Indeed, the entire optimal growth curve is well approximated by the piecewise continuous curve consisting of a linear branch at short-times that is independent of $Ra$ and an exponentially decreasing branch at large times given by $G \sim Re^2 \left|\exp{(2 \omega_1 t)}\right|$, where $\omega_1$ is the complex eigenvalue of the $RB$-mode.}

\begin{figure}[h]
\begin{flushleft}
\epsfig{file=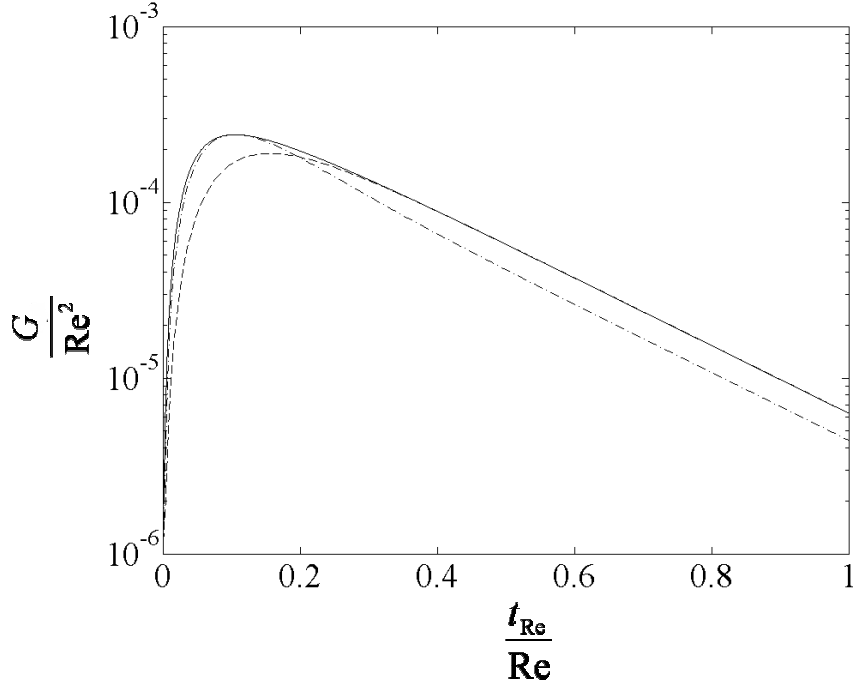,width=0.6\textwidth,keepaspectratio=true}
\end{flushleft}
\caption{Time history of the growth $\frac{\left\| \textbf{\textit{q}} \right\|^2_{RB}}{\left\| \textbf{\textit{q}}_{0} \right\|^2_{RB}}$ with $\textbf{\textit{q}}_{0}$ being the optimal input at each time horizon $\left( \textbf{------} \right)$; optimal input corresponding to $G_{max}$ $\left( \cdot-\cdot-\cdot \right)$ and dominant-adjoint-mode input $\left( --- \right)$ at $Re = 1000$, $Ra = 1000$, $Pr = 1$, $\alpha = 0$ and $\beta = 1.558$.}
\label{fig:adjoint_mode_inpt_resp}
\end{figure}

Figure \ref{fig:adjoint_mode_inpt_resp} compares optimal gain curve versus time (continuous line) with the response to different inputs at $Re = 1000$ and $Ra = 1000$: optimal input corresponding to $G_{max}$ (dot-dashed line) and the classical $RB$-mode without streamwise velocity which is the adjoint dominant eigenmode. Both responses exhibit transient growth $\mathcal{O}\left(G_{max}\right)$ at $t_{Re} \approx \frac{1}{4}Re$ and eventually decay monotonically. This implies that the adjoint dominant eigenmode is a good approximation to the optimal initial condition at all times. The transient growth mechanism is similar at all times and it is well-approximated by pure Rayleigh-B\'enard rolls. Thus, \textit{the dominant optimal growth in the presence of a cross-stream temperature gradient is due to streamwise vortices in the form of Rayleigh-Bénard convection rolls that act in tandem with the inviscid lift-up mechanism to produce large streamwise streaks $\mathcal{O}(Re)$ which eventually decay exponentially in time.}

%The optimal input for the lift-up mechanism in pure shear flows is in the form of streamwise-uniform vortex motion with a streamwise velocity $\mathcal{O}\left( \frac{1}{Re} \right)$. This suggests that the optimal streamwise-uniform perturbation could possibly be in the form of buoyancy-induced vortex motions, namely, streamwise-uniform convection rolls. Since the streamwise velocity is not explicitly forced by temperature, it can be expected to be $\mathcal{O}\left( \frac{1}{Re} \right)$ even when $Ra \neq 0$, similarly to the case of pure shear flows. Consider figure \ref{fig:adjoint_mode_inpt_resp} where the continuous line denotes the optimal transient growth curve, the dot-dashed line denotes the evolution of the optimal streamwise-uniform perturbation and the dashed line denotes the response to the dominant-adjoint-eigenmode-input of the $RBP$ flow at $Re = 1000$, $Ra = 1000$ and $Pr = 1$. The response exhibits transient growth up to $t_{Re} \approx \frac{1}{4}Re$ and then decays monotonically at a rate corresponding to the dominant eigenvalue which depends only on Rayleigh number. Its maximum transient growth and the time at which it occurs are $\mathcal{O}\left(G_{max}\right)$ and $\mathcal{O}\left(t_{max}\right)$, respectively. This implies that the dominant adjoint eigenmode is a good first approximation to the optimal initial condition. Hence, the dominant optimal transient growth process is in the form of streamwise-uniform convection rolls that supplement the lift-up mechanism thereby increasing the production of streamwise fluid motion.
\subsection{Effect of Prandtl number}
\label{sec:Pr_Effect}
%%\\
%%\begin{figure}[h]
%%\centering
%%\includegraphics[width = 0.7 \textwidth, keepaspectratio]{./xtras/liftup_scaling_Pr_0p1.eps}
%%\caption{Optimal Transient Growth at different Reynolds numbers in RBP when $Ra = 1000$, $Pr = 10^{-1}$, $\alpha = 0$ and $\beta = 1.558$}
%%\label{fig:liftup_scaling_Pr_0p1}
%%\end{figure}\\

%The Prandtl number is the property of a fluid that indicates its heat diffusing capacity against its ability to diffuse fluid momentum or vice versa. An increase or decrease in $Pr$ can result in decreased or increased heat diffusion, respectively, at a given viscous momentum diffusion. Hence, it can equivalently delay or advance the appearance of the buoyancy-induced convective motion that is responsible for the increased transient growth in $RBP$ and $RBC$ flows.

It was shown in section \ref{sec:NMS2}, when $ Pr = 1$, that the standard large Reynolds number scaling law of streamwise-uniform disturbances in pure shear flows is also satisfied by Boussinesq fluids in the presence of a constant cross-stream temperature gradient for all Rayleigh numbers. The same result (not presented here) has also been verified for various Prandtl numbers.

\begin{figure}[h]
\begin{flushleft}
\epsfig{file=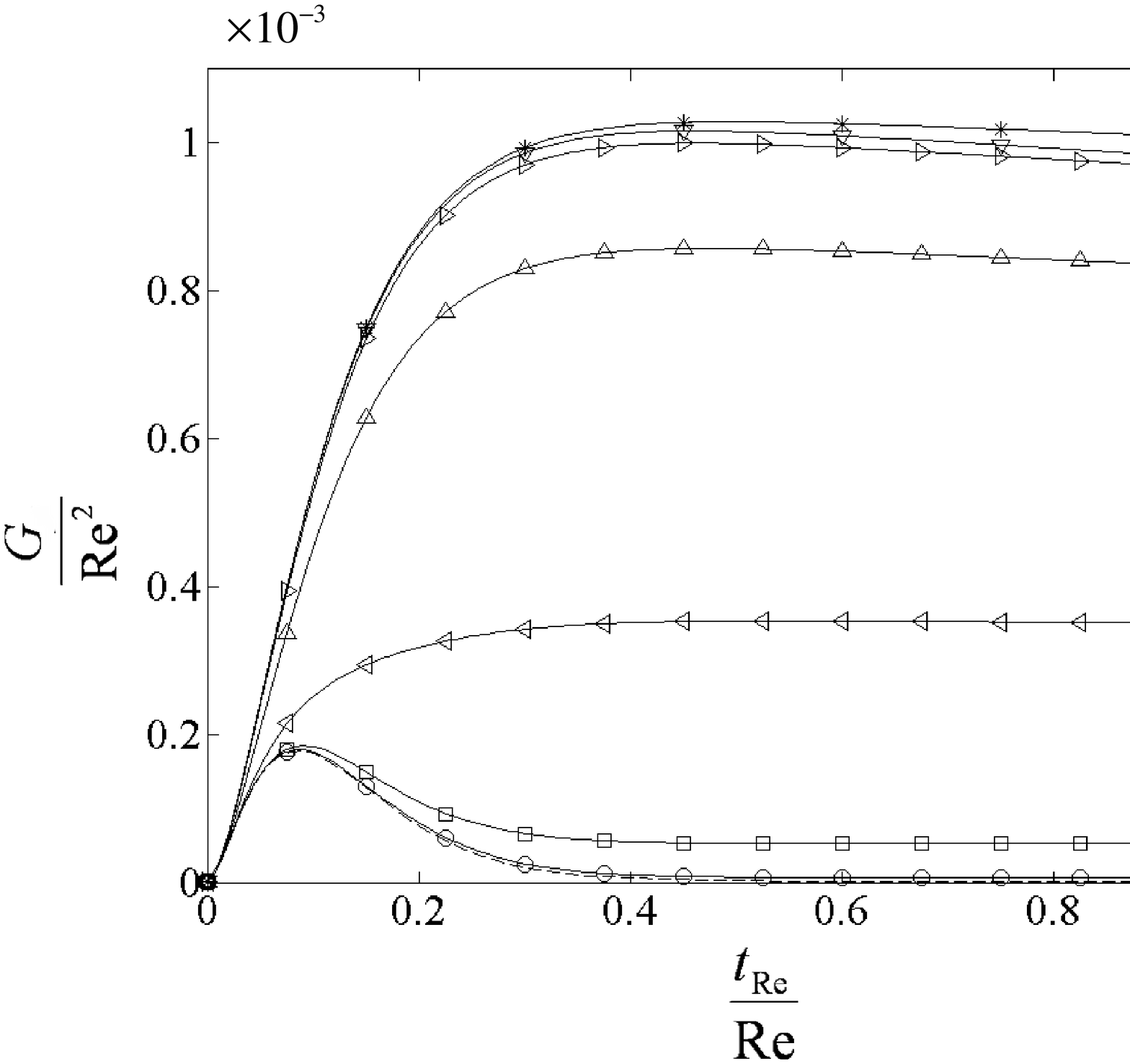,width=0.6\textwidth,keepaspectratio=true}
\end{flushleft}
\caption{Optimal gain at various Prandtl numbers when $Re = 1000$, $\alpha = 0$ and $\beta = 1.558$ for  $Ra = 0$ $(----)$ and $Ra = 1700$ $(\textbf{------})$: $\circ$ $ Pr = 10^2$, $\square$ $ Pr = 10$, $\triangleleft$ $Pr = 1$, $\triangle$ $Pr = 10^{-1}$, $\triangleright$ $Pr = 10^{-2}$, $\triangledown$ $Pr = 10^{-3}$ and $*$ $Pr = 10^{-4}$.}
\label{fig:effect_of_Pr.eps}
\end{figure}

In $RBP$ flow, the effect of Prandtl number on the transient growth of streamwise-uniform perturbations at a fixed $Ra$ is shown in figure \ref{fig:effect_of_Pr.eps} where, as previously discussed, the optimal growth $G$ and advective time scale $t_{Re}$ have been scaled with $Re^2$ and $Re$, respectively. The different symbols and the continuous line correspond to the optimal growth curves for various Prandtl numbers at large Reynolds numbers (here, $Re = 1000$) when $Ra = 1700$.  The dashed curve, which is almost identical with the continuous curve with the least maximum optimal gain ($\circ$ $ Pr = 10^2$), represents the case when $Ra = 0$. Transient growth exists for all Prandtl numbers and  the maximum optimal transient growth markedly increases with decreasing Prandtl number. As $Pr \rightarrow 0$, ${G_{max}}/{Re^2}$ asymptotically reaches a maximum about an order of magnitude larger than for $Ra = 0$. Meanwhile, the time at which it occurs, say $t_{Re}^{max}$, increases by the same factor. As $Pr \rightarrow \infty$ the maximum optimal transient growth $G_{max}$ asymptotically reaches the value for the case without temperature gradient ($Ra = 0$) at the same $Re$.
\begin{figure}[h]
\begin{flushleft}
\epsfig{file=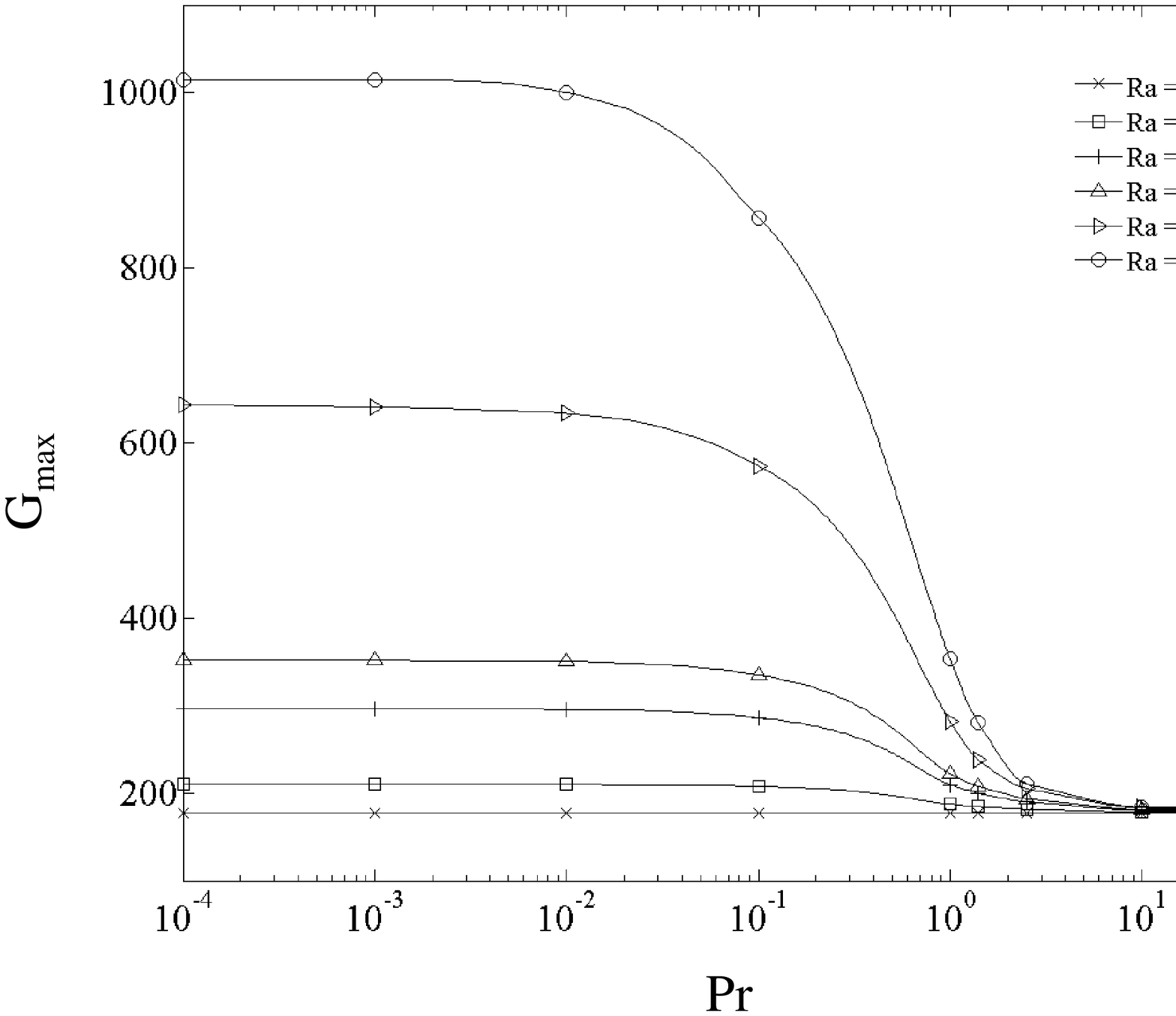,width=0.6\textwidth,keepaspectratio=true}
\end{flushleft}
\caption{Effect of Prandtl number on $G_{max}$ $\left( Re = 1000 \mbox{, } \alpha = 0 \mbox{ and } \beta = 1.558 \right)$}
\label{fig:effect_of_Pr_Gmax_beta_1p558.eps}
\end{figure}

These features are more vividly illustrated in figure \ref{fig:effect_of_Pr_Gmax_beta_1p558.eps} wherein the maximaum optimal gain $G_{max}$ for various Rayleigh numbers has been plotted against Prandtl number at $Re = 1000$ for $\beta = 1.558$. Above $Pr = 1$, all the curves collapse on the curve $Ra = 10^{-3}$ whereas for vanishing Prandtl numbers the curves are well-separated, $G_{max}$ being larger for large Rayleigh numbers. This suggests that, in a Boussinesq fluid of sufficiently large Prandtl number, the temperature gradients have negligible influence on the transient growth of a parallel shear flow.

%It is to be noted, however, that a large Prandtl number quenches only the effect of buoyancy on the transient growth of a shear flow with thermal stratification and not the inherent transient growth of the pure shear flow itself. In effect, it acts as a coupling agent between buoyancy and shear flow transient growth mechanisms.

\begin{figure}[h]
\begin{flushleft}
\epsfig{file=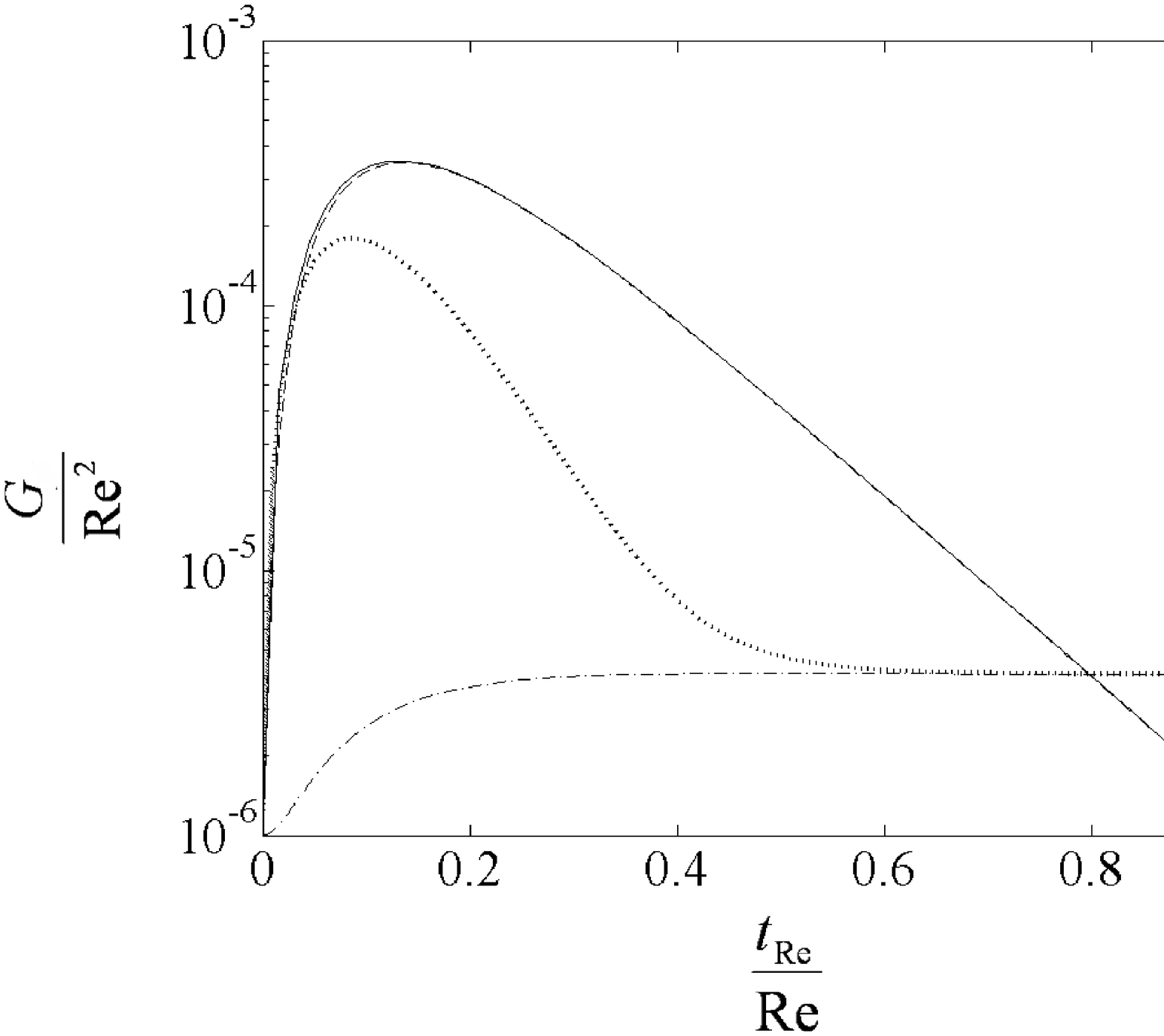,width=0.6\textwidth,keepaspectratio=true}
\end{flushleft}
\caption{Effect of Prandtl number on the optimal gain curves $\left( Pr = 10^{-3}\textbf{------}; Pr = 10^2 \cdots \cdots \right)$; Time evolution of the growth of perturbations $\frac{\left\| \textbf{\textit{q}} \right\|^2_{RB}}{\left\| \textbf{\textit{q}}_{0} \right\|^2_{RB}}$, where $\textbf{\textit{q}}_{0}$ is the normalised adjoint of the leading eigenmode $\left( Pr = 10^{-3}---; Pr = 10^2 -\cdot\cdot- \right)$ at $Re = 1000 \mbox{, } \alpha = 0 \mbox{ and } \beta = 2.04$.}
\label{fig:effect_of_Pr_Gmax_beta_2p04_adjoint_mode_inpt_resp_low_n_high_Pr}
\end{figure}

The effect of Prandtl number on the dominant transient growth mechanism in $RBP$ flow can be further illustrated by comparing the optimal gain curve with the response to the adjoint of the leading eigenmode input at large and small Prandtl numbers as shown in figure \ref{fig:effect_of_Pr_Gmax_beta_2p04_adjoint_mode_inpt_resp_low_n_high_Pr}. Here, the optimal gain curve $G(t)$ is represented by a continuous line when $Pr = 10^{-3}$ and by a dotted line when $Pr = 10^2$. The dashed line ($Pr = 10^{-3}$) and dot-dashed line ($Pr = 10^{2}$) represent the time evolution of the energy from the normalized adjoint to the leading eigenmode. This eigenmode is always the Rayleigh-B\'{e}nard convection roll irrespective of the Prandtl number. While both responses exhibit transient growth, the maximum growth exhibited by the Rayleigh-B\'{e}nard convection roll at $Pr = 10^2$ is an order of magnitude less than $G_{max}$ at the same Prandtl number. At $Pr = 10^{-3}$, the response to the Rayleigh-B\'{e}nard convection roll is amplified as much as the maximum optimal growth, although it is not the optimal initial condition at $t_{max}$. Thus, at large $Pr$, the Rayleigh-B\'{e}nard convection roll is not effectively amplified by the lift-up mechanism and vice versa at small Prandtl numbers. In effect, the Prandtl number acts as a coupling agent between buoyancy and shear flow transient growth mechanisms, as shown below.

As explained in section \ref{sec:domi_T}, the leading order transient growth process is due to the inviscid lift-up mechanism acting in tandem with the convective motion to produce large streamwise streaks $\mathcal{O}(Re)$ together with Rayleigh-B\'{e}nard rolls. Such streaks and convection rolls ultimately decay exponentially in time. The time scale at which the viscous and thermal dissipative motion can occur is $\mathcal{O}\left( l^2/\nu^* \right)$ and $\mathcal{O}\left( l^2/\kappa^* \right)$, respectively, where $l$ is the characteristic length scale (here, $l = h/2$). Note that the Rayleigh and Prandtl number may be written as 
\begin{eqnarray}
Ra = \frac{\tau_{\nu^*} \tau_{\kappa^*}}{\tau_{b}} \mbox{ and } Pr = \frac{\tau_{\kappa^*}}{\tau_{\nu^*}},
\label{eq:Rayleigh_number}
\end{eqnarray}
where $\tau_b = \sqrt{l/\alpha^* \Delta T g}$ is the buoyancy time scale, $\tau_{\nu^*} = l^2/\nu^*$ is the viscous momentum diffusion time scale and $\tau_{\kappa^*} = l^2/\kappa^*$ is the thermal diffusion time scale. Hence, at a fixed $Ra$, if $Pr << 1$ ($\tau_{\kappa^*} << \tau_{\nu^*}$), $\tau_b$ is much smaller than $\tau_{\nu^*}$ and vice versa when $Pr >> 1$. At large $Pr$ the presence of any thermal disturbance cannot be communicated swiftly across the channel before viscous dissipation begins to act and therefore, the  convective motion can no longer take place before viscous momentum diffusion has invaded the channel. As a result, any convective motion brought-in by the presence of a thermal perturbation cannot effectively complement the production and/or sustenance of streamwise motion. Overall, the effect of large $Pr$ is to hamper the influence of any temperature perturbation on the lift-up mechanism. At low Prandtl numbers, however, any temperature disturbance can be swiftly conveyed across the channel and, since $\tau_b << \tau_{\nu^*}$, a convective motion can be set-up immediately which produces streamwise velocity through the lift-up mechanism. 

Viscous forces become active at non-dimensional times $\mathcal{O}\left(Re \right)$ for all $Pr$ and hence, in the stable region of the $Re$-$Ra$ plane, any perturbation should decay at large time. Thus, in spite of the potential enhancement effect of the buoyancy-induced convection, at $t \sim \mathcal{O}\left( Re \right)$ any perturbation should eventually decay under the action of viscous forces. Viscous forces and thermal diffusion then lead to the dissipation of the convective motion. Furthermore, viscous diffusion also dissipates the so-formed streamwise streaks. The reason why $t_{Re}^{max}$ increases only marginally with decreasing Prandtl number is owing to the fact that the increased production of streamwise streaks induced by the coupling between the convective motion and the lift-up mechanism is always overtaken by viscous dissipation at a finite non-dimensional time $t \sim \mathcal{O}\left( Re \right)$.
\subsection{Effect of the norm $\left\| \textbf{\textit{q}}\right\|_{\gamma}$}
\label{norm_effect}
In this section the transient growth computations for the class of norms $\gamma \neq \sqrt{|Ra_{\mbox{\tiny{\textsl{h/2}}}}|Pr}$ are considered. Only a few key results are discussed.

\begin{figure}[h]	
	\begin{flushleft}
	\epsfig{file=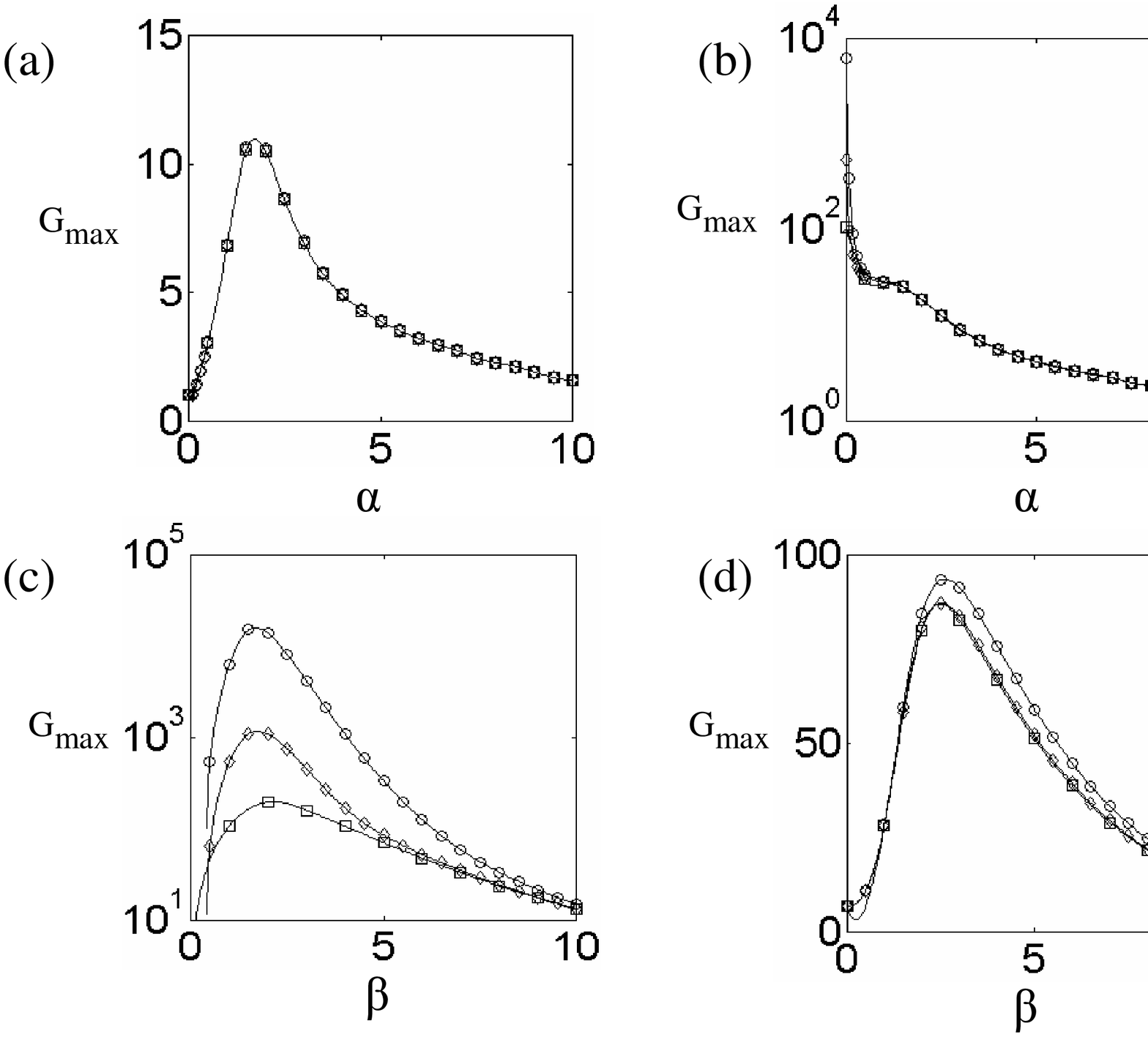,width=0.6\textwidth,keepaspectratio=true}
	\end{flushleft}
	\caption{Optimal gain computations for the norm $\left\| \textbf{\textit{q}} \right\|_{\gamma}$ with $\gamma = 1$ at $Re = 1000$, $Pr = 1$ and $\square$ $ Ra = 0$, $\diamond$ $ Ra = 500$, $\circ$ $ Ra = 1500$: (a) $\beta = 0$, (b) $\beta = 1$, (c) $\alpha = 0$ and (d) $\alpha = 1$}
	\label{fig:Gmax_vs_alpha_beta_L2}
\end{figure}

Let us consider the case when $\gamma = 1$. Figure \ref{fig:Gmax_vs_alpha_beta_L2} displays the effect of Rayleigh number on the optimal gain for the norm $\left\| \textbf{\textit{q}} \right\|_{\gamma = 1}$. Results are shown for various streamwise wavenumbers at $\beta = 0$ (figure \ref{fig:Gmax_vs_alpha_beta_L2}a) and $\beta = 1$ (figure \ref{fig:Gmax_vs_alpha_beta_L2}b) and various spanwise wavenumbers at $\alpha = 0$ (figure \ref{fig:Gmax_vs_alpha_beta_L2}c) and $\alpha = 1$ (figure \ref{fig:Gmax_vs_alpha_beta_L2}d). The symbols correspond to different Rayleigh numbers ($\square$ $ Ra = 0$, $\diamond$ $ Ra = 500$ and $\circ$ $ Ra = 1500$). In figures \ref{fig:Gmax_vs_alpha_beta_L2}a and \ref{fig:Gmax_vs_alpha_beta_L2}b these symbols collapse onto a single curve indicating that the Rayleigh number has very little effect on such perturbations. When $\alpha = 0$ and $\alpha = 1$, however, $G_{max}$ is larger at large Rayleigh number for a range of spanwise wavenumbers. On comparing with the results for the case $\gamma = \sqrt{|Ra_{\mbox{\tiny{\textsl{h/2}}}}|Pr}$ (figure \ref{fig:Gmax_contourplot}), the maximum optimal gain is larger for $\gamma = 1$. However, as in the case $\gamma = \sqrt{|Ra_{\mbox{\tiny{\textsl{h/2}}}}|Pr}$, the effect of Rayleigh number is primarily limited to streamwise-uniform and nearly-streamwise-uniform disturbances only.

%For $\gamma = 1, \sqrt{10}$, the computations showed qualitatively similar results as presented in sections \ref{sec:NMS1} \& \ref{sec:NMS2} for $\gamma = \sqrt{|Ra_{\mbox{\tiny{\textsl{h/2}}}}|Pr}$. For example, see figure \ref{fig:Gmax_vs_alpha_beta_L2} wherein the maximum optimal transient growth $G_{max}$ is displayed as a function of the streamwise and spanwise wavenumbers for different Rayleigh numbers at $Re = 1000$. Here, $\gamma = 1$. It is seen that $G_{max}$ does not vary with $Ra$ if $\alpha \neq 0$. By contrast, the growth $S$ is larger than that obtained when $\left\| \textbf{\textit{q}} \right\|_{RB}$ is the norm selected. As observed for the case $\gamma = \sqrt{|Ra_{\mbox{\tiny{\textsl{h/2}}}}|Pr}$, however, the effect of Rayleigh number is primarily limited to streamwise-uniform and weakly-streamwise-uniform disturbances only.

\begin{figure}[h]
\vspace{5pt}
\begin{flushleft}
\epsfig{file=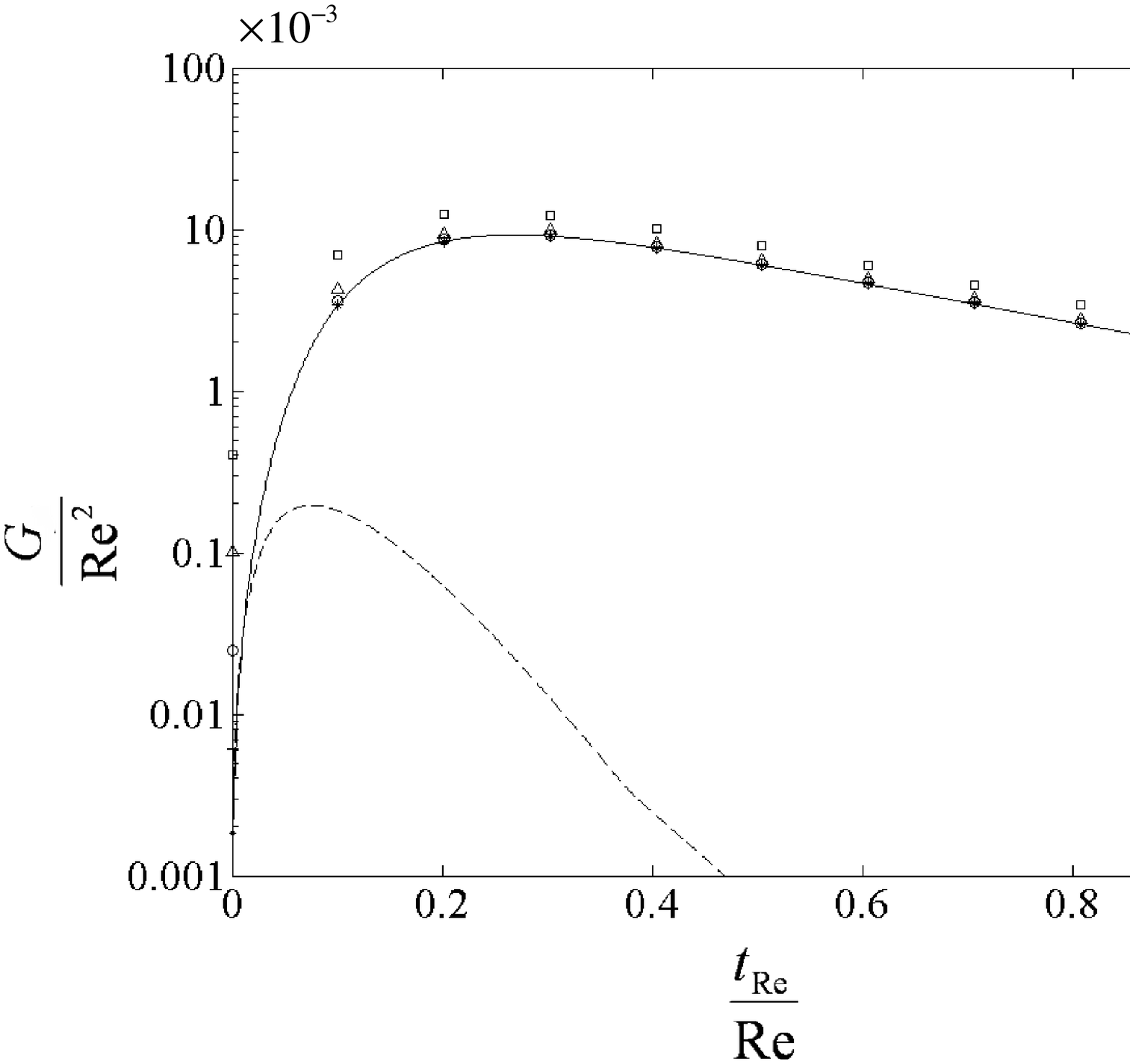,width=0.6\textwidth,keepaspectratio=true}
\end{flushleft}
\caption{Optimal gain in RBP as computed for the norm $\left\| \textbf{\textit{q}} \right\|_{\gamma}$ for $\gamma = 1$ at $Ra = 1300$ ($\alpha = 0$, $\beta = 2.04$): $\textbf{------}$ $Re = 5000$, $\times$ $ Re = 2000$, $\cdot$ $ Re = 1000$,  $+$ $ Re = 500$, $\triangle$ $ Re = 200$, $\circ$ $ Re = 100$, $\square$ $Re = 50$ and $---$ $Ra = 0$, $Re = 5000$}
\label{fig:lift_up_scaling_Poiseuille_normL2}
\end{figure}

Figure \ref{fig:lift_up_scaling_Poiseuille_normL2} shows the variations of optimal growth $G(t)$, with $\gamma = 1$, of streamwise-uniform disturbances for different Reynolds numbers ($\times$ $ Re = 2000$, $\cdot$ $ Re = 1000$,  $+$ $ Re = 500$, $\triangle$ $ Re = 200$, $\circ$ $ Re = 100$ and $\square$ $Re = 50$) at $Ra = 1300$ and $Pr = 1$. The continuous line and the dashed line correspond to the case $Re = 5000$ at $Ra = 0$ and $Ra = 1300$, respectively. When compared with the case of zero temperature difference (dashed line) the maximum transient growth is seen to be almost an order of magnitude larger at $Ra = 1300$. The collapse of all the symbols onto a single continuous curve ($Re = 5000$) as the Reynolds number increases, demonstrates that the large Reynolds number scaling holds also for the case $\gamma = 1$. This was observed for various non-zero values of the weight $\gamma$ (data not shown). A comparison between figure \ref{fig:lift_up_scaling_Poiseuille_normL2} and figure \ref{fig:lift_up_scaling_Poiseuille} confirms that the optimal gain is larger when $\gamma = 1$.
Note that, when $\gamma = \sqrt{|Ra_{\mbox{\tiny{\textsl{h/2}}}}|Pr}$, the coupled operator governing the wall-normal velocity component $\tilde{v}(y, t)$ and the temperature $\tilde{\theta} (y, t)$ for a streamwise-uniform perturbation ($\alpha = 0$) is normal with respect to the scalar product \eqref{eq:scalar_product}. If $Ra < Ra_c^{RB}$, its spectrum lies in the lower half-plane for all wavenumbers and hence, the Hille-Yosida theorem \cite{Reddy_n_Henningson_1993} states that the vector $\begin{bmatrix} \tilde{v}(y, t), \tilde{\theta} (y, t) \end{bmatrix}^T$ cannot exhibit transient growth. Whereas, for any $\gamma \neq \sqrt{\left| Ra_{\mbox{\tiny{\textsl{h/2}}}} \right| Pr}$, this operator is no longer normal with respect to the scalar product \eqref{eq:scalar_product}. Since the spectrum lies in the lower half-plane for all wavenumbers, the increase in the optimal transient growth $G(t)$ when $\gamma \neq \sqrt{\left| Ra_{\mbox{\tiny{\textsl{h/2}}}} \right| Pr}$ can only come from the non-normal block of system \eqref{eq:LOBE_matrix}. Thus, when $\gamma \neq \sqrt{\left| Ra_{\mbox{\tiny{\textsl{h/2}}}} \right| Pr}$, $G(t)$ can be very large depending on the weight $\gamma$ and the increase in $G(t)$ is due to the presence of off-diagonal terms corresponding to the forcing of wall-normal velocity by temperature perturbations, and vice versa, that render the governing equations explicitly non-normal.

Using the transformation
\begin{equation}
\begin{bmatrix}
	\hat{v}_{A}\\
	\hat{\theta}_{*A}\\
	\hat{\eta}_{A}
\end{bmatrix}
 = 
\begin{bmatrix}
	1 	&0 																						&0\\
	0 	&\frac{\left| Ra_{\mbox{\tiny{\textsl{h/2}}}} \right| Pr}{\gamma} 	&0\\
	0 	&0																						&1
\end{bmatrix}
\begin{bmatrix}
	\hat{v}_{A}\\
	\frac{\gamma \hat{\theta}_{*A}}{\left| Ra_{\mbox{\tiny{\textsl{h/2}}}} \right| Pr}\\
	\hat{\eta}_{A}
\end{bmatrix},
\label{eq:trans_vari_gamma}
\end{equation}
the adjoint system \eqref{eq:LOBEadjoint_matrix} for streamwise-uniform perturbations becomes
\begin{multline}
		-i\omega^*
		\begin{bmatrix}
			-D_\beta^2	&0	&0\\
			0	&1	&0\\
			0	&0	&1
		\end{bmatrix} 
		\begin{bmatrix}
			\hat{v}_{A}\\
			\frac{\gamma \hat{\theta}_{*A}}{\left| Ra_{\mbox{\tiny{\textsl{h/2}}}} \right| Pr}\\
			\hat{\eta}_{A}
		\end{bmatrix}
		\\=
		\begin{bmatrix}
			Pr D_\beta^4		&-\beta^2 \left| Ra_{\mbox{\tiny{\textsl{h/2}}}} \right| Pr 	&-i\beta \left( Re Pr \right)\frac{dU_0}{dy}\\
			-1							&-D_\beta^2															&0\\
			0 							&0																			&-Pr D_\beta^2
		\end{bmatrix}
		\begin{bmatrix}
			\hat{v}_{A}\\
			\frac{\gamma \hat{\theta}_{*A}}{\left| Ra_{\mbox{\tiny{\textsl{h/2}}}} \right| Pr}\\
			\hat{\eta}_{A}
		\end{bmatrix},
	\label{eq: L_longi_gamma}
\end{multline}
%the linear operator \eqref{eq: L_longi_gamma}, becomes the same as that in LOBE \eqref{eq:LOBE_fourier1}, \eqref{eq:LOBE_fourier2} and \eqref{eq:LOBE_fourier3}.
where $D_\beta^2 = D^2-\beta^2$. Thus, if $\textbf{\textit{q}}_{\left( RB \right)} = \left[ \hat{v}_{\left( RB \right)}, \hat{\theta}_{\left( RB \right)}, 0 \right]^T$ denotes the leading Rayleigh-B\'{e}nard mode of the pure conduction problem, the adjoint of the leading eigenmode, for any arbitrary norm $\left\| \textbf{\textit{q}} \right\|_{\gamma}$, is given by
\begin{equation}
\textbf{\textit{q}}_{A\left( 1 \right)}^{\left( \gamma \right)}
 = 
\begin{bmatrix}
	\hat{v}_{\left( RB \right)}\\
	\frac{1}{\gamma_0^2}\hat{\theta}_{\left( RB \right)}\\
	0
\end{bmatrix}, \mbox{ where } \gamma_0 = \frac{\gamma}{\sqrt{\left| Ra_{\mbox{\tiny{\textsl{h/2}}}} \right| Pr}}.
\label{eq:domi_adj_gamma}
\end{equation}
The dominant adjoint velocity and temperature eigenfunctions are seen to be identical to the Rayleigh-B\'{e}nard mode except for a multiplicative constant $\gamma_0$ in the temperature eigenfunction. Note that $\gamma_0$ depends on the type of norm through $\gamma$ and it is equal to unity for the norm $\left\| \textbf{\textit{q}} \right\|_{RB}$ which makes the governing equations self-adjoint at $Re = 0$.

\begin{figure}[h]
\begin{flushleft}
\epsfig{file=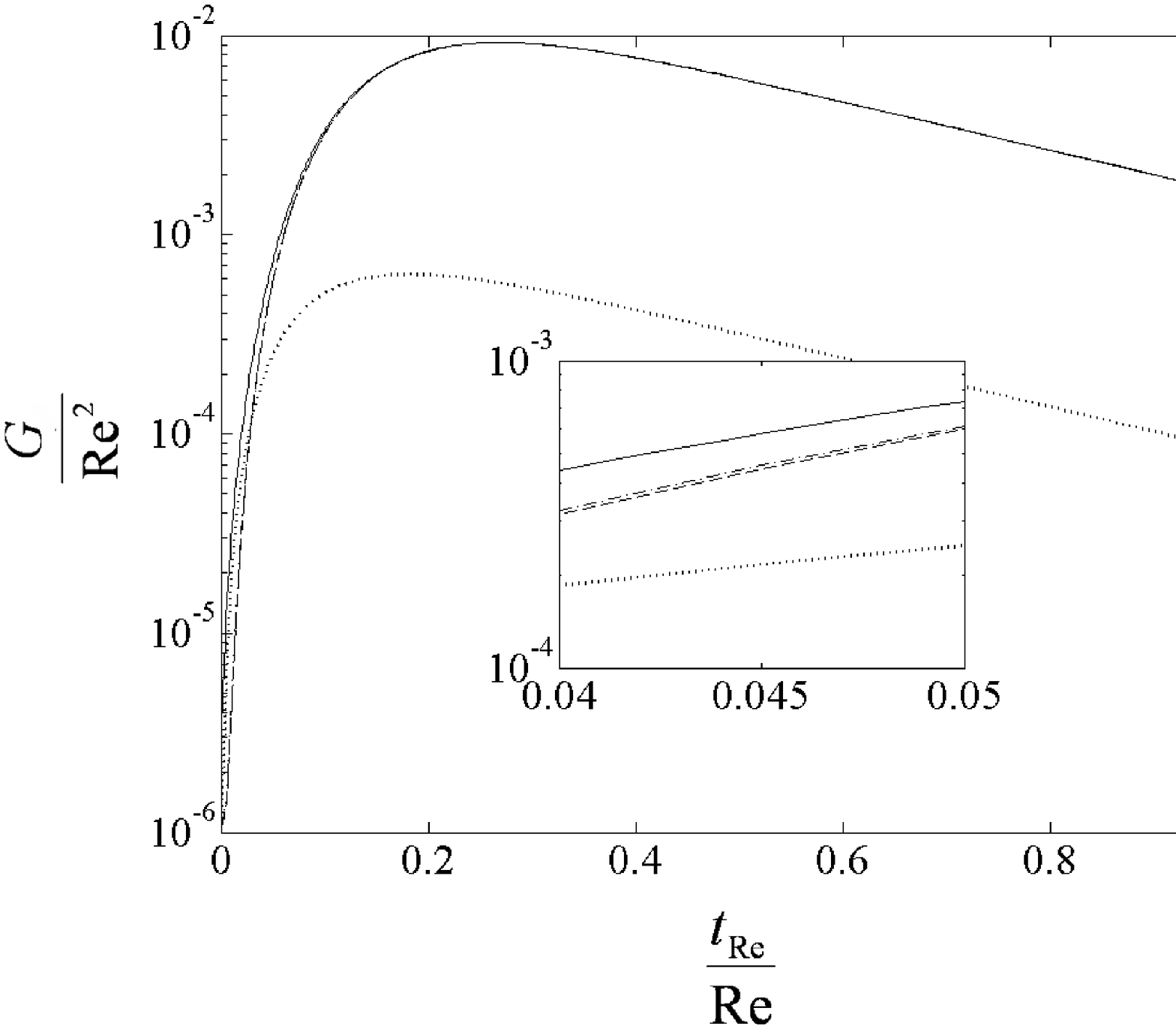,width=0.6\textwidth,keepaspectratio=true}
\end{flushleft}
\caption{Optimal gain curve $\left( \textbf{------} \right)$ and time evolution of the growth of perturbations $\frac{\left\| \textbf{\textit{q}} \right\|^2_{\gamma = 1}}{\left\| \textbf{\textit{q}}_{0} \right\|^2_{\gamma = 1}}$, where $\textbf{\textit{q}}_{0}$ is the normalized adjoint of the leading eigenmode $\left(  ---- \right)$, the optimal initial condition corresponding to the maximum optimal gain $\left(-\cdot-\cdot-\right)$ and the normalized Rayleigh-B\'{e}nard mode without its streamwise velocity component $\left( \cdots \cdots \right)$ at $Re = 1000$, $Ra = 1300$, $Pr = 1$, $\alpha = 0$ and $\beta = 2.04$ for $\gamma = 1$.}
\label{fig:adjoint_mode_inpt_resp_L2}
\end{figure}

The response of $RBP$ flow to various inputs at $Re = 1000$, $Ra = 1300$ and $Pr = 1$ is shown in figure \ref{fig:adjoint_mode_inpt_resp_L2}. The continuous line denotes the optimal gain curve, the dashed line denotes the evolution of the optimal streamwise-uniform perturbation which grows up to $G_{max}$, the dot-dashed line denotes the response to  the adjoint of the leading eigenmode \eqref{eq:domi_adj_gamma} and the dotted line denotes the response to the Raleigh-B\'{e}nard mode of the pure conduction problem. Here, $G(t)$ has been computed based on the norm $\left\| \textbf{\textit{q}} \right\|_{\gamma}$ with $\gamma = 1$. All the initial conditions display transient growth and the dominant-adjoint-mode is amplified as much as the optimal input. Again,  the adjoint of the leading eigenmode is a good approximation to the optimal initial condition. Note that the same conclusion was reached in section \ref{sec:domi_T} where the norm $\left\| \textbf{\textit{q}} \right\|_{RB}$ was selected to compute the optimal response. This has also been verified for several values of $\gamma$ (not presented here). 

Thus, in general, \textit{the dominant optimal transient growth mechanism, irrespective of the selected norm $\left\| \textbf{\textit{q}} \right\|_{\gamma}$, consists of two processes. The short-time optimal is due to the convective vortex motion, in the form of a ``modified" Rayleigh-B\'{e}nard mode given by \eqref{eq:domi_adj_gamma}, which acts in tandem with the inviscid lift-up mechanism, thereby resulting in large streamwise velocity streaks.} The long-time optimal simply consists of the transiently amplified Rayleigh-B\'{e}nard convection roll. It either decays or grows in time depending on the magnitude of $Ra$.
\section{Summary and Conclusion}
\label{sec:Rem_n_Sum}
The effect of unstable thermal stratification on the linear stability properties of plane Poiseuille flow and plane Couette flow has been summarized for a wide range of Reynolds  numbers $Re$ and Rayleigh numbers $Ra$. The scaling laws for spanwise-uniform modes governing the dependence of the critical Rayleigh number and critical wavenumber on the corresponding critical Reynolds number have been obtained.

It was demonstrated that these flows are susceptible to large and sustained transient growth for a wide range of Reynolds and Rayleigh numbers at all Prandtl numbers. It was observed that unstable stratification, in RBP and RBC flows, increases the maximum optimal transient growth and maintains such a growth over a longer period of time. The maximum optimal transient growth $G_{max}$ and the corresponding time at which it occurs $t_{max}$ remain of the same order of magnitude as in the case of pure shear flows. In particular, the increase in $G_{max}$ is more effective for streamwise-uniform disturbances. Unlike the computations by Sameen et. al. \cite{Sameen_n_Govindarajan_2007}, spanwise-uniform disturbances were never observed to be the dominant optimal behavior in $RBP$ flow at any Rayleigh or Prandtl number. The optimal spanwise wavenumber varies between the value for pure shear flow and that for the most unstable Rayleigh-B\'{e}nard convection mode as the Rayleigh number increases towards the critical Rayleigh number in $RBI$. The large Reynolds number scaling laws, such as, $G_{max} \propto Re^2$ and $t_{max} \propto Re$, were shown to remain valid in both $RBP$ and $RBC$ flows for all Rayleigh and Prandtl numbers.

The associated dominant growth mechanisms for the production of streamwise velocity streaks in the presence of an unstable temperature gradient were identified. A 3D vector model of the governing equations was used to demonstrate that the \textit{short-time} behavior is governed by the inviscid lift-up mechanism and that the effect of Rayleigh number on this mechanism is secondary and negligible. By contrast, the optimal initial condition for the largest \textit{long-time} response is given by the Rayleigh-B\'{e}nard mode without its streamwise velocity component. It was established that such a disturbance sets up streamwise-uniform convection rolls with no streamwise velocity component which act in tandem with the inviscid lift-up mechanism to produce and sustain streamwise motion in the form of streaks. A good approximation to the optimal initial condition was shown to be the dominant-adjoint-eigenmode, namely, the $RB$ mode with zero streamwise velocity.

It was shown that the Prandtl number $Pr$ of a Boussinesq fluid plays an important role in the coupling between temperature perturbations and the lift-up mechanism. At large $Pr$ for a given Rayleigh number $Ra$, the convection rolls cannot take place before the viscous diffusion process and hence, the short-time optimal transient growth is similar to the case without cross-stream temperature gradient: the classical inviscid lift-up mechanism without the thermal convective motion. Whereas, at small $Pr$, the convection rolls can effectively couple with the lift-up mechanism, thereby resulting in large transient growth.

An analysis of the direct and adjoint equations \eqref{eq:LOBE_matrix} and \eqref{eq:LOBEadjoint_matrix} revealed that the resulting transient growth depends on the type of norm selected. Thus, for the norm $\left\| \textbf{\textit{q}} \right\|_\gamma$, optimal growth $G(t)$ can vary largely as a function of $\gamma$ and, when $\gamma \neq \sqrt{\left| Ra_{\mbox{\tiny{\textsl{h/2}}}} \right| Pr}$, the increase in $G(t)$ is due to the off-diagonal terms that render the governing equations explicitly non-normal. It was shown, however, that the dominant mechanism of transient growth is independent of the norm used to quantify it. 

Experimental estimates of the transient growth in a horizontal fluid layer heated from below in the presence of laminar shear flow are not presently available and the present work is expected to motivate such experiments.
\begin{acknowledgments}
Special thanks to Peter Schmid, Yongyun Hwang, Xavier Garnaud and Cristobal Arratia at LadHyX for many helpful discussions. J J S J gratefully acknowledges the financial support of the ``Direction des Relations Ext\'{e}rieures" of \'Ecole Polytechnique.
\end{acknowledgments}

\section*{Appendix: Short-time dynamics and the lift-up mechanism}
\label{sec:Appendix}
It is shown that the short-time evolution of streamwise perturbation velocity is linear in time and is independent of Rayleigh and Prandtl numbers at very large Reynolds numbers. The arguments presented here are similar to those in Ellingsen et al \cite{Ellingsen_n_Palm_1975}.

If the advective time scale (as in eqn. \eqref{eq:time_scaling}) had been used, instead of the time scale and velocity scale based on the thermal diffusion coefficient, eqns. \eqref{eq:LOBE1} - \eqref{eq:LOBE3} would have been
\begin{eqnarray}
\nabla\cdot\vec{\breve{u}}& =0,
\label{eq:LOBE1_new}
\end{eqnarray}
\begin{eqnarray}
\left(\frac{\partial}{\partial \breve{t}}+ U_0 \frac{\partial}{\partial x} \right)\vec{\breve{u}} + \breve{v} \frac{dU_0}{dy}\vec{e}_x = -\nabla \breve{p} + \frac{Ra_{\mbox{\tiny{\textsl{h/2}}}}/Pr}{Re^2} \breve{ \theta} \vec{e}_y +  \frac{1}{Re}\nabla^2 \vec{\breve{u}} ,
\label{eq:LOBE2_new}
\end{eqnarray}
\begin{eqnarray}
\left(\frac{\partial}{\partial \breve{t}}+ U_0 \frac{\partial}{\partial x} \right)\breve{\theta} + \breve{v} \frac{d \Theta_0}{dy} &= \frac{1}{Re Pr} \nabla^2 \breve{\theta},
\label{eq:LOBE3_new}
\end{eqnarray}
where the new variables are $\breve{t} = (Re Pr) t$, $\vec{\breve{u}} = \frac{1}{Re Pr}\vec{{u}}$, $\breve{\theta} = \theta$, and $\breve{p} = \frac{1}{(Re Pr)^2} p$. Note that $\breve{t}$ is the same as the time $t_{Re}$ as used in section \ref{sec:NMS2}. If $\breve{\psi}(y, z; t)$ represents the stream function in the horizontal $y$-$z$ plane then
\begin{eqnarray}
\breve{v} = -\frac{\partial \breve{\psi}}{\partial z} \mbox{ and } \breve{w} = \frac{\partial \breve{\psi}}{\partial y}.
\label{eq:v_n_w_in_psi}
\end{eqnarray}
The governing equation of the streamwise velocity component written in terms of the stream function becomes
\begin{eqnarray}
\frac{\partial}{ \partial \breve{t}} \nabla_h^2 \breve{\psi} = -\frac{Ra_{\mbox{\tiny{\textsl{h/2}}}}/Pr}{Re^2} \frac{\partial \breve{ \theta}}{\partial z} +  \frac{1}{Re}\nabla_h^4 \breve{\psi},
\label{eq:u_in_psi}
\end{eqnarray}
where $\nabla_h^2 = \frac{\partial^2}{\partial y^2} + \frac{\partial^2}{\partial z^2}$. When $Re >> 1$, the R.H.S. of eqn. \eqref{eq:u_in_psi} becomes negligibly small $\mathcal{O}(1/Re)$. Thus, in an inviscid flow (or equivalently for $\breve{t} << Re$) $\breve{\psi}$ is independent of time. This implies that the rescaled wall-normal velocity $\breve{v}$ and spanwise velocity $\breve{w}$ are constant for all $\breve{t} << Re$. The streamwise velocity grows linearly with time. Also, $\breve{v}$ and  $\breve{w}$ do not depend on any control parameters, namely, Reynolds number, Rayleigh number and Prandtl number. Hence, \textit{at short times, the linear growth in streamwise velocity is directly related to the classical lift-up mechanism as in pure shear flows}. To compute the inviscid optimal growth curves, one can either pose a separate eigenvalue problem without any control parameters (as in section V A of Malik et al.\cite{Malik_Dey_n_Alam_2008} for the case of compressible plane Couette flow) or simply increase the Reynolds number and consider the asymptotic large Reynolds number growth curve. In the results presented in figures \ref{fig:shape_of_opt_curve_Re_1000_Ra_various} and \ref{fig:shape_of_opt_curve_Re_2000_Ra_various_RBC}, the latter approach is used to numerically compute the short time inviscid optimal growth.

\nocite{*}
\bibliography{Transient_Growth_in_Rayleigh_Benard_Poiseuille_Couette_convection_aip_preprint}
\end{document}